%% file: main.tex
\documentclass[11pt]{article}

\usepackage[a4paper, margin=1in]{geometry}
\usepackage{amsmath, amssymb}
\usepackage{graphicx}
\usepackage{booktabs}
\usepackage{longtable}
\usepackage{array}
\usepackage{caption}
\usepackage{subcaption}
\usepackage{authblk}
\usepackage[hidelinks]{hyperref}
\usepackage{tabularx}
\usepackage[table]{xcolor}
\usepackage{multirow}
\usepackage{float}
\usepackage[most]{tcolorbox}
\usepackage{fvextra}
\usepackage{placeins}

\definecolor{TblHeader}{gray}{0.9}
\definecolor{TblGroup}{gray}{0.93}
\definecolor{TblStripe}{gray}{0.97}

\title{ECG-LLM: Foundation Model for ECG-Based Cardiac Reasoning}
\author[1]{Alexander Selivanov\thanks{\texttt{alexander.selivanov@tum.de}}}
\author[1,3]{Friederike Jungmann}
\author[2]{Jan Kehrer}
\author[2]{Karl-Ludwig Laugwitz}
\author[2]{Eimo Martens}
\author[1,4,5]{Daniel Rueckert}

\affil[1]{Chair for AI in Healthcare and Medicine, Technical University of Munich (TUM) and TUM University Hospital, Germany}
\affil[2]{Department of Cardiology, TUM University Hospital, Germany}
\affil[3]{Institute for Diagnostic and Interventional Radiology, TUM University Hospital, Technical University of Munich (TUM), Munich, Germany}
\affil[4]{Department of Computing, Imperial College London, UK}
\affil[5]{Munich Center for Machine Learning (MCML), Germany}
\date{\vspace{-5ex}}

\begin{document}
\maketitle

\begin{abstract}
\input{sections/abstract}
\end{abstract}

\section{Introduction}
\input{sections/introduction}

\section{Results}
\input{sections/results}

\section{Discussion}
\input{sections/discussion}

\section{Methods}
\input{sections/methods}

\section*{Data availability}
\input{sections/data_availability}

\section*{Code availability}
\input{sections/code_availability}

\section*{Acknowledgments}
\input{sections/acknowledgments}

\section*{Author contributions}
\input{sections/author_contributions}

\section*{Competing interests}
\input{sections/competing_interests}

\bibliographystyle{unsrt}
\bibliography{references}

\newpage
\input{sections/supplementary}
\end{document}

%% file: sections/abstract.tex
Electrocardiography (ECG) is an inexpensive, standard-of-care test for cardiac symptoms, but front-line triage often lacks immediate access to definitive imaging such as echocardiography (ECHO) or cardiac magnetic resonance (CMR). Furthermore, most existing ECG-AI systems are limited to fixed diagnostic labels or automated reports, constraining their use for patient-specific clinical reasoning. To address this gap, we introduce ECG-LLM, an ECG-conditioned large language model trained across four cohorts comprising 679,112 ECG studies from 186,409 patients. Using a novel multimodal-to-language supervision strategy, ECG-LLM is trained on clinically structured question-answer pairs derived from ECG signals, clinical context, CMR, and ECHO. This unified approach enables the model to answer diverse cardiovascular questions from a 12-lead ECG alone, spanning both conventional interpretation and phenotypes not directly visible on standard ECGs. ECG-LLM successfully recovers conventional ECG measurements, such as heart rate, and strongly predicts complex CMR-derived phenotypes, including ventricular and atrial volumes and ventricular function. Crucially, it detects vital echocardiographic phenotypes, including increased LV wall thickness, aortic stenosis, and right-ventricular systolic dysfunction. On standard ECG understanding tasks, ECG-LLM matches or exceeds existing baselines for diagnostic report generation and the ECG-QA benchmark. By moving beyond fixed-label prediction, this multimodal framework provides clinically valuable, question-driven cardiovascular reasoning to support general practitioner and front-line triage decisions when specialist review is delayed.

%% file: sections/introduction.tex
Cardiovascular diseases (CVDs) remain the leading cause of death worldwide~\cite{cvd1, cvd2}, yet early detection in front-line care and general practitioner (GP) practice is often constrained by limited access to advanced diagnostic tools such as imaging. In primary care settings, triage decisions frequently rely on clinical history and a 12-lead electrocardiogram (ECG) as the first objective cardiac test~\cite{ecg_freq}. While imaging modalities such as echocardiography (ECHO) and cardiac magnetic resonance (CMR) provide the "gold standard" quantitative assessment of cardiac structure and function~\cite{echo_hf_lv_sd, echo_pulm_htn, echo_rhf_review, Salerno2017-cmr, Mayala2019-cmr, echo_cmr_hcm_review}, they require specialized equipment, trained operators, and expert interpretation. They are also often limited by high costs, long waiting lists, and, in the case of CMR, longer workflows and patient contraindications. This creates a practical clinical limitation: first-line triage often relies on the immediately available ECG, while the imaging-grade information needed for clinical decisions may be delayed until scheduled echocardiography or CMR is performed.

\begin{figure*}[t]
    \centering
    \includegraphics[width=\textwidth]{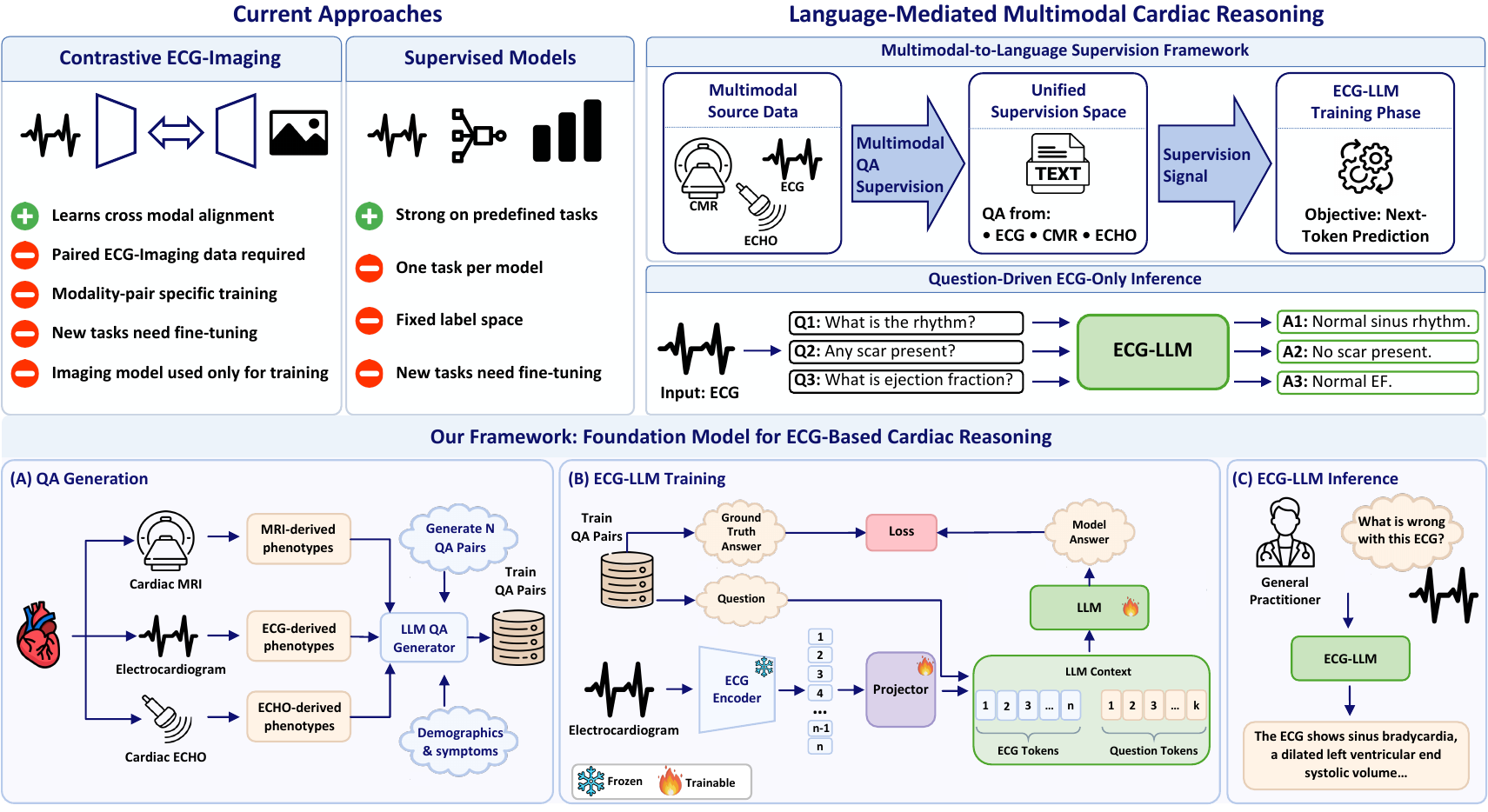}
    \caption{
    Overview of language-mediated multimodal cardiac reasoning. 
    Upper panels: existing ECG AI approaches and our proposed multimodal QA supervision framework. 
    Lower panels: (A) generation of QA pairs from ECG, CMR, ECHO, and clinical metadata; (B) ECG-LLM training with ECG tokens and question--answer supervision; (C) ECG-only inference for free-text clinical reasoning.
    }
    \label{fig:concept_overview}
\end{figure*}

In this work, we address this gap with ECG-LLM, an ECG-conditioned large language model (LLM) that operates directly on the raw multichannel 12-lead ECG waveform, rather than on a rendered ECG image, and enables free-text cardiac question answering from the ECG alone. Our central hypothesis is that ECG waveforms, echocardiography, CMR, and clinical labels/reports are different observations of a shared underlying cardiovascular state. We therefore propose to use natural language as a unified supervision space for representing this state. Natural language provides a dense and flexible supervision format: it can describe clinically meaningful findings that rely on multiple tests, make heterogeneous information comparable through question--answer (QA) pairs, and match the natural way clinicians ask questions and communicate diagnoses. Instead of training separate models to predict a limited set of predefined categories, as in standard supervised approaches, we use language-based QA supervision to train a single ECG-conditioned LLM to answer clinical questions about both ECG findings and broader imaging-related cardiac phenotypes. Figure~\ref{fig:concept_overview} summarizes this transition from conventional ECG AI toward language-mediated multimodal cardiac reasoning. The upper panels contrast existing AI approaches with our multimodal QA supervision framework, while the lower panels show how multimodal cardiac information is converted into QA supervision, used to train ECG-LLM, and applied for ECG-only clinical reasoning at inference time. 

We build this framework across four large cohorts: UK Biobank~\cite{ukbiobank}, MIMIC-IV-ECG~\cite{mimic_ecg}, PTB-XL~\cite{ptbxl}, and EchoNext~\cite{echonext}. For each ECG study, we collect structured metadata describing all available ECG measurements and labels, imaging-derived phenotypes, patient demographics, symptoms, and clinical context when available. We then use this metadata to generate over 5 million clinically grounded QA pairs, formulated as natural GP-style queries about ECG findings and broader cardiac phenotypes. Using these QA pairs, we train ECG-LLM, an autoregressive large language model conditioned on a 12-lead ECG raw signal and available patient context, to answer cardiac questions in free text. At inference time, ECG-LLM operates solely on the 12-lead ECG and patient context, without echocardiography, CMR, or task-specific classification heads, supporting ECG-based triage and second-opinion use in settings where imaging is delayed or unavailable. We describe the overall framework in more detail in Section~\ref{sec:method}, the construction of clinical QA supervision from cardiac records in Section~\ref{sec:qa_gen}, and we show representative examples of ECG-LLM answering diverse questions about ECG findings and CMR- and ECHO-derived cardiac phenotypes in Section~\ref{sec:qualitative_evaluation}.

The motivation for ECG-LLM is grounded in a broader shift in ECG AI: deep learning models are no longer limited to detecting visible waveform abnormalities, but can also infer clinically relevant cardiac phenotypes that are not directly measured on the ECG. Deep learning has substantially improved automated ECG interpretation, evolving from direct waveform analysis toward the inference of hidden phenotypes. Early models reached expert-level performance in arrhythmia detection and multi-label diagnosis~\cite{hannun2019arrhythmia, ribeiro2020diagnosis, ecg_diagnosis}. Building on this, supervised approaches demonstrated that the 12-lead ECG can detect "indirect" structural diseases—such as aortic stenosis, reduced LVEF, and hypertrophic cardiomyopathy, as well as future arrhythmia risk~\cite{attia2019lvef, raghunath2021af, cohenshelly2021as_ecg_ai, adedinsewo2020lv_dysfunction_ed, ko2020hcm_ecg_cnn, tison2022hcm_treatment_response}. Further studies indicate that ECG representations can recover diastolic function, support unsupervised disease profiling, and act as digital biomarkers for hypertension-related risk~\cite{lee2024diastolic, friedman2025disease, alalusi2025hypertension}.

Despite this progress, most supervised ECG models are designed to answer only the specific diagnostic questions defined before training. For example, a model trained to detect reduced left ventricle ejection fraction (LVEF) or aortic stenosis usually returns only that predefined output, rather than answering a clinician's broader question in natural language. This makes such models difficult to use for open-ended triage, where the relevant question may vary across patients and clinical settings. Multimodal contrastive methods partially address this limitation by aligning ECG and imaging representations to transfer broader cardiovascular knowledge across modalities~\cite{selivanov2025ecg_cmr_contrastive, GaoYua_EchoingECG_MICCAI2025}. However, these methods usually learn representations rather than directly answering clinical questions. In practice, they still require additional training of a prediction model to extract each specific phenotype of interest. This limits their ability to support direct, question-driven inference at the point of care. Their scalability is also constrained by reliance on imaging data, which remains costly to acquire, difficult to standardize, and often restricted from sharing due to privacy and infrastructure limitations.

In parallel, multimodal vision--language models (VLMs) have moved medical AI toward conversational reasoning and free-text generation~\cite{retvlm_curriculum, chexagent2024, llavamed2023, medflamingo2023, gu2026cardiac}. Recent ECG-focused vision--language models shift toward language-based interpretation~\cite{ecgchat, PULSE_ECG, ecgdoctor, GEM_ECG}, representing an important step toward clinically usable ECG-language systems. In practice, however, these models are primarily trained to reproduce ECG reports or generate textual interpretations of waveform-level findings, tasks that largely mirror existing automated reporting systems installed in ECG machines. As a result, their outputs remain focused on ECG-derived observations and do not directly target the broader cardiac phenotypes typically assessed using imaging, nor the targeted clinical questions encountered during triage. We build on this progress by extending ECG-language supervision beyond standard ECG machine-derived output: ECG-LLM uses multimodal cardiac information as dense QA supervision so that an ECG-conditioned LLM can answer flexible, question-driven queries about both ECG findings and imaging-related phenotypes from the raw ECG signal alone. The goal is to support triage and provide a second-opinion tool in settings where imaging is unavailable, delayed, or time- or access-constrained.

In summary, our contributions are:
\begin{itemize}
\item A multimodal-to-language supervision strategy that shifts ECG-language training beyond reproducing ECG-machine outputs toward question-driven reasoning over both ECG-derived and imaging-related cardiac phenotypes.
\item A large-scale 12-lead ECG instruction-tuning resource and QA-generation pipeline, built from unified ECG measurements, diagnostic reports, ECHO and CMR phenotypes, clinical context, and 5M+ clinically realistic GP-style QA pairs. We plan to release the QA-generation pipeline and all shareable derived resources, subject to cohort-specific data-use agreements.
\item ECG-LLM, an open-source ECG-conditioned autoregressive model that directly processes raw 12-lead ECG waveforms and answers free-text cardiac questions from ECG alone, without rendered ECG images, imaging at inference, or task-specific classification heads.
\item A systematic evaluation across UK Biobank, MIMIC-IV-ECG, PTB-XL, and EchoNext, covering ECG interpretation, CMR- and ECHO-derived phenotype inference, the ECG-QA benchmark, and diagnostic report generation.
\end{itemize}

%% file: sections/results.tex
\subsection{Evaluation framework}

ECG-LLM is an autoregressive large language model (LLM) conditioned on 12-lead ECGs and trained with multimodal question--answer supervision, as described in Methods (Section~\ref{sec:method}, Figure~\ref{fig:concept_overview}). At inference, the model receives a 5-s centre crop of the 12-lead ECG recording and a natural-language question, then generates a free-text answer (Figure~\ref{fig:concept_overview}C). We evaluate this ECG-to-text LLM without retrieval-augmented generation, auxiliary classification heads, or direct access to imaging measurements. Predictions are therefore conditioned only on the learned ECG representation and the question. We intentionally exclude retrieval-augmented generation because it can shift ECG interpretation from the ECG-conditioned language model to an external retrieval or prediction module. This setup tests how much clinically relevant information can be represented and queried through a generative ECG-conditioned language model. We evaluate five capabilities: (i) ECG-related question answering, (ii) CMR-related phenotype question answering from ECG only, (iii) echocardiography-related phenotype question answering from ECG only, (iv) ECG report generation, and (v) standard benchmark performance on ECG-QA~\cite{ecgqa}.

\subsection{ECG-LLM answers ECG and CMR phenotype questions from ECG alone}

We evaluated whether ECG-LLM can answer clinical questions about both conventional ECG-derived phenotypes and CMR-derived imaging phenotypes using the ECG as the only input. This setting reflects a first-line care use case in which a clinician may ask targeted questions about rhythm, conduction, repolarisation, intervals, and imaging-derived structural or functional phenotypes, such as chamber size, ventricular function, myocardial wall thickness, strain, and possible remodelling, before deciding whether further imaging is needed. Existing ECG-language models such as ECG-Chat~\cite{ecgchat}, PULSE-ECG~\cite{PULSE_ECG}, and ECG-Doctor~\cite{ecgdoctor} represent important steps toward ECG-language interpretation, but they primarily focus on diagnostic interpretation or report generation. Here, the model is instead queried in free text about both ECG measurements and imaging-derived cardiac phenotypes. We evaluated this task on 1{,}000 UK Biobank~\cite{ukbiobank} test ECGs sampled from unique patients with non-trivial ECG device diagnoses, defined as any diagnosis other than the normal pair ''Normal sinus rhythm'' and ''Normal ECG''. This sampling avoids an evaluation dominated by completely normal ECGs, although class imbalance remains.

We organized ECG-derived and CMR-derived targets into clinically connected question groups, each with paraphrased natural-language questions and prespecified phenotypes. For example, a heart-rate group can target both heart rate and RR interval. For each ECG and question group, we sampled one question, generated an answer, and extracted only the phenotypes assigned to that group. This kept the evaluation aligned with the clinical intent of the question and avoided rewarding or penalizing unrelated statements. The ECG-derived phenotypes include heart rate, RR/PP/PQ intervals, P duration, QRS duration, QTc interval, and frontal P/R/T axes. The CMR-derived fields, derived from short-axis and long-axis CMR views, included ventricular and atrial volumes, ejection fractions, stroke volumes, LV myocardial mass, global wall thickness, strain, cardiac output, cardiac index, and aortic distensibility. The complete list of evaluated fields is provided in Supplementary Table~\ref{tab:supp_training_fields}, and the question groups, questions per group, and target phenotypes are listed in Supplementary Table~\ref{tab:supp_ukb_question_bank}. Because ECG-LLM generates free-text answers, we mapped each answer to the predefined category space using a separate schema-constrained extractor LLM. Details and the prompt used are provided in Supplementary Section~\ref{sec:supp_ukb_phenotype_extraction}. The extracted category for each phenotype was compared with the ground-truth category, treating each field as a single-label multi-class classification task. We report class-macro F1 as the primary metric because the category distributions are strongly imbalanced, with many fields dominated by normal values. Accuracy, balanced accuracy, weighted F1, coverage, missing rate, majority-class baseline performance, number of classes, and full class distributions are reported in Supplementary Table~\ref{tab:supp_ukb_full_results}.

\begin{figure*}[!t]
    \centering
    \includegraphics[width=\textwidth]{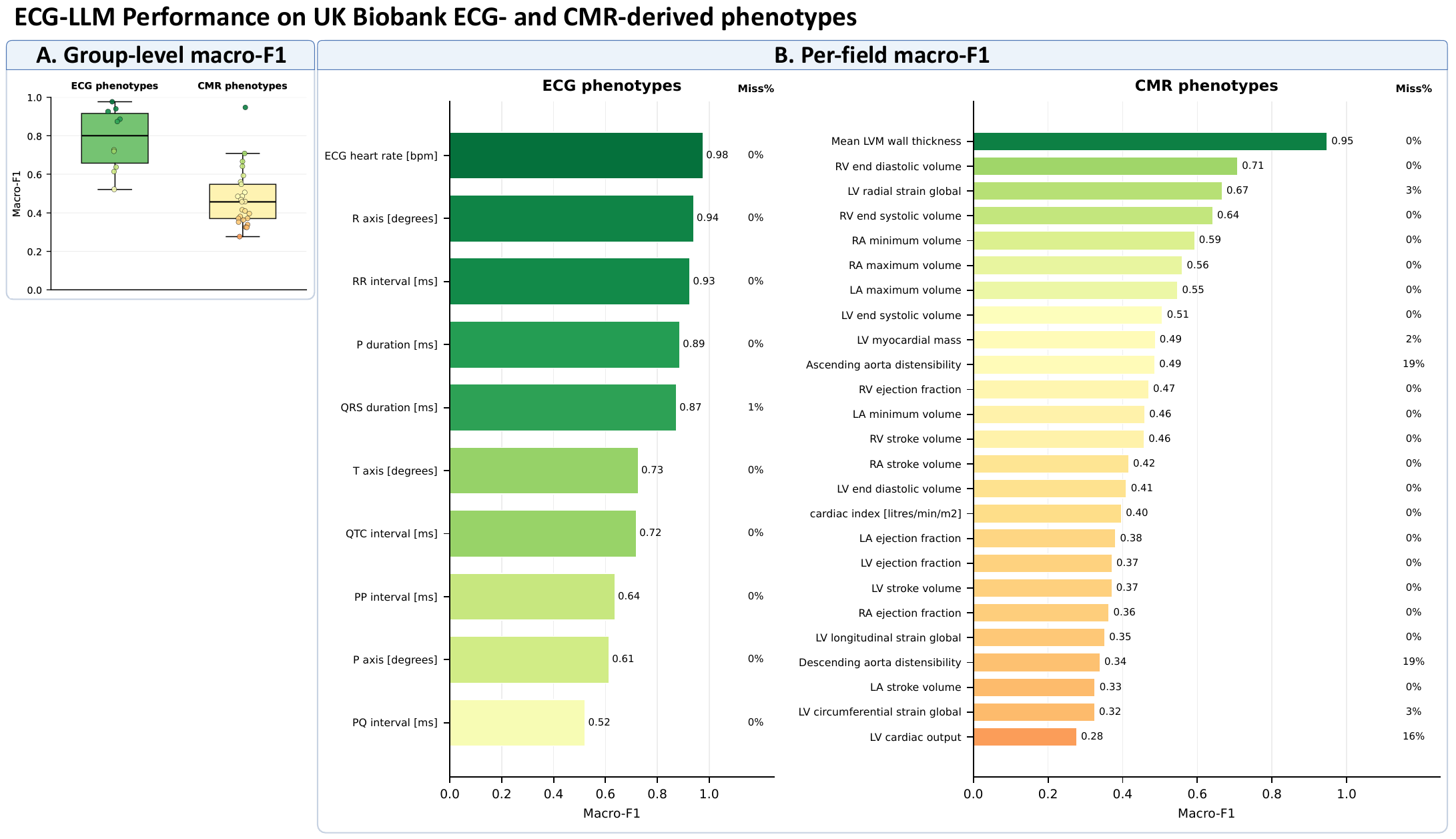}

    \caption{
    ECG-only question-answering performance for ECG-derived and CMR-derived phenotype categories in UK Biobank.
    The figure shows group-level and per-field class-macro F1 for ECG and CMR-derived phenotypes. Miss \% indicates the fraction of cases in which no valid category was extracted from the generated answer for that phenotype.
    }
    \label{fig:ukb_category_f1}
\end{figure*}

Figure~\ref{fig:ukb_category_f1} shows a higher class-macro F1 for ECG-derived fields than for CMR-derived fields. The strongest ECG-derived results are observed for directly ECG-visible properties, including heart rate, RR interval, R axis, P duration, and QRS duration. T axis and QTc interval show intermediate performance, while P axis and PP/PQ intervals are weaker, suggesting that atrial timing and atrial orientation are less consistently recovered than rate-, axis-, and ventricular conduction-related fields. Several CMR-derived phenotypes nevertheless show a clear signal above the majority baseline, especially LV mean myocardial wall thickness, RV end-diastolic volume, LV radial strain, RV end-systolic volume, RA minimum and maximum volume, LA maximum volume, and LV end-systolic volume. The strong result for LV wall thickness may partly reflect the training setup, because ECG-LLM was trained on QA pairs from ECG, CMR, and ECHO phenotypes, and wall-thickness-related remodelling appears in both the CMR and ECHO phenotype tasks. In contrast, weaker performance is observed for cardiac output, cardiac index, aortic distensibility, longitudinal and circumferential strain, and stroke-volume-related fields. These variables reflect hemodynamic function, vascular properties, or derived mechanical measurements rather than features directly measured on the ECG. For some phenotypes, performance remains close to the majority baseline, suggesting a weak ECG signal or that the model may partially collapse toward the most common category, leveraging its prior textual knowledge.

\begin{figure*}[!t]
    \centering
    \includegraphics[width=\textwidth]{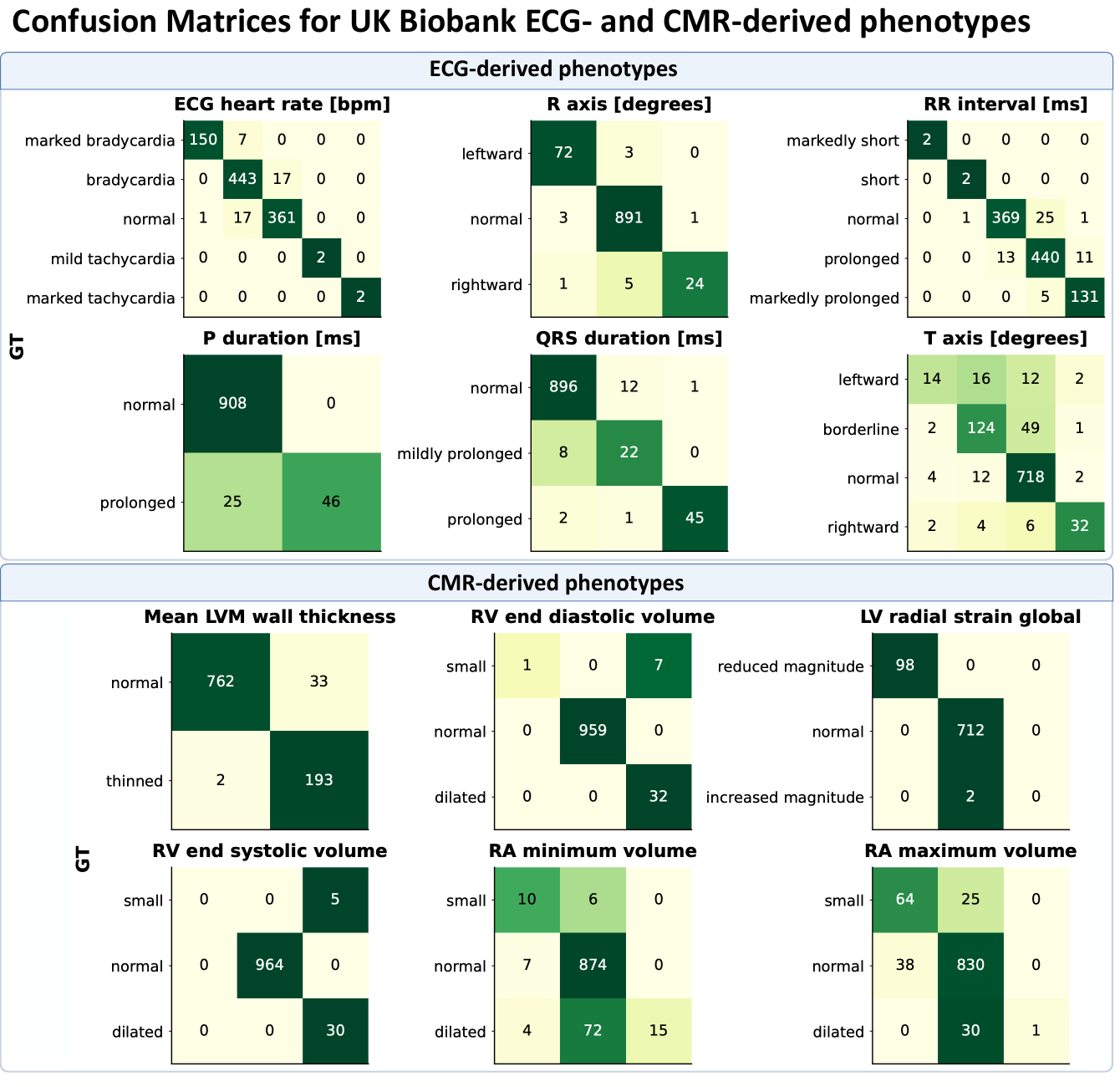}

    \caption{
    Confusion matrices for the top six ECG-derived and top six CMR-derived phenotype fields ranked by macro-F1 in the UK Biobank free-form phenotype extraction evaluation.
    Rows show ground-truth categories and columns show predicted categories.
    }
    \label{fig:ukb_category_confusions_top6}
\end{figure*}

The confusion matrices in Figure~\ref{fig:ukb_category_confusions_top6}A--B show errors for top-performing ECG-derived and CMR-derived fields. For ECG-derived variables, the strongest fields show clear diagonal structure, especially for heart rate, RR interval, R axis, P duration, and QRS duration. Errors are mainly concentrated in neighbouring or clinically related categories, although T-axis categories remain less stable. For CMR-derived variables, the matrices are more heterogeneous. Mean LVM wall thickness shows strong separation between normal and thinned categories, while ventricular volume fields show useful separation of dominant normal or dilated categories. However, rare classes are less stable and are sometimes shifted toward more common categories. This partly reflects the difficulty of forcing continuous phenotypes, particularly values close to category boundaries, into strict discrete labels. In such borderline cases, the model may assign an incorrect adjacent category. However, because the answer is generated in free text, the model can still mention that the finding is borderline or that the prediction is uncertain. Thus, the CMR results support the presence of coarse structural signal in the ECG representation, while also showing that rare categories and precise severity boundaries remain challenging.

Taken together, this evaluation shows that ECG-LLM can answer free-text clinical questions about both conventional ECG measurements and CMR-derived imaging phenotypes from the ECG representation. The strongest performance is observed for ECG-proximal measurements, confirming that the model has learned clinically meaningful ECG concepts, while the above-baseline CMR results show that it also captures coarse structural signals that are not part of standard ECG-machine output. This distinguishes ECG-LLM from report-centred ECG-language systems: it is not limited to describing the ECG itself, but can describe imaging-related cardiac phenotypes that may help determine whether further cardiac imaging is warranted.

\FloatBarrier

\subsection{ECG-only question answering for echocardiographic phenotypes}

To test whether ECG-LLM can reason about echocardiographic phenotypes from ECG alone, we evaluated it on the public EchoNext ECG--echocardiogram dataset~\cite{echonext}. The evaluation used 5{,}442 paired ECG--ECHO studies from the EchoNext test set. For each study, the model received only the 12-lead ECG and a phenotype-specific natural-language question. Because ECG-LLM was trained to produce free-text clinical answers rather than fixed binary predictions, we used a fixed two-turn protocol: first, the model answered the clinical question in free text. Second, within the same conversation, it was asked to provide its final answer as "yes" or "no". This preserved the model's natural response format while producing a standardized binary label for evaluation. The final answer was parsed and compared with the echocardiographic ground truth. The exact questions used for each phenotype are provided in the Supplementary Table~\ref{tab:supp_echonext_questions}. We evaluated eleven ECHO-derived binary labels covering clinically thresholded or report-graded abnormalities (Supplementary Table~\ref{tab:supp_training_fields}): reduced left ventricular ejection fraction (LVEF $\leq 45\%$), increased left ventricular wall thickness (LVWT $\geq 1.3$ cm), moderate-or-greater right-ventricular systolic dysfunction, moderate-or-greater aortic stenosis, aortic regurgitation, mitral regurgitation, tricuspid regurgitation, pulmonary regurgitation, moderate-or-large pericardial effusion, elevated pulmonary artery systolic pressure (PASP $\geq 45$ mmHg), and elevated tricuspid regurgitation maximum velocity (TR Vmax $\geq 3.2$ m/s).

\input{tables/echonext_results}
\FloatBarrier

As shown in Table~\ref{tab:echonext_ecgllm_binary}, ECG-LLM recovered a clear ECG-derived signal for several structural and functional echocardiographic phenotypes. The strongest result was observed for increased LV wall thickness, with an F1-score of 0.77, precision of 0.79, and recall of 0.75. The model also performed well for moderate-or-greater aortic stenosis, reaching an F1-score of 0.67 and a recall of 0.86 despite a positive prevalence of only 5.3\%. Ventricular systolic dysfunction was also captured: moderate-or-greater right-ventricular systolic dysfunction reached an F1-score of 0.61 with balanced precision and recall, while reduced LVEF reached an F1-score of 0.55 with high recall of 0.78. Pulmonary-pressure-related phenotypes showed moderate performance. PASP $\geq 45$ mmHg reached an F1-score of 0.58, with a precision of 0.70 and a recall of 0.50, whereas TR Vmax $\geq 3.2$ m/s reached an F1-score of 0.39. Regurgitant valve phenotypes were more difficult. Moderate-or-greater mitral and tricuspid regurgitation had high recall, 0.79 and 0.84, respectively, but low precision around 0.24, resulting in F1-scores of 0.37. Aortic regurgitation reached an F1-score of 0.31. Thus, for several valve-related phenotypes, ECG-LLM often identified abnormal cases but also overcalled positives. The weakest results were observed for pulmonary regurgitation and pericardial effusion. These phenotypes were rare in the test set, with prevalences of 0.4\% and 1.3\%, and achieved F1-scores of 0.04 and 0.10, respectively. Accuracy was high for these tasks, but this was driven mainly by the dominant negative class. We also trained a supervised linear probe baseline using the same frozen ECG encoder as in ECG-LLM. On the EchoNext test set, this supervised ViT achieved a macro-AUROC of 0.79, macro-AUPRC of 0.24, and macro-F1 of 0.27, confirming that the encoder itself captures the echocardiographic signal. At the phenotype level, it performed best on reduced LVEF, increased LV wall thickness, and right-ventricular systolic dysfunction, consistent with the strongest ECG-LLM phenotypes.

\input{tables/echonext_comparison}
\FloatBarrier

The EchoNext benchmark was released together with the Columbia Mini-Model, whose performance on the same public test set was reported in the original study~\cite{echonext}. As summarized in Table~\ref{tab:echonext_component_compare}, ECG-LLM achieved the highest F1-score for 10 of 11 phenotypes among the evaluated approaches, including LV wall thickness, aortic stenosis, right-ventricular systolic dysfunction, PASP elevation, TR Vmax elevation, mitral regurgitation, tricuspid regurgitation, aortic regurgitation, pericardial effusion, and pulmonary regurgitation. The only phenotype for which the supervised ViT achieved a higher F1-score was reduced LVEF, reaching 0.60 compared with 0.55 for ECG-LLM. Together, these results show that ECG-only echocardiographic question answering captures clinically meaningful signals for selected phenotypes, especially LV wall thickness, aortic stenosis, right-ventricular systolic dysfunction, PASP elevation, and reduced LVEF. Performance was weaker but still non-random for mitral and tricuspid regurgitation, and remained unreliable for rare or weakly ECG-expressed findings such as pulmonary regurgitation and pericardial effusion. These findings strengthen ECG-LLM as an ECG-based screening and reasoning model: from the ECG alone, it can identify several ECHO-derived structural and functional signals that are relevant for triage and follow-up imaging decisions. Echocardiography nevertheless remains necessary for definitive assessment of cardiac structure, ventricular function, and valve disease.

\subsection{ECG-LLM answers clinically relevant questions across diverse ECG concepts}
We next tested whether ECG-LLM understands a broad range of ECG concepts and can answer clinically relevant questions about raw ECG recordings. ECG-QA is a benchmark for question answering over ECG signals, built from ECG-expert-validated templates covering clinically relevant findings~\cite{ecgqa}. The benchmark includes questions about individual ECGs and comparisons between ECG recordings, with answer formats that test verification, choice between candidate findings, and open recovery of relevant findings. We evaluated ECG-LLM on the single-ECG subset from PTB-XL and MIMIC-IV-ECG, focusing on Single-Verify, Single-Choose, and Single-Query questions. These formats reflect increasing difficulty: Single-Verify asks whether a specific finding is present, Single-Choose requires discrimination between candidate findings, and Single-Query requires recovering the full set of relevant findings for a broader question. From a first-line clinical perspective, Single-Verify and Single-Choose are closest to the focused questions a general practitioner or triage clinician may ask about a suspected ECG abnormality. Single-Query is still useful as a benchmark of broad, report-like recovery, but it is less central to our intended use case because it asks the model to enumerate a complete multi-label answer set rather than answer a targeted clinical question. The evaluated concepts include, but are not limited to, rhythm disorders, conduction disease, ischemic or infarction-related changes, hypertrophy, chamber enlargement, repolarization abnormalities, axis deviation, pacing, ectopy, interval abnormalities, and signal quality. Although ECG machines already provide intervals, axes, and rule-based diagnostic statements, these concepts remain important in ECG-QA because they test whether the model has learned the basic ECG descriptors required for broader ECG understanding. During evaluation, ECG-LLM received the question and candidate options, generated a free-text answer, and this answer was parsed back into the ECG-QA option space using regular-expression matching.

\FloatBarrier
\input{tables/ecgqa_ptbxl_mimic}

\FloatBarrier
\input{tables/ecgqa_ptbxl_baselines}

ECG-LLM performed consistently above the per-question-type majority baseline on both datasets (Table~\ref{tab:ecgqa_ptbxl_mimic}). Overall, exact match improved from the majority baseline of 39 to 58 EM on PTB-XL and from 25 to 44 EM on MIMIC-IV-ECG. Unless otherwise stated, paired values report PTB-XL/MIMIC-IV-ECG. Option-level F1 was higher than exact match, indicating that many errors were partial rather than completely wrong interpretations (66/59 $\mu$F1). Performance was strongest for targeted questions: Single-Verify reached 77/81 EM and Single-Choose reached 68/68 EM. Compared with published PTB-XL ECG-QA baselines, ECG-LLM was substantially stronger on the targeted ECG-QA tasks despite using a generative formulation rather than supervised classification over a fixed answer space (Table~\ref{tab:ecgqa_ptbxl_baselines}). It achieved the best reported result on Single-Verify (76.71 vs.\ 74.60 EM for M3AE~\cite{M3A}) and Single-Choose (68.32 vs.\ 57.10 EM for M3AE), giving the strongest overall performance on the clinically focused ECG-QA formats. On the harder Single-Query task, ECG-LLM ranked second (38.26 vs.\ 41.00 EM for M3AE) while still exceeding Fusion~\cite{ecgqa} and MedViLL~\cite{medViLL}. The main remaining gap was Single-Query exact match, where the model must recover a complete multi-label answer set. This metric is strict for free-text generation: missing one correct label or adding one extra label in the free-text answer makes the answer incorrect. Accordingly, Single-Query option-level F1 was substantially higher than exact match on both datasets. On PTB-XL, Single-Query reached 38 EM and 58 $\mu$F1. On MIMIC-IV-ECG, it reached 20 EM and 54 $\mu$F1, showing that many errors reflected incomplete answer sets rather than entirely incorrect ECG interpretations.

\begin{figure*}[p]
    \centering
    \includegraphics[width=\textwidth]{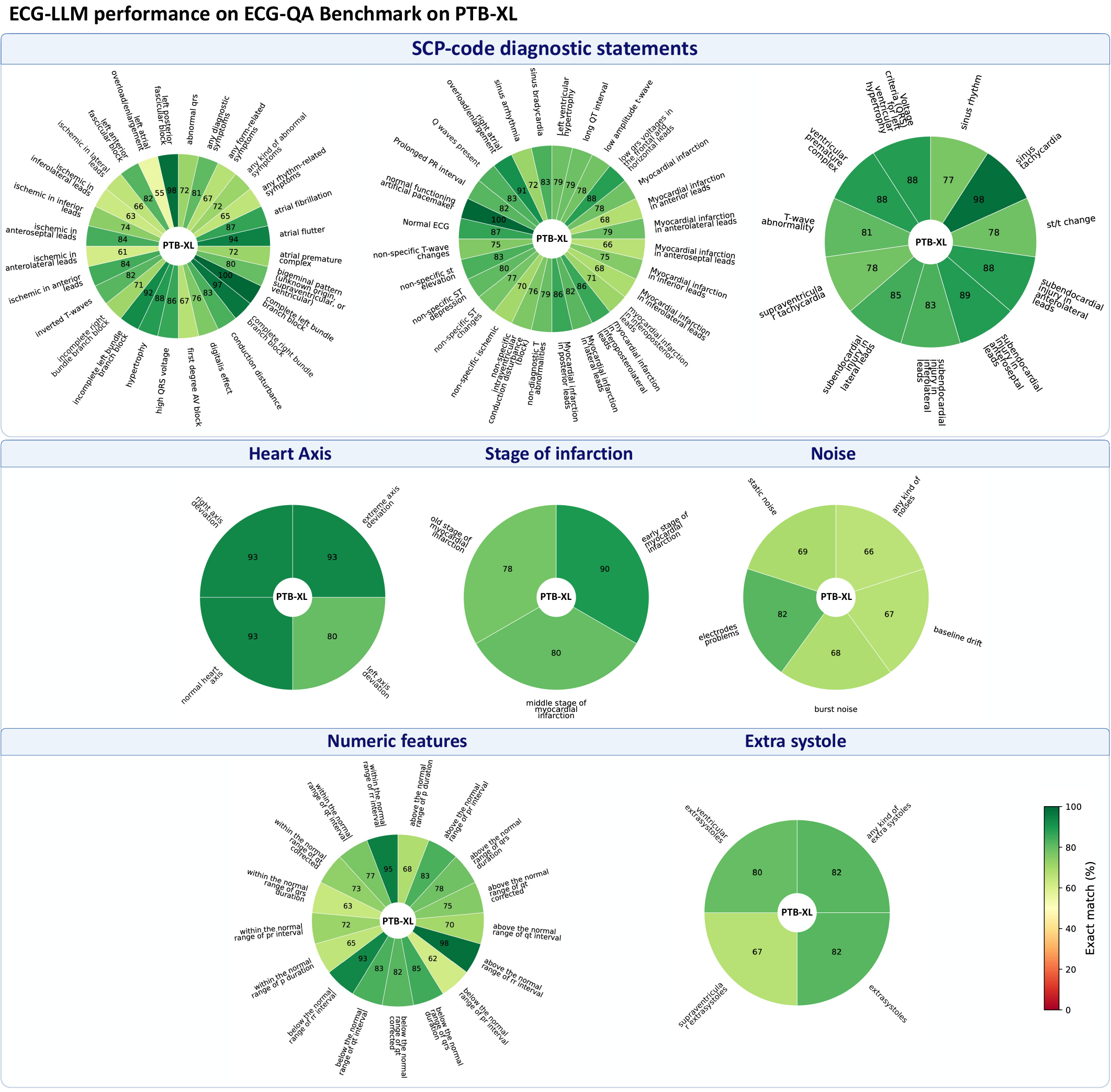}

    \caption{
    Single-Verify ECG-QA exact-match (EM) performance across ECG concepts on PTB-XL.
    Results include SCP-code diagnostic statements, heart axis, stage of infarction, noise, numeric features, and extra systoles.
    Sector values indicate exact-match performance for each attribute. The full table is provided in the Supplementary Table~\ref{tab:supp_ecgqa_single_verify_attribute}.
    }
    \label{fig:ecgqa_single_verify_ptbxl}
\end{figure*}

\begin{figure*}[p]
    \centering
    \includegraphics[width=\textwidth]{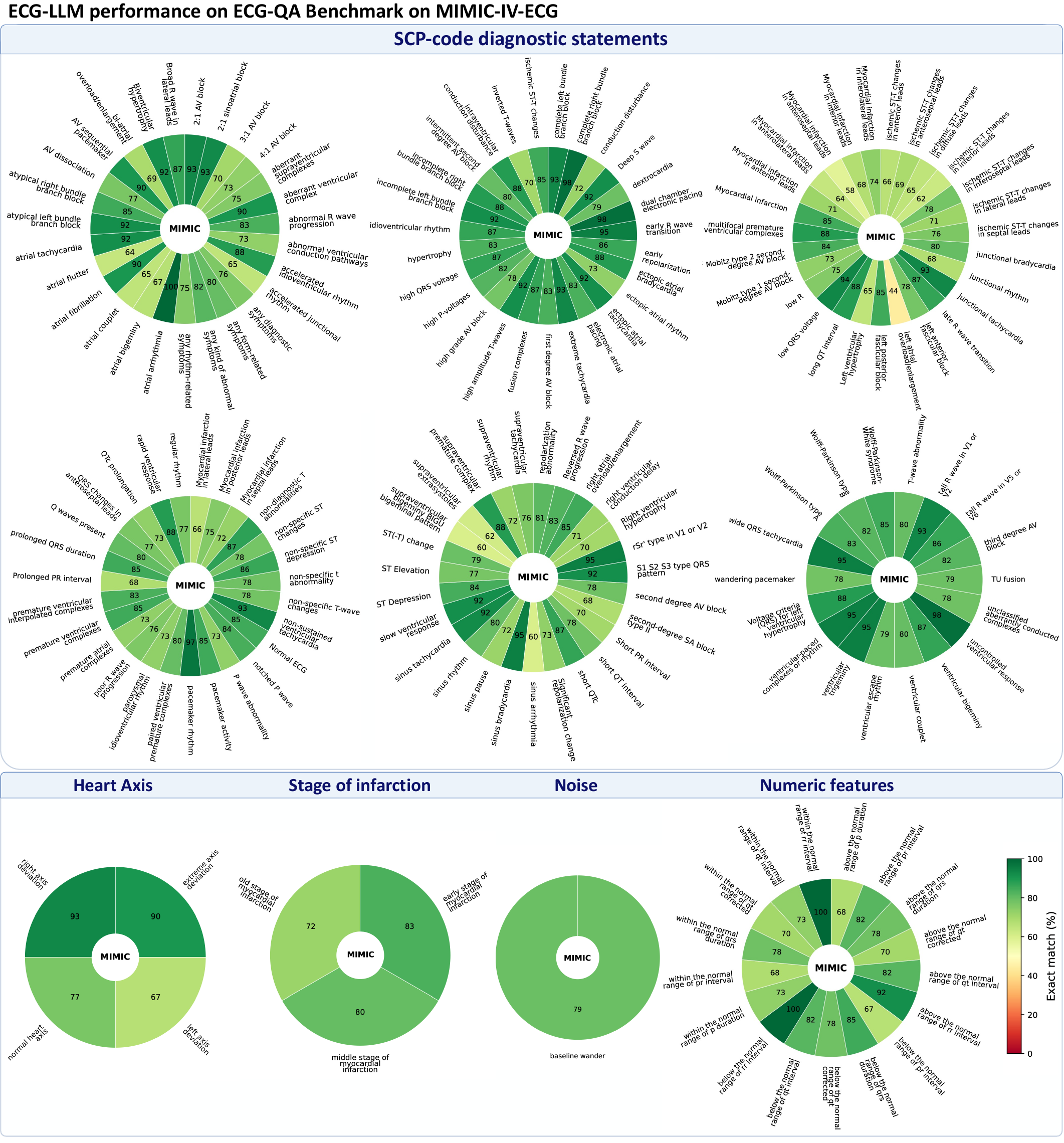}

    \caption{
    Single-Verify ECG-QA exact-match (EM) performance across ECG concepts on MIMIC-IV-ECG.
    Results include SCP-code diagnostic statements, heart axis, stage of infarction, noise, and numeric features.
    Sector values indicate exact-match performance for each attribute. The full table is provided in the Supplementary Table~\ref{tab:supp_ecgqa_single_verify_attribute}.
    }
    \label{fig:ecgqa_single_verify_mimic}
\end{figure*}

Stratifying the results by ECG-QA attribute type shows that the strong targeted-question performance is not driven by one narrow label group, but extends across many clinically interpretable ECG concepts (Supplementary Tables~\ref{tab:supp_ptbxl_ecgqa_qtype_attribute_type}--\ref{tab:supp_mimic_ecgqa_qtype_attribute_type}). In Single-Verify questions, exact match reached 68--90 EM across heart axis, infarction stage, extra systoles, SCP diagnostic statements, numeric features, and noise-related questions on PTB-XL, and 78--82 EM across all corresponding MIMIC-IV-ECG groups available in the benchmark (SCP diagnostic statements, heart axis, infarction stage, noise-related questions, and numeric features). In Single-Choose questions, ECG-LLM answered multiple concept groups: SCP-code diagnostic statements reached 69/68 EM (75/74 $\mu$F1), heart-axis questions 75/50 EM, infarction-stage questions 56/78 EM, and numeric-feature questions 61/61 EM. Noise and extra-systole questions were lower on PTB-XL (43 and 50 EM).

For individual findings (Figures~\ref{fig:ecgqa_single_verify_ptbxl} and~\ref{fig:ecgqa_single_verify_mimic},
Supplementary Table~\ref{tab:supp_ecgqa_single_verify_attribute}), we focused on clinically salient Single-Verify attributes because these most directly correspond to targeted yes/no questions about a suspected ECG finding. ECG-LLM was strong for major conduction disease, including complete left and right bundle branch block (100/93 and 97/98 EM, respectively), incomplete left bundle branch block (92/92 EM), left posterior fascicular block (98/85 EM), and first-degree AV block (67/83 EM). It also performed well for rhythm and pacing findings, including atrial fibrillation (87/90 EM), sinus tachycardia (98/93 EM), sinus bradycardia (83/95 EM), normal functioning artificial pacemaker (100 EM on PTB-XL), pacemaker rhythm (97 EM on MIMIC-IV-ECG), and ventricular-paced complexes or rhythm (95 EM on MIMIC-IV-ECG). Structural, infarction, and repolarization concepts were also recovered, including hypertrophy (88/83 EM), right atrial enlargement (91/85 EM), long QT interval (79/88 EM), T-wave abnormality (81/80 EM), Q waves (83/80 EM), and myocardial infarction as a broad diagnostic concept (78/85 EM). Additional high-performing clinically important labels included atrial flutter (94 EM on PTB-XL), atrial tachycardia (92 EM on MIMIC-IV-ECG), low QRS voltage (94 EM on MIMIC-IV-ECG), biventricular hypertrophy (92 EM on MIMIC-IV-ECG), and rapid ventricular-response or ventricular tachycardia-related labels such as non-sustained ventricular tachycardia, wide QRS tachycardia, and uncontrolled ventricular response (93--98 EM on MIMIC-IV-ECG). Axis and signal-quality questions were also clinically useful, with axis-deviation findings reaching 67--93 EM and noise-related Single-Verify performance reaching 68/79 EM at the group level. The main limitations were more granular tasks. Localized infarction or ischemia labels, which require assigning abnormalities to specific lead territories such as anterior, inferior, lateral, or septal regions, generally ranged from 58--86 EM. Left atrial overload/enlargement remained difficult (55/44 EM). Interval-range questions varied (62--98 EM on PTB-XL and 67--100 EM on MIMIC-IV-ECG). Broad Single-Query questions remained harder (38/20 EM, despite 58/54 $\mu$F1) because exact match penalizes incomplete multi-label answers. Overall, this ECG-QA evaluation shows that ECG-LLM has learned a broad ECG vocabulary and can answer focused, clinically relevant questions from the raw signal, while complete report-like recovery, fine-grained localization, subtle atrial findings, and exact interval reasoning remain important limitations.

\subsection{ECG-LLM generates free-text ECG diagnostic summaries}

We next evaluated whether ECG-LLM can generate free-text ECG diagnostic summaries from raw ECGs. ECG machines already provide automated measurements and diagnostic statements, and many prior ECG-language and ECG-image models use report generation or diagnostic interpretation as a primary endpoint. In our study, this task serves a different purpose: it tests whether the ECG-conditioned language model has learned the mapping between raw ECG morphology and conventional diagnostic language. Diagnostic reports provide dense weak supervision because they summarize rhythm, conduction, repolarisation, ischemia/infarction, pacing, and other ECG abnormalities in compact clinical language. Solving this task, therefore, demonstrates that the model can associate ECG signal patterns with clinically named ECG concepts, strengthening the ECG-language grounding needed for broader question answering over ECG-derived and imaging-related cardiac phenotypes. For each ECG and each evaluated model, one diagnostic-report question was randomly selected from a fixed prompt set (Supplementary Table~\ref{tab:supp_diagnostic_report_questions}), and the model was asked to produce a concise free-text ECG diagnosis. The generated report was compared with the reference diagnosis by an LLM judge scoring correctness, completeness, and specificity.

We sampled 1{,}000 non-trivial ECG studies from unique patients in each of UK Biobank~\cite{ukbiobank}, PTB-XL~\cite{ptbxl}, and MIMIC-IV-ECG~\cite{mimic_ecg}. Non-trivial records were defined using dataset-specific normal-report rules. In the UK Biobank, an ECG was considered non-trivial if the device diagnosis was not exactly the pair ''Normal sinus rhythm'' and ''Normal ECG''. In MIMIC-IV-ECG, an ECG was considered non-trivial if the device diagnosis was not exactly ''Sinus rhythm; Normal ECG''. In PTB-XL, an ECG was considered non-trivial if its SCP-code set was not exactly \{NORM, SR\}. For each selected ECG, ECG-LLM was asked to produce a concise free-text ECG diagnosis. Ground truth was obtained from dataset-specific diagnostic annotations: device diagnoses for UK Biobank and MIMIC-IV-ECG, and report-derived diagnostic annotations for PTB-XL (the original German/Swedish report and its English translation).

We compared ECG-LLM with three publicly available baseline models: MedGemma-4B~\cite{medgemma,medgemma_deepmind,medgemma_hf,medgemma_developers}, ECG-Chat~\cite{ecgchat}, and PULSE-7B~\cite{PULSE_ECG,pulse_ecg_hf}. ECG-Chat is the closest raw-signal ECG-language baseline. It's published system combines an ECG encoder and LLM with additional components for report generation, including diagnosis-driven prompting, GraphRAG-based retrieval~\cite{GraphRAG}, DSPy-based prompt tuning~\cite{dspy}, and a structured LaTeX report-generation pipeline~\cite{ecgchat}. Here, we evaluate ECG-Chat without GraphRAG retrieval or downstream report-generation modules to focus on ECG-conditioned generation rather than a retrieval-augmented reporting system, although results with GraphRAG may be higher because it uses an additional supervised ECG interpretation method. PULSE-7B is a strong ECG-image baseline trained for ECG image interpretation using ECGInstruct, a large-scale ECG image-text instruction dataset~\cite{PULSE_ECG,pulse_ecg_hf,ecginstruct_hf}. For this comparison, raw ECG signals were rendered as ECG images using ECG-Image-Kit~\cite{ecg_image_kit_physmeas,ecg_image_kit_2024}. UK Biobank is not among the reported training sources of PULSE-7B, so its UK Biobank evaluation should be interpreted as out-of-domain. MedGemma-4B was included to show the current performance of a general medical vision-language model on the same rendered ECG images. Although MedGemma-4B has been trained with medical image data that includes ECG images, it is not an ECG-specialised model like PULSE-7B. Thus, the baselines cover both raw-signal and image-based ECG-language settings, while all models are assessed with the same diagnostic-generation prompt, reference reports, and LLM-judge protocol.

Standard NLP string-overlap metrics provide an incomplete evaluation of free-text ECG reports. The same clinical finding can be expressed with different wording, while similar wording can correspond to different clinical meanings. Moreover, the severity of an error depends on the finding, so missing a major rhythm, conduction, or ischemic abnormality is not equivalent to omitting a minor or borderline statement. We therefore used a locally deployed GPT-OSS-20B model~\cite{openai_gptoss_blog,openai_gptoss_hf,openai2025gptoss120bgptoss20bmodel} as a reasoning-capable LLM judge, following the broader use of LLM-as-judge evaluation for open-ended model outputs~\cite{zheng2023judging} and recent work applying LLM judges to medical report generation~\cite{wang2024llmradjudge}. The judge acted as an automated clinical adjudicator: it compared the generated report with the reference diagnosis, assigned structured scores for correctness, completeness, and specificity, and provided notes as its reasoning. Correctness measures whether the main ECG interpretation is clinically accurate overall, completeness measures whether important ground-truth findings are recovered, and specificity measures whether the generated findings are supported by or compatible with the reference report. Each score ranges from 0 to 2, where 0 indicates poor performance, 1 indicates partial performance, and 2 indicates good performance. The judge's prompt allowed clinically equivalent wording while penalising major rhythm, conduction, ischemia/infarction, pacing, and unsupported-diagnosis errors. The full judge prompt is provided in Supplementary Section~\ref{sec:llm_as_judge_sec}.

\begin{figure*}[!t]
\centering
\includegraphics[width=0.99\textwidth]{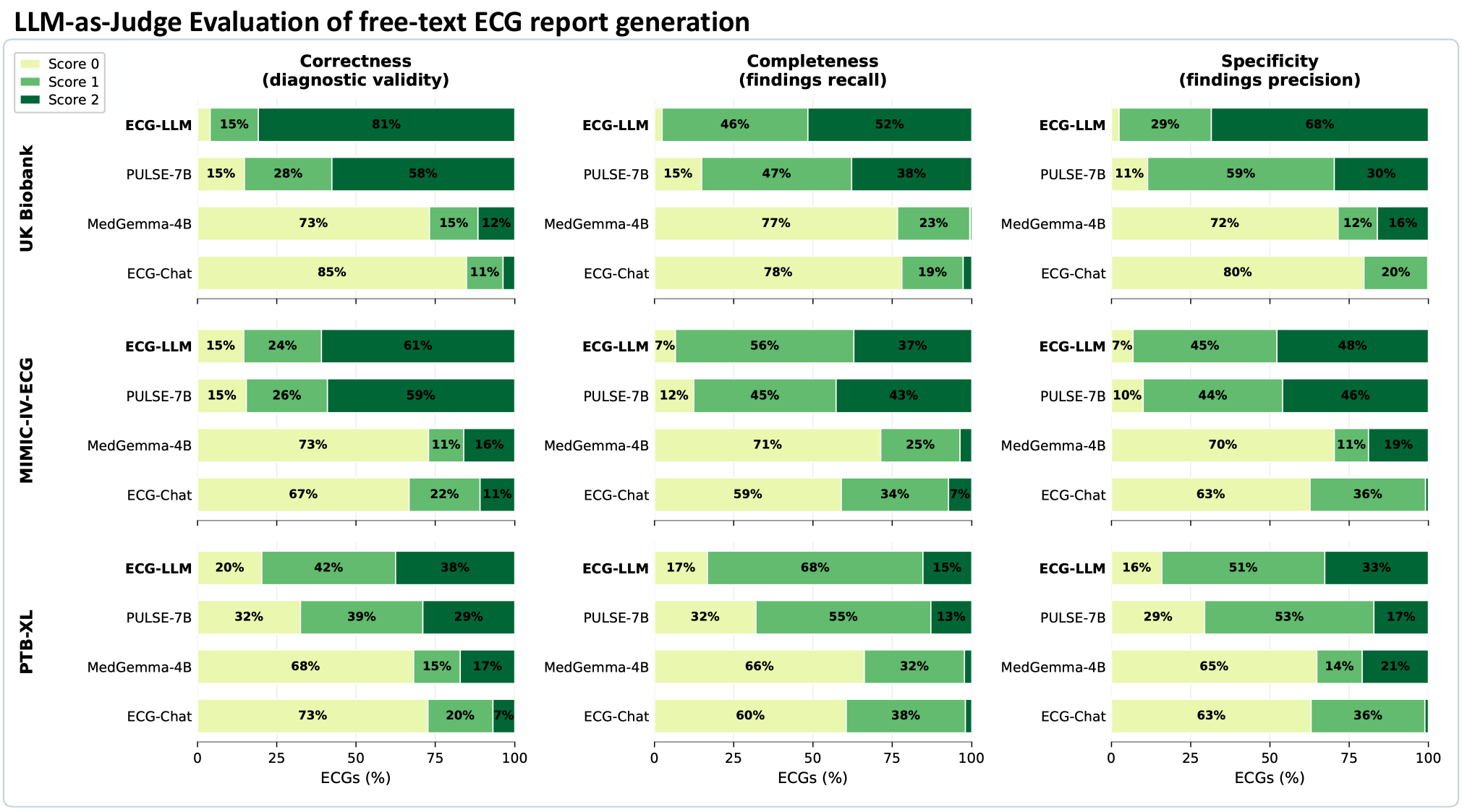}
\caption{
    LLM-judge evaluation of free-text ECG diagnostic summaries across UK Biobank, MIMIC-IV-ECG, and PTB-XL. A locally deployed GPT-OSS-20B judge compares each generated diagnosis with the ground-truth diagnostic report and scores correctness, completeness, and specificity on a three-level scale. Score 0 denotes poor performance, score 1 partial performance, and score 2 good performance. Bars show the percentage of ECGs assigned to each score level.
}
\label{fig:judge}
\end{figure*}

Figure~\ref{fig:judge} shows that ECG-LLM and PULSE-7B are the strongest models in this evaluation. On UK Biobank, ECG-LLM achieves the clearest advantage, with a substantially larger fraction of score-2 reports for correctness and specificity. PULSE-7B nevertheless remains a strong baseline, especially because UK Biobank is out-of-domain for this model and because it receives the ECG only as a rendered image. On MIMIC-IV-ECG, ECG-LLM and PULSE-7B are broadly comparable: both models obtain similar score distributions for correctness and specificity, while PULSE-7B shows slightly higher completeness. On PTB-XL, ECG-LLM shows higher fractions of score-2 correctness and specificity than PULSE-7B, although both models frequently receive partial completeness scores, indicating that open-ended reports often recover some but not all reference findings. For PTB-XL, the LLM judge used the original PTB-XL report as the primary reference. The English translation was provided only to help interpret the non-English report text and was not treated as an additional list of required findings, as specified in the judge prompt. MedGemma-4B and ECG-Chat perform less consistently under this protocol. For MedGemma-4B, this likely reflects the difficulty of ECG interpretation for a general medical vision-language model, even when ECG images are included in its broader medical-image training data. For ECG-Chat, the result should be interpreted as a model-level raw-signal comparison rather than as an evaluation of its full published ECG-Chat system. It's published report-generation pipeline explicitly supplements the ECG-conditioned model with diagnosis-driven prompting, retrieval-augmented knowledge with GraphRAG, prompt optimization with DSPY, and structured report formatting. Therefore, the lower performance observed here is consistent with the smaller scale and narrower coverage of the underlying ECG-language instruction data, and with the possibility that part of ECG-Chat's long-form reporting capability depends on its retrieval and report-generation modules rather than on the bare ECG-conditioned generator alone.

Overall, these results show that ECG-LLM learns a useful mapping from raw ECG morphology to conventional ECG diagnostic language. This report-generation task is not the main clinical aim of ECG-LLM: unlike prior ECG-language systems that focus primarily on reproducing diagnostic reports, our central goal is flexible question answering over ECG-derived and imaging-related cardiac phenotypes. Nevertheless, strong performance on diagnostic summaries is an important intermediate learning step, because it confirms that the ECG-conditioned language model grounds signal patterns in clinically named ECG concepts. ECG-LLM remains competitive with the strongest ECG-image baseline and outperforms the general medical image-text baseline and the raw-signal ECG-language baseline in most settings, while the partial completeness scores observed across models highlight the continuing difficulty of open-ended report generation, especially for secondary findings and dataset-specific report annotations.

\subsection{Qualitative evaluation of ECG-LLM}
\label{sec:qualitative_evaluation}
We assessed whether ECG-LLM produces ECG-conditioned outputs while preserving general language competence. When no ECG waveform was provided, the model declined to make a diagnosis and requested the missing trace. When the ECG input was replaced by random noise, the output lost clinical coherence, suggesting that meaningful predictions depend on structured ECG information rather than language priors alone. For non-cardiac general-knowledge prompts without ECG tokens, the model behaved similarly to the underlying instruction-tuned LLM, indicating no apparent catastrophic forgetting after ECG-specific fine-tuning. 

\begin{figure*}[!t] 
\centering 
\includegraphics[width=\textwidth]{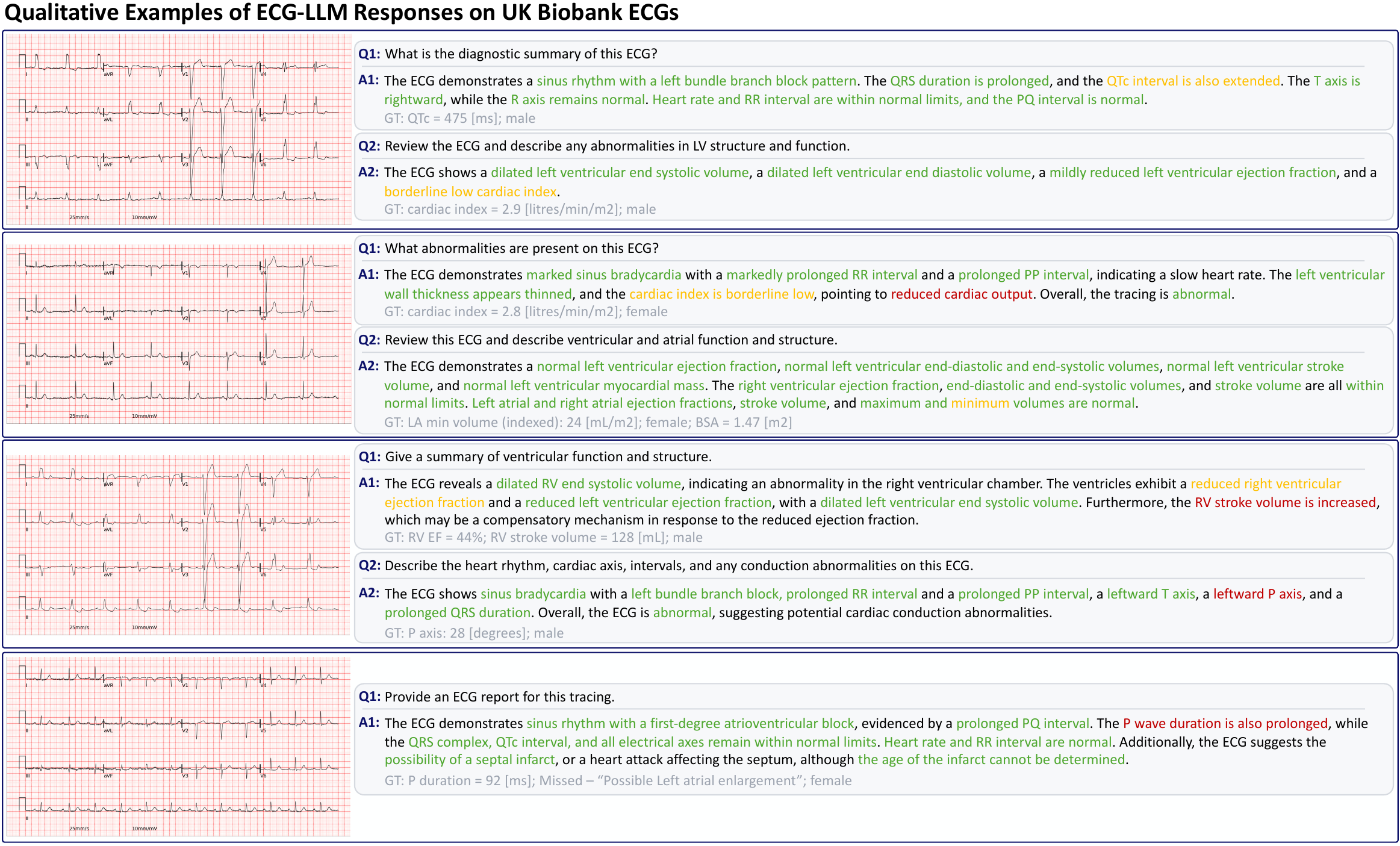} 
\caption{
Qualitative ECG-LLM examples from UK Biobank for ECG-derived findings and imaging-derived cardiac phenotypes. Representative cases show the ECG input, the user question, the model answer, and the corresponding ground-truth label or measurement. Green highlights denote exact agreement with the ground truth, yellow highlights denote clinically close or partially correct predictions, and red highlights denote incorrect predictions. For non-exact predictions, the relevant ground-truth value or label is shown below the model answer.
} 
\label{fig:qualitative_examples_ukb} 
\end{figure*}

\begin{figure*}[!t] 
\centering 
\includegraphics[width=\textwidth]{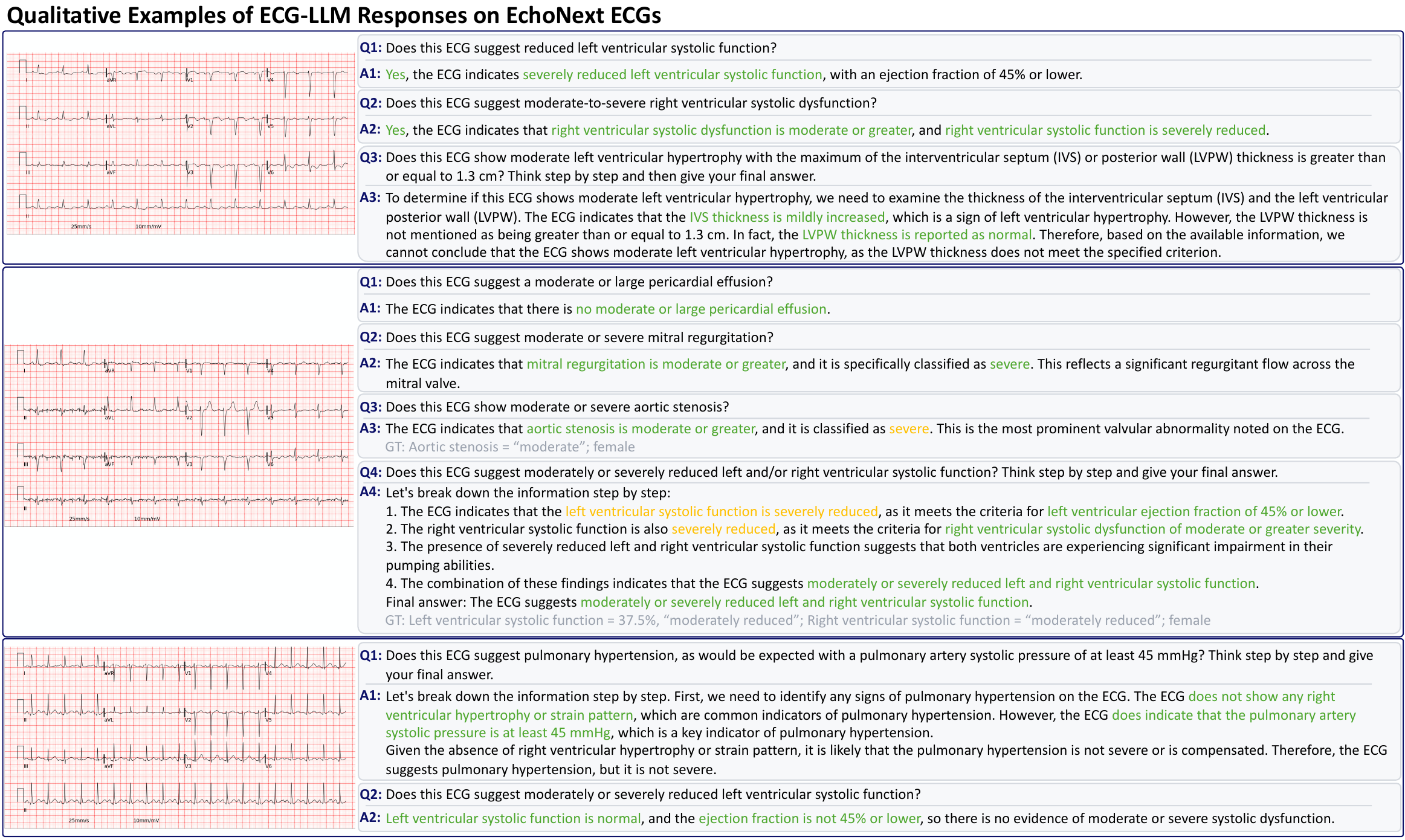} 
\caption{
Qualitative ECG-LLM examples from EchoNext for ECG-derived findings and imaging-derived cardiac phenotypes. Representative cases show the ECG input, the user question, the model answer, and, where applicable, the corresponding ground-truth diagnosis or measurement. Green highlights denote exact agreement with the ground truth, yellow highlights denote clinically close or partially correct predictions, and red highlights denote incorrect predictions. For non-exact predictions, the relevant ground-truth value or label is shown below the model answer.
} 
\label{fig:qualitative_examples_echo} 
\end{figure*}

\begin{figure*}[!t] 
\centering 
\includegraphics[width=\textwidth]{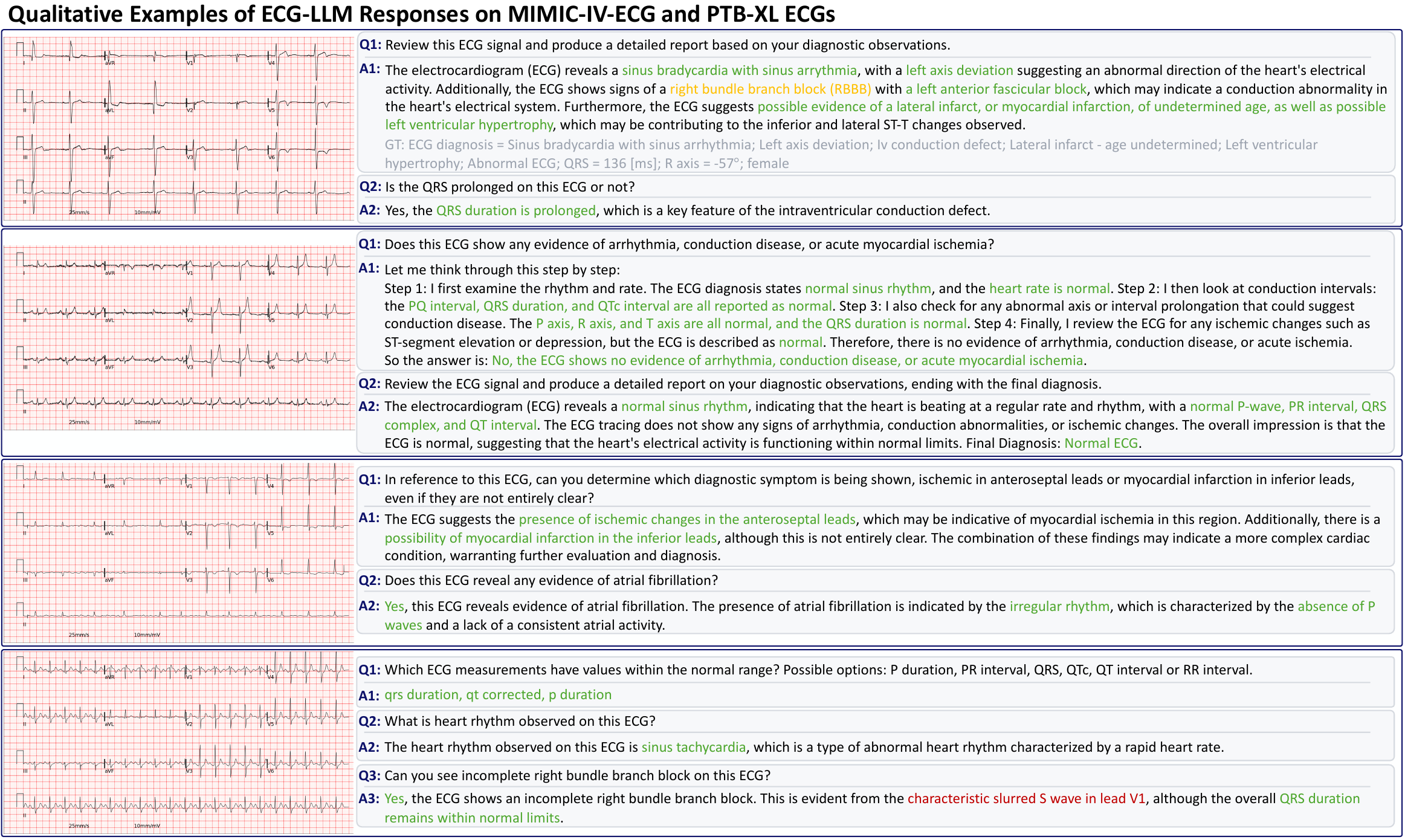} 
\caption{
Qualitative ECG-LLM examples from MIMIC-IV-ECG and PTB-XL for direct ECG interpretation. Representative cases show the ECG input, the user question, the model answer, and, where applicable, the corresponding ground-truth diagnosis or measurement. The examples cover diagnostic report generation and questions related to rhythm, interval measurements, conduction abnormalities, and ischemic findings. Green highlights denote exact agreement with the ground truth, yellow highlights denote clinically close or partially correct predictions, and red highlights denote incorrect predictions. For non-exact predictions, the relevant ground-truth value or label is shown below the model answer.
} 
\label{fig:qualitative_examples_mimic_ptbxl} 
\end{figure*}

We present qualitative examples from UK Biobank~\cite{ukbiobank}, EchoNext~\cite{echonext}, MIMIC-IV-ECG~\cite{mimic_ecg}, and PTB-XL~\cite{ptbxl} (Figures~\ref{fig:qualitative_examples_ukb}--\ref{fig:qualitative_examples_mimic_ptbxl}). The UK Biobank and EchoNext examples include both ECG-derived findings, such as rhythm, intervals, axis, and conduction abnormalities, and ECG-predicted imaging-derived phenotypes, including ventricular function, chamber volumes, valvular disease, and pulmonary pressure-related findings. The MIMIC-IV-ECG and PTB-XL examples focus on direct ECG interpretation, including rhythm, interval measurements, conduction abnormalities, ischemic changes, and overall diagnostic impressions. 

%% file: tables/echonext_results.tex
\begin{table}[!htbp]
\centering
\small
\rowcolors{3}{TblStripe}{white}
\caption{
ECG-LLM performance on EchoNext binary echocardiographic phenotypes.
All tasks were evaluated on 5{,}442 ECG--echocardiogram pairs using a fixed ECG-only yes/no question-answering protocol.
Prev. denotes positive-class prevalence; Pos. rate denotes the fraction of ECG-LLM predictions parsed as positive.
}
\label{tab:echonext_ecgllm_binary}
\begin{tabular}{lrrrrrr}
\toprule
\rowcolor{TblHeader}
\textbf{Phenotype} & \textbf{Prev.} & \textbf{Pos. rate} & \textbf{Acc.} & \textbf{Prec.} & \textbf{Rec.} & \textbf{F1} \\
\midrule
LVWT $\geq 1.3$ cm                        & 19.5 & 18.5 & 0.911 & 0.787 & 0.748 & 0.767 \\
AS $\geq$ moderate                        & 5.3  & 8.2  & 0.956 & 0.552 & 0.857 & 0.671 \\
RV systolic dysf. $\geq$ moderate         & 7.7  & 7.5  & 0.940 & 0.615 & 0.599 & 0.607 \\
PASP $\geq 45$ mmHg                       & 12.8 & 9.2  & 0.908 & 0.697 & 0.498 & 0.581 \\
LVEF $\leq 45\%$                          & 17.7 & 32.6 & 0.774 & 0.424 & 0.783 & 0.550 \\
TR Vmax $\geq 3.2$ m/s                    & 6.9  & 13.5 & 0.875 & 0.293 & 0.573 & 0.388 \\
MR $\geq$ moderate                        & 6.2  & 20.1 & 0.834 & 0.242 & 0.786 & 0.370 \\
TR $\geq$ moderate                        & 6.5  & 23.0 & 0.813 & 0.236 & 0.836 & 0.368 \\
AR $\geq$ moderate                        & 1.2  & 0.9  & 0.985 & 0.367 & 0.273 & 0.313 \\
Pericardial eff. $\geq$ moderate/large    & 1.3  & 4.5  & 0.948 & 0.061 & 0.217 & 0.095 \\
PR $\geq$ moderate                        & 0.4  & 0.5  & 0.992 & 0.037 & 0.050 & 0.043 \\
\bottomrule
\end{tabular}
\end{table}

%% file: tables/echonext_comparison.tex
\begin{table*}[t]
\centering
\scriptsize
\setlength{\tabcolsep}{3.5pt}
\rowcolors{3}{TblStripe}{white}
\begin{tabularx}{0.99\textwidth}{l *{3}{>{\centering\arraybackslash}X} *{3}{>{\centering\arraybackslash}X} *{3}{>{\centering\arraybackslash}X}}
\toprule
\rowcolor{TblHeader}
& \multicolumn{3}{c}{Columbia Mini-Model~\cite{echonext}} & \multicolumn{3}{c}{Supervised ViT} & \multicolumn{3}{c}{\textbf{ECG-LLM}} \\
\cmidrule(lr){2-4}\cmidrule(lr){5-7}\cmidrule(lr){8-10}
\rowcolor{white}
Phenotype & AUROC & AUPRC & F1 & AUROC & AUPRC & F1 & P & R & F1 \\
\midrule
LVWT $\geq 1.3$ cm              & 0.73 & 0.37 & 0.45 & 0.75 & 0.42 & 0.47 & 0.79 & 0.75 & \textbf{0.77} \\
AS $\geq$ moderate              & 0.86 & 0.25 & 0.25 & 0.79 & 0.19 & 0.21 & 0.55 & 0.86 & \textbf{0.67} \\
RV dysf $\geq$ moderate         & 0.87 & 0.43 & 0.37 & 0.86 & 0.43 & 0.38 & 0.62 & 0.60 & \textbf{0.61} \\
PASP $\geq 45$ mmHg             & 0.77 & 0.36 & 0.39 & 0.73 & 0.29 & 0.36 & 0.70 & 0.50 & \textbf{0.58} \\
LVEF $\leq 45\%$                & 0.85 & 0.60 & 0.55 & 0.87 & 0.62 & \textbf{0.60} & 0.42 & 0.78 & 0.55 \\
TR Vmax $\geq 3.2$ m/s          & 0.75 & 0.21 & 0.24 & 0.71 & 0.15 & 0.23 & 0.29 & 0.57 & \textbf{0.39} \\
MR $\geq$ moderate              & 0.81 & 0.23 & 0.26 & 0.80 & 0.24 & 0.27 & 0.24 & 0.79 & \textbf{0.37} \\
TR $\geq$ moderate              & 0.83 & 0.30 & 0.29 & 0.80 & 0.23 & 0.28 & 0.24 & 0.84 & \textbf{0.37} \\
AR $\geq$ moderate              & 0.74 & 0.04 & 0.05 & 0.74 & 0.03 & 0.05 & 0.37 & 0.27 & \textbf{0.31} \\
Pericardial eff.\ $\geq$ mod    & 0.77 & 0.08 & 0.05 & 0.75 & 0.04 & 0.06 & 0.06 & 0.22 & \textbf{0.10} \\
PR $\geq$ moderate              & 0.83 & 0.15 & 0.03 & 0.88 & 0.03 & 0.03 & 0.04 & 0.05 & \textbf{0.04} \\
\bottomrule
\end{tabularx}
\caption{
Per-phenotype comparison on the EchoNext public test set.
Columbia Mini-Model results are taken from the original EchoNext study~\cite{echonext}.
Supervised ViT denotes a linear probe trained on top of the same frozen ECG encoder used in ECG-LLM.
ECG-LLM predicts binary labels through ECG-only free-text question answering followed by a yes/no final-answer prompt and post hoc parsing.
For Columbia Mini-Model and Supervised ViT, we report AUROC, AUPRC, and F1.
For ECG-LLM, we report precision, recall, and F1 from parsed yes/no outputs.
Bold indicates the highest F1-score for each phenotype.
}
\label{tab:echonext_component_compare}
\end{table*}

%% file: tables/ecgqa_ptbxl_mimic.tex
\begin{table}[H]
\centering
\small

\begin{minipage}{0.48\columnwidth}
\centering
\rowcolors{3}{TblStripe}{white}
\begin{tabularx}{\linewidth}{l *{4}{>{\centering\arraybackslash}X}}
\toprule
\rowcolor{TblHeader}
\multicolumn{5}{c}{\textbf{PTB-XL}} \\
\midrule
Type & EM & $\mu$F1 & MF1 & Maj \\
\midrule
Overall & 57.71 & 65.60 & 67.86 & 39.28 \\
Single-Verify & 76.71 & 76.71 & 76.71 & 67.73 \\
Single-Choose & 68.32 & 74.08 & 74.70 & 31.15 \\
Single-Query & 38.26 & 58.35 & 57.75 & 23.19 \\
\bottomrule
\end{tabularx}
\end{minipage}
\hfill
\begin{minipage}{0.48\columnwidth}
\centering
\rowcolors{3}{TblStripe}{white}
\begin{tabularx}{\linewidth}{l *{4}{>{\centering\arraybackslash}X}}
\toprule
\rowcolor{TblHeader}
\multicolumn{5}{c}{\textbf{MIMIC-IV-ECG}} \\
\midrule
Type & EM & $\mu$F1 & MF1 & Maj \\
\midrule
Overall        & 43.75 & 59.36 & 64.10 & 24.83 \\
Single-Verify  & 81.06 & 81.06 & 81.06 & 66.45 \\
Single-Choose  & 68.37 & 73.49 & 74.57 & 30.91 \\
Single-Query   & 19.69 & 53.70 & 53.39 &  5.61 \\
\bottomrule
\end{tabularx}
\end{minipage}

\caption{
ECG-QA performance on PTB-XL and MIMIC-IV-ECG test sets.
EM is exact-match accuracy on the set of selected options, $\mu$F1 is micro-averaged F1 over all question--option decisions, MF1 is macro-averaged F1, and Maj is a majority baseline that always predicts the most frequent option set for each question type.
}
\label{tab:ecgqa_ptbxl_mimic}
\end{table}

%% file: tables/ecgqa_ptbxl_baselines.tex
\begin{table}[H]
\centering
\small
\rowcolors{3}{TblStripe}{white}
\begin{tabularx}{0.95\columnwidth}{l *{5}{>{\centering\arraybackslash}X}}
\toprule
\rowcolor{TblHeader}
\multicolumn{6}{c}{\textbf{PTB-XL}} \\
\midrule
Type & Majority & Fusion~\cite{ecgqa} & MedViLL~\cite{medViLL} & M3AE~\cite{M3A} & \textbf{ECG-LLM} \\
\midrule
Single-Verify & 67.73 & 72.10 & 73.90 & 74.60 & \textbf{76.71} \\
Single-Choose & 31.15 & 49.20 & 54.40 & 57.10 & \textbf{68.32} \\
Single-Query & 23.19 & 32.00 & 38.20 & \textbf{41.00} & 38.26 \\
\bottomrule
\end{tabularx}
\caption{Exact-match comparison on PTB-XL Single-Verify, Single-Choose, and Single-Query. Baseline values (Fusion, MedViLL~\cite{medViLL}, M3AE~\cite{M3A}) are taken from the ECG-QA benchmark and correspond to supervised multi-label classifiers over a fixed ECG-QA answer set. ECG-LLM instead generates free-text answers, which are parsed into label sets and evaluated with the same exact-match metric. Best score in each row is bolded.}
\label{tab:ecgqa_ptbxl_baselines}
\end{table}

%% file: sections/discussion.tex
This work introduces ECG-LLM, an ECG-conditioned language model for flexible cardiac question answering from a 12-lead ECG. Our central contribution is to use language as a common supervision space for heterogeneous cardiovascular observations. ECG measurements and reports, CMR-derived phenotypes, echocardiographic phenotypes, and diagnostic labels can all be expressed as clinically grounded question--answer pairs and used to train a single ECG-conditioned model. The work is motivated by broader shifts in ECG AI: from conventional ECG-derived interpretation toward language-based ECG assistants, and from ECG-only phenotype prediction toward cross-modal learning of cardiac phenotypes that are not directly measured on the ECG~\cite{ecgchat,PULSE_ECG,ecgdoctor,selivanov2025ecg_cmr_contrastive,GaoYua_EchoingECG_MICCAI2025}. Our emphasis is to combine these directions. ECG-LLM uses question--answer supervision to connect raw ECG morphology with both conventional ECG concepts and imaging-derived structural and functional phenotypes, including ventricular function, chamber size, wall thickness, and valve- or pressure-related findings. In a front-line setting, this creates an ECG-only interface through which clinicians can ask focused questions about the ECG and broader cardiovascular state to support triage, follow-up imaging, or referral prioritization.

This design has several practical advantages. With the same model, a single ECG representation can support open-ended ECG questions, imaging-related phenotype queries, benchmark-style QA, and report-style summaries without changing the architecture for each task. This is clinically attractive because the relevant question often depends on symptoms, context, and uncertainty, and may not be known when a model is designed. The language-supervision framework also supports scalability. Many cohorts may not be able to share raw CMR or echocardiography images, but structured reports, measurements, or extracted phenotypes can often be represented as text; once expressed in this form, they can enter the same QA-generation pipeline. Thus, ECG-LLM can be trained from ECG waveforms paired with text-derived supervision from multiple modalities, without requiring raw imaging data during instruction tuning. By including both ECG-derived and imaging-derived phenotypes in QA generation, the model is encouraged to learn associations between ECG morphology, conventional ECG interpretation, and broader cardiovascular status rather than only ECG-machine-style outputs. The ECG-QA benchmark and report-generation experiments therefore serve mainly as grounding checks, showing that the model has learned ECG concepts and a mapping between ECG signal patterns and conventional ECG language, while the broader aim is flexible second-opinion support for triage and referral prioritization.

Several limitations remain. First, ECG-LLM is still an LLM-based system, so hallucination and overconfident generation remain important risks. These risks can arise during free-text answering, structured extraction, and LLM-based evaluation, and they are especially important in clinical settings where plausible wording can mask incorrect content. Prior work has shown that medical LLMs can encode useful clinical knowledge while still producing factual errors and potentially harmful recommendations, with performance sensitive to prompting and input formulation~\cite{Hager2024}. ECG-LLM should therefore be viewed as clinician-facing decision support, not as an autonomous diagnostic system or a patient-facing replacement for expert interpretation. Clinical human evaluation also remains necessary before deployment. We partly addressed the absence of large-scale expert review by using an LLM judge for scalable report-generation assessment, but this should be viewed as a screening and comparison tool rather than a substitute for cardiologist evaluation, especially for safety-critical errors, subtle findings, and clinical actionability.

Second, some evaluated phenotypes are continuous measurements that were converted into discrete categories. This approach facilitates evaluation and is better suited to language-model supervision, as textual categories are generally more natural targets for generative models than exact numerical regression. However, discretization introduces boundary effects, whereby values close to a threshold may be clinically borderline but still counted as categorical errors. The free-text formulation partly alleviates this issue by allowing the model to describe a finding as borderline or uncertain, although future work should more directly incorporate continuous uncertainty, calibrated numerical estimates, and clinically meaningful threshold handling. Dataset imbalance represents an additional limitation, particularly in UK Biobank~\cite{ukbiobank}, where many cardiac phenotypes are dominated by normal values. This imbalance can reduce the representation of abnormal findings during training and evaluation. We partly mitigate this imbalance by incorporating datasets with a higher prevalence of abnormal findings and by enriching the UK Biobank training data during QA generation. For studies containing at least one abnormal field, we retained the available abnormal findings and generated additional QA pairs to increase the model's exposure to non-normal phenotypes. 

Third, ECG-LLM infers imaging-derived cardiovascular phenotypes from patterns represented in the ECG rather than directly measuring cardiac structure or function. Previous studies have demonstrated that ECGs contain predictive information about ECHO- and CMR-derived structural and functional phenotypes, supporting the feasibility of learning cross-modal cardiovascular information from ECG signals~\cite{echonext,selivanov2025ecg_cmr_contrastive}. Nevertheless, the strength of this association varies across phenotypes, and some imaging-derived findings may be more readily inferred from ECG morphology than others. 

Architecturally, ECG-LLM currently passes all ECG encoder tokens to the language model through the projector, allowing the model to retain the full encoded ECG representation. Future work could explore more compact conditioning strategies, such as learnable pooling or attention-based token selection, to summarize the ECG sequence into a smaller set of task-relevant latent tokens before passing it to the LLM. Such approaches may improve computational efficiency and reduce context length while maintaining the diagnostically relevant information contained in the ECG.

Overall, ECG-LLM shows that knowledge derived from multiple cardiac modalities can be used as supervision for an ECG-conditioned question-answering model. By representing cardiovascular knowledge from imaging and non-imaging tests as language supervision, the approach turns the widely available 12-lead ECG into a more flexible interface for front-line cardiovascular reasoning when imaging or specialist review is delayed or unavailable. This expands the role of ECG-language modelling from generating reports toward answering clinically meaningful questions about a patient's broader cardiovascular status.

%% file: sections/methods.tex
\subsection{Overview}
\label{sec:method}
ECG-LLM follows the common vision--language model design used in systems such as MiniGPT-4 and LLaVA~\cite{zhu2024minigpt,llava,llava_15}: a pretrained signal encoder extracts modality-specific tokens, a lightweight projector maps these tokens into the language-model embedding space, and an autoregressive LLM generates the text output. In our case, the visual encoder is replaced by a self-supervised ECG encoder, and the resulting model is trained to answer cardiac questions conditioned on the ECG.
More specifically, given a 12-lead ECG $x \in \mathbb{R}^{C \times T}$ (with $C=12$ leads and $T$ time samples) and a question $q$, the ECG encoder produces latent representations $h = f_{\text{ecg}}(x)$, which are mapped into the LLM embedding space by a projector $z = g_{\text{proj}}(h)$. The LLM $f_{\text{llm}}$ then generates an answer token sequence $y = (y_1,\dots,y_N)$ autoregressively conditioned on $(z,q)$. We train the model using the standard causal language modelling (next-token prediction) loss on the ground-truth answer tokens:
\begin{equation}
\label{eq:causal_llm}
\mathcal{L} = -\sum_{n=1}^{N} \log p\left(y_n \,\middle|\, y_{<n}, q, z\right), \qquad z = g_{\text{proj}}(f_{\text{ecg}}(x)).
\end{equation}

Our aim is to study how much clinically relevant information can be transferred into an ECG-conditioned language model through multimodal language supervision, while keeping inference strictly ECG-only. We therefore avoid task-specific ECG heads, retrieval pipelines, and prompt augmentation with precomputed ECG measurements at test time. Our model receives only the raw 12-lead ECG and the user's question. This design allows us to isolate what can be inferred directly from the ECG representation under a unified free-text question-answering setup.

\subsection{Architecture}
\label{sec:architecture}

\paragraph{ECG encoder.}
We use a self-supervised ECG encoder $f_{\text{ecg}}$ pretrained with a masked autoencoder (MAE) objective~\cite{MAE}. The encoder follows a ViT-MAE design adapted to 12-lead ECGs. A patch embedding layer tokenizes each ECG into non-overlapping temporal patches, learnable positional embeddings are added, and a random subset of tokens is masked during pretraining. A Transformer~\cite{transformer} encoder processes the visible tokens, and a lightweight Transformer decoder is trained to reconstruct the masked patches. Let $\mathcal{M}$ denote the set of masked token positions, $x_i$ the original ECG patch at masked position $i$, and $\hat{x}_i$ the corresponding reconstructed patch. The MAE reconstruction loss is:
\begin{equation}
\mathcal{L}_{\mathrm{MAE}}
=
\frac{1}{|\mathcal{M}|}
\sum_{i \in \mathcal{M}}
\left\|
\hat{x}_i - x_i
\right\|_2^2.
\label{eq:mae_loss}
\end{equation}
Here, $|\mathcal{M}|$ is the number of masked tokens, and the loss is computed only over masked patches. After pretraining, we discard the decoder and retain only the encoder for ECG-LLM training.

We use MAE pretraining rather than ECG--text or ECG--imaging contrastive pretraining to keep the ECG encoder general-purpose and independent of paired reports or imaging data. This objective is well suited to ECGs because 12-lead traces contain local temporal structure and inter-lead redundancy, allowing masked segments to be reconstructed from surrounding waveform information.

\paragraph{Projector.}
The projector $g_{\text{proj}}$ maps ECG encoder tokens into the LLM input embedding space. We write it as:
\begin{equation}
z = g_{\text{proj}}(h) = g_{\text{map}}(g_{\text{agg}}(h)),
\label{eq:projector}
\end{equation}
where $g_{\text{agg}}$ controls token aggregation and $g_{\text{map}}$ performs the embedding-space mapping. In this work, we use identity aggregation, so all ECG tokens are preserved. The mapping module is a token-wise two-layer MLP with an activation function between the two linear layers.

\paragraph{Language model.}
The language component of ECG-LLM is an autoregressive LLM, Llama-3.1-8B-Instruct~\cite{meta_llama3_1_blog,meta_llama3_1_8b_instruct}. The LLM receives the projected ECG tokens together with the tokenized question and predicts the answer tokens using the causal language-modelling objective (Eq.~\ref{eq:causal_llm}). Each example is formatted as a chat conversation: a fixed system prompt defines the model as a cardiology assistant, the user message begins with "Here is the ECG:" followed by the ECG tokens and the clinical question, and the assistant message contains the target answer. The full chat template is provided in the Supplementary Section~\ref{sec:supp_ecgllm_chat_template}. During ECG-LLM training, the original LLM parameters are kept frozen, and only LoRA adapters~\cite{Lora} are trained in selected self-attention projections. This reduces the risk of catastrophic forgetting and helps preserve the model's broad general knowledge while still allowing adaptation to ECG-conditioned question answering. All implementation details are provided in Section~\ref{sec:training_implementation} and Supplementary Section~\ref{sec:implementation_details}.

\subsection{ECG cohorts and preprocessing}
We use four 12-lead ECG cohorts to train and evaluate ECG-LLM: UK Biobank~\cite{ukbiobank}, MIMIC-IV-ECG~\cite{mimic_ecg}, PTB-XL~\cite{ptbxl}, and EchoNext~\cite{echonext}. These cohorts were chosen because they provide complementary views of cardiac status. MIMIC-IV-ECG and PTB-XL support conventional ECG interpretation through measurements, diagnostic statements, reports, and structured diagnostic labels. UK Biobank links ECGs to CMR-derived cardiac structure and function. EchoNext links ECGs to echocardiographic measurements and structural heart disease labels. Together, they allow the model to learn from both ECG-visible findings and imaging-related cardiac phenotypes. In total, the study includes 918{,}819 ECG recordings from 251{,}492 unique patients. The training split contains 679{,}112 ECGs from 186{,}409 patients. The validation split contains 68{,}151 ECGs from 18{,}615 patients. The test split contains 171{,}556 ECGs from 46{,}468 patients. Dataset split statistics are provided in Supplementary Table~\ref{tab:ecg_dataset_split_stats}. For MIMIC-IV-ECG and PTB-XL, we adopt the official ECG-QA~\cite{ecgqa} splits to preserve benchmark comparability. For EchoNext, we use the official dataset splits. For UK Biobank, we construct a subject-wise split with $85\%$ of patients assigned to the training set and $15\%$ to the test set, which avoids patient overlap between training and testing.

For all datasets, we harmonise ECG waveforms to a shared 12-lead representation for the common ECG encoder. UK Biobank, MIMIC-IV-ECG, and PTB-XL waveforms are retained in physical millivolt units, while EchoNext waveforms are first denormalised using the released per-lead normalisation parameters and converted back to millivolts. We then reorder all ECGs to the PTB-XL lead order [I, II, III, aVL, aVR, aVF, V1--V6]. EchoNext ECGs are additionally resampled from 250~Hz to 500~Hz to match the common temporal grid. When baseline correction has not already been applied by the acquisition system, slow non-physiological baseline drift is removed lead-wise using a level-10 Daubechies-2 discrete wavelet transform by discarding the approximation coefficients before signal reconstruction, thereby reducing baseline wander while preserving diagnostically relevant ECG morphology~\cite{cuesta2000ecg_baseline_wavelet}.

\subsection{Clinical question-answer supervision from cardiac records}
\label{sec:qa_gen}

To cover diverse cardiac concepts and support targeted questions that arise in general practice or front-line care, we use question-answer (QA) supervision. This format expresses dense cardiac information in natural clinical language and matches how clinicians ask patient-specific questions. QA pairs are also a standard format for instruction tuning multimodal assistants~\cite{llava,llavamed2023}. In our setting, this strategy is useful because large-scale ECG datasets rarely provide diverse text instructions for question-driven cardiac reasoning.

\begin{figure}[!t]
\centering
\includegraphics[width=\textwidth]{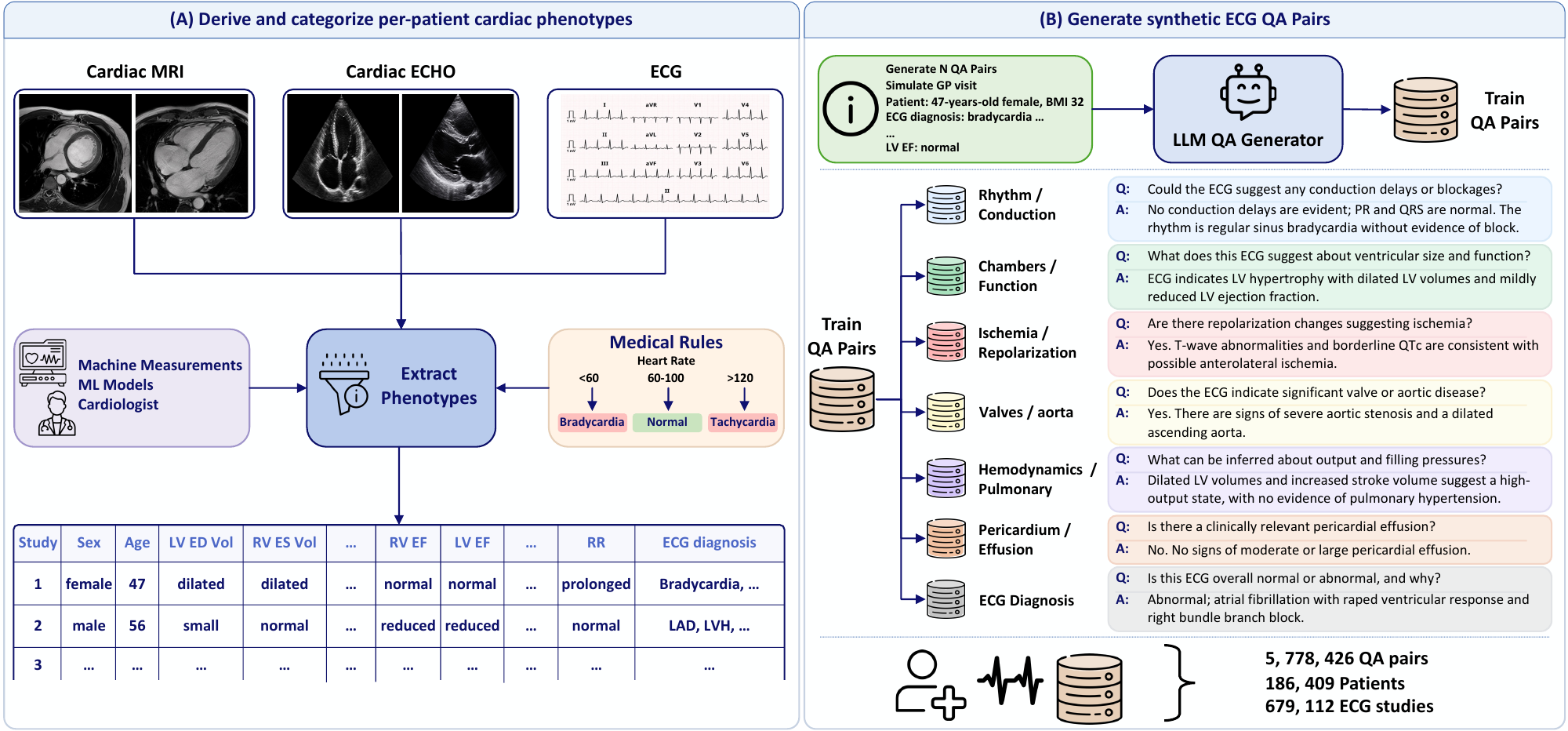}
\caption{
Clinical QA supervision from cardiac records.
(A) ECG, CMR, echocardiography, demographic, and symptom information are converted into structured per-study cardiac records.
(B) A locally hosted instruction-tuned LLM uses each structured record to generate ECG-focused QA pairs that simulate front-line clinical questions.
}
\label{fig:qa_gen}
\end{figure}

We therefore construct ECG-specific instruction-tuning data from structured cardiac records (Figure~\ref{fig:qa_gen}). We refer to these generated examples as structured-record QA, because the answer source is a per-study cardiac record created in this work. For each ECG study, we collect the available information linked to that ECG, including ECG measurements, diagnostic statements or reports, imaging-derived phenotypes when available, and patient demographics or symptoms when available. These data are converted into a structured text record. A QA generator then uses the record to produce clinically grounded front-line cardiac questions and answers. The generator does not see the raw ECG waveform. After generation, each QA pair is paired with the corresponding ECG waveform for ECG-LLM training, so that ECG-LLM learns to answer from ECG tokens and question tokens at inference time.

The structured records combine complementary information from the four cohorts. MIMIC-IV-ECG~\cite{mimic_ecg} provides device-derived ECG measurements and diagnostic statements, while PTB-XL~\cite{ptbxl} provides cardiologist ECG reports and structured SCP-ECG diagnostic labels. UK Biobank~\cite{ukbiobank} provides device-derived ECG measurements together with CMR-derived cardiac phenotypes extracted from long-axis and short-axis cardiac MRI using a machine-learning segmentation pipeline~\cite{Bai2020-id}. EchoNext~\cite{echonext} provides device-derived ECG measurements together with echocardiographic measurements and binary structural heart disease labels. The complete list of dataset-specific ECG, CMR, echocardiographic, demographic, symptom, and report-derived fields used to construct the structured records is provided in Supplementary Table~\ref{tab:supp_training_fields}. 

Before prompting the QA generator, continuous measurements are converted into clinically meaningful categorical descriptors. This makes the records closer to clinical language, reduces dependence on exact numerical values, and avoids asking the generator to perform thresholding inside the prompt. Prior work has shown that language models can be unreliable on exact numerical reasoning compared with symbolic or textual representations~\cite{google_transformers_count}. We therefore apply rule-based categorization across all four cohorts using clinical ECG thresholds, sex-specific CMR reference ranges, ventricular-function thresholds, wall-thickness categories, strain reference ranges, and age- and sex-specific aortic-distensibility ranges. When sex, age, or body surface area is available, the corresponding sex-specific, age-specific, or indexed reference rule is applied. The complete rule set is provided in Supplementary Table~\ref{tab:supp_categorization_rules}.

Each structured record is prompted independently using GPT-OSS-20B~\cite{openai_gptoss_blog,openai_gptoss_hf,openai2025gptoss120bgptoss20bmodel}. The prompt describes a front-line care scenario in which a clinician asks clinically meaningful ECG-based questions and an expert ECG interpreter answers using the cardiac record. The generator is instructed to ask ECG-focused questions grounded in the record, express imaging-derived descriptors as ECG-inferable cardiac findings without explicitly mentioning CMR, echocardiography, or imaging, and produce answers faithful to the supplied information. This design lets imaging-derived supervision teach the ECG-conditioned model about broader cardiac phenotypes while keeping inference ECG-only. To encourage lexical diversity, we randomize the order of phenotype fields and add a study-specific randomization token before prompting the QA generator. The generation procedure is applied independently to each ECG study and yields triplets $(x,q,a)$, where $x$ is the ECG waveform, $q$ is a front-line cardiac question, and $a$ is the corresponding grounded free-text answer.

For UK Biobank and EchoNext, where ECGs are paired with imaging-derived phenotypes, we also use the same structured-record setup to generate chain-of-thought QA pairs. These examples are intended to encourage clinically plausible intermediate reasoning between ECG-visible findings and downstream structural or functional cardiac phenotypes, while keeping the final answer grounded in the same record. Prior work has shown that intermediate reasoning traces can improve complex reasoning in large language models and can serve as useful supervision during rationale-based distillation~\cite{chain_of_thought,chain_of_thought_2}. For the UK Biobank specifically, where most participants are normal for many phenotypes, we additionally generate an abnormal-enriched subset. We retain only studies with at least one abnormal imaging-derived cardiac phenotype. For each retained study, we create an abnormal-focused record by keeping only non-normal findings, including abnormal CMR phenotype descriptors, abnormal ECG measurements, and ECG diagnosis text when available. This increases exposure to clinically important abnormal findings and partially mitigates the dominance of normal cases in UK Biobank. 

In addition to structured-record QA, we include two external ECG-language sources to strengthen conventional ECG interpretation. First, we use ECG-QA training examples~\cite{ecgqa}, which provide established ECG-conditioned question-answering tasks with candidate options. These examples anchor the model to standard ECG measurement and diagnostic QA. Second, we use selected items from the ECGInstruct dataset~\cite{ecginstruct_hf}. We retain open-ended and close-ended questions from feature, rhythm, morphology, and report tasks because they correspond to clinically interpretable ECG findings, rhythm assessment, waveform morphology, and summary reporting. These tasks complement the structured-record QA without shifting the model away from front-line cardiac question answering. ECG-Instruct prompts are lightly cleaned to remove image placeholders and role-play preambles and to replace image-specific wording with ECG-specific wording. All external QA examples are converted to the same ECG-LLM chat format used for structured-record QA (Supplementary Section~\ref{sec:supp_ecgllm_chat_template}). The final instruction-tuning corpus composition by dataset, QA source, and QA subset or task is provided in Supplementary Table~\ref{tab:ecg_instruction_corpus_composition}.

\subsection{Training and implementation details}
\label{sec:training_implementation}

We train ECG-LLM on the ECG-specific QA corpus described above. The ECG encoder described in Section~\ref{sec:architecture} is first pretrained as a ViT-MAE~\cite{MAE} on ECGs from the four training cohorts: UK Biobank~\cite{ukbiobank}, MIMIC-IV-ECG~\cite{mimic_ecg}, PTB-XL~\cite{ptbxl}, and EchoNext~\cite{echonext}. During MAE pretraining, ECGs are randomly cropped to 2{,}500 time steps and augmented with Fourier-transform surrogate perturbation, jitter, and rescaling. The augmented ECGs are tokenized using patches of size $1 \times 100$, yielding 300 ECG tokens before masking. The MAE objective masks 75\% of tokens and reconstructs the masked patches. The pretraining batch size is 768, and optimisation uses bfloat16 mixed precision. Full ECG encoder pretraining details are provided in Supplementary Table~\ref{tab:supp_ecg_mae_pretraining_config}.

After ECG encoder pretraining, we train ECG-LLM for ECG-conditioned question answering. The pretrained ECG encoder and original LLM backbone remain frozen. The trainable parameters are the ECG projector and LoRA adapter weights~\cite{Lora}. LoRA adapters are attached to the query, key, value, and output projections of self-attention layers, with rank 128, scaling factor 128, and dropout 0.1. We use a two-phase curriculum to stabilise ECG--language alignment. The first phase lets the projector move ECG tokens into an approximate region of the LLM embedding space before LLM-side adaptation begins. This reduces the chance that early joint updates force the LoRA adapters to compensate for poorly scaled ECG tokens or rely mainly on the LLM's text priors instead of learning ECG-conditioned representations. For the first 1{,}500 optimiser steps, only the projector is trained, after which the projector and LoRA adapters are trained jointly under the causal language-modelling objective. We optimise trainable parameters with paged AdamW 8-bit~\cite{paged_adam_8bit}. The projector learning rate is $10^{-4}$, the LoRA learning rate is $10^{-5}$, and weight decay is $10^{-3}$. We use linear warm-up over 3\% of optimiser steps followed by cosine decay to 0.3 of the peak learning rate. ECG-LLM training is performed for one epoch using bfloat16 precision on four NVIDIA A100-SXM4-80GB GPUs. The per-device batch size is 8 with gradient accumulation over 4 steps, giving an effective batch size of 128 ECG--question--answer examples. During ECG-LLM training, ECGs are randomly cropped to 2{,}500 time steps, and no additional ECG augmentations are applied. The maximum tokenized sequence length is 1{,}200 tokens, covering 99.5\% of training examples. Longer examples are truncated. Full implementation details are provided in Supplementary Table~\ref{tab:supp_ecg_llm_instruction_config}.

%% file: sections/data_availability.tex
The source datasets analysed in this study are available through public or controlled-access repositories. UK Biobank data are available to approved researchers following application through the UK Biobank Research Analysis Platform (\href{https://www.ukbiobank.ac.uk/enable-your-research/apply-for-access}{link}). For this study, permission to access and analyze the UK
Biobank data was approved under application 87802, with initial approval
granted in July 2022. MIMIC-IV-ECG is available to credentialed users through PhysioNet under the applicable data-use agreement (\href{https://doi.org/10.13026/4nqg-sb35}{link}). PTB-XL is publicly available through PhysioNet (\href{https://doi.org/10.13026/kfzx-aw45}{link}), and EchoNext is available through PhysioNet under its applicable access conditions (\href{https://doi.org/10.13026/3ykd-bf14}{link}). The original source data cannot be redistributed by the authors because they remain subject to the access conditions and data-use agreements of the respective data providers. Derived ECG question--answer supervision data, cohort-specific metadata mappings, and evaluation splits generated in this study are planned for release upon publication to the extent permitted by the corresponding data-use agreements.

%% file: sections/code_availability.tex
The underlying code for this study, including preprocessing, question--answer generation, model training and fine-tuning, inference, evaluation, and reproducibility scripts, will be available in the GitHub repository at \href{https://github.com/placeholder/}{https://github.com/placeholder/} upon publication. Model checkpoints for the main experiments will also be made available through the repository, subject to the applicable licences and data-use agreements.

%% file: sections/acknowledgments.tex
This research has been conducted using the UK Biobank Resource under Application Number 87802. This work is funded by the European Research Council (ERC) project Deep4MI (884622).

%% file: sections/author_contributions.tex
A.S. and D.R. conceived and designed the study. A.S. developed the methodology, implemented the models, performed the experiments and analyses, and wrote the original manuscript. F.J., J.K., K.-L.L., and E.M. provided clinical expertise and contributed to the interpretation of the results. D.R. supervised the study and provided technical and scientific guidance. All authors critically reviewed, revised, and approved the final manuscript.

%% file: sections/competing_interests.tex
The authors declare no competing interests.

%% file: sections/supplementary.tex
\section*{Supplementary Information}

\setcounter{section}{0}
\renewcommand{\thesection}{\arabic{section}}

\newcommand{\SuppSection}[1]{%
    \clearpage
    \section{#1}%
}

\section{Dataset statistics}

Table~\ref{tab:ecg_dataset_split_stats} shows the number of ECG recordings and patients in each dataset split. Table~\ref{tab:ecg_instruction_corpus_composition} summarizes the final ECG-LLM instruction-tuning corpus.

\input{tables/ecg_dataset_stats}
\input{tables/ecg_instruction_corpus_stats}

\SuppSection{UK Biobank ECG and CMR phenotype question bank}
\input{tables/ukb_question_bank}

\SuppSection{UK Biobank phenotype extraction protocol}
\label{sec:supp_ukb_phenotype_extraction}

Because ECG-LLM was trained to generate free-text answers, the UK Biobank phenotype evaluation required an additional mapping step from generated text to the predefined phenotype-category schema. We used the Google LangExtract framework~\cite{langextract,google_langextract} to apply a schema-constrained extraction prompt to ECG-LLM answers. LangExtract is a Python library for extracting structured information from unstructured text using LLMs and user-defined extraction instructions. In our setup, the backend model was a locally deployed GPT-OSS-20B instance~\cite{openai_gptoss_blog,openai_gptoss_hf,openai2025gptoss120bgptoss20bmodel}.

For each ECG and question group, the extractor received the generated answer, the target phenotype fields assigned to that question group, and the allowed category labels for each target phenotype. It was instructed to return allowed category labels only for those target fields found in the ECG-LLM answers. This group-specific extraction kept the evaluation aligned with the clinical intent of the question: the model was neither rewarded nor penalized for mentioning unrelated phenotypes outside the asked group. This step was necessary because a free-text answer could describe several related phenotypes together, refer to multiple measurements collectively, or express a category without repeating the exact phenotype name used in the evaluation schema. Therefore, simple string matching or regular expressions would not reliably map answers to field-specific category labels. The extractor served only as a mapping layer from free text to the predefined category space used for evaluation.

\begin{tcolorbox}[
    title=Prompt for schema-constrained phenotype extraction,
    colback=gray!5,
    colframe=gray!50,
    boxrule=0.5pt,
    arc=2pt,
    left=2pt,right=2pt,top=2pt,bottom=2pt,
    fonttitle=\bfseries,
    breakable
]
\label{box:supp_ukb_extractor_prompt}
\footnotesize
\begin{Verbatim}[breaklines, breakanywhere]
Extract discrete cardiac phenotype categories.
Use only the phenotype field names listed below as possible extraction classes.
Use the allowed category words as the extraction text.
Do not extract numeric values or explanations.
If the answer says several named phenotypes are normal, extract normal for each named phenotype.
Sometimes, phenotype fields may not contain dimension information in square brackets, this is still valid.
If the answer says one phenotype is abnormal but does not give a discrete allowed category, do not extract it.
ALLOWED PHENOTYPES AND CATEGORIES:
- <phenotype field 1>: <allowed category 1>, <allowed category 2>, ...
- <phenotype field 2>: <allowed category 1>, <allowed category 2>, ...
- ...
\end{Verbatim}
\end{tcolorbox}

\FloatBarrier

\SuppSection{UK Biobank phenotype extraction results}
\input{tables/ukb_full_results}

\SuppSection{EchoNext ECG-only binary phenotype questions}
\input{tables/echonext_questions}

\SuppSection{PTB-XL ECG-QA performance by question format and attribute}
\label{sec:supp_ptbxl_ecgqa_qtype_attribute_type}
\input{tables/supp_ptbxl_ecgqa_qtype_attribute_type}

\SuppSection{MIMIC-IV-ECG ECG-QA performance by question format and attribute}
\label{sec:supp_mimic_ecgqa_qtype_attribute_type}
\input{tables/supp_mimic_ecgqa_qtype_attribute_type}

\SuppSection{ECG-QA Single-Verify performance by attribute}
\label{sec:supp_ecgqa_single_verify_attribute}
\input{tables/supp_ecgqa_single_verify_attribute}

\SuppSection{Fields used for structured-record QA generation}
\label{sec:supp_training_fields}
\input{tables/supp_training_fields}

\SuppSection{Clinical categorization rules}
\input{tables/supp_categorization_rules}

\FloatBarrier

\SuppSection{ECG-LLM Chat Template}
\label{sec:supp_ecgllm_chat_template}
During ECG-LLM instruction tuning and evaluation, each example was serialized as a chat conversation with a fixed system message, a user message containing the ECG token placeholder and clinical question, and an assistant message containing the target answer. The ECG token placeholder denotes the projected ECG-token sequence inserted into the LLM input embedding stream.

\begin{tcolorbox}[
    title=ECG-LLM chat template,
    colback=gray!5,
    colframe=gray!50,
    boxrule=0.5pt,
    arc=2pt,
    left=2pt,right=2pt,top=2pt,bottom=2pt,
    fonttitle=\bfseries,
    breakable
]
\footnotesize
\begin{Verbatim}[breaklines, breakanywhere]
System:
You are a cardiology assistant. The user will provide ECG data after the phrase "Here is the ECG:" followed by a clinical question. Answer the question directly using only findings supported by the ECG and available patient context. Do not invent unsupported findings, reason only on the provided ECG.

User:
Here is the ECG: <ecg_tokens>
<question_tokens>

Assistant:
<answer_tokens>
\end{Verbatim}
\end{tcolorbox}

\SuppSection{Question Answer Generation Prompts}
\label{sec:supp_qa_prompts}
The following prompts were used to generate question--answer pairs for training. Dataset-specific prompt variants were used for UK Biobank, EchoNext, MIMIC-IV-ECG, and PTB-XL according to the structured fields available in each cohort. For UK Biobank and EchoNext, we additionally generated reasoning-style QA pairs. In these examples, the generator returned separate \texttt{reasoning} and \texttt{answer} fields. During training, the model input was the generated question, and the target response was formed by concatenating the reasoning field with the final answer field.

\begin{tcolorbox}[
    title=Prompt for standard QA generation with UK Biobank data,
    colback=gray!5,
    colframe=gray!50,
    boxrule=0.5pt,
    arc=2pt,
    left=2pt,right=2pt,top=2pt,bottom=2pt,
    fonttitle=\bfseries,
    breakable
]
\footnotesize
\begin{Verbatim}[breaklines, breakanywhere]
Create N question-answer pairs from this text for LLM training in English.

BEHAVIOR:
- Simulate an interaction via QA pairs between a General Practitioner (GP) and a perfect AI system for ECG interpretation.
- The GP asks questions about this ECG, and the AI answers.
- The GP sees only the ECG and asks clinically meaningful, open-ended questions about this patient to obtain a comprehensive understanding of the patient's cardiac condition and abnormalities.
- The AI may internally use all provided information from the text.
- The AI must always answer as if the insights come purely from ECG interpretation.
- Answers should use as much provided information as possible, combining several fields to provide a comprehensive picture of the patient's cardiac status and abnormalities.
- Do not invent diagnoses, numbers, thresholds, or advice not supported by the provided fields.
- These QA pairs are intended for LLM fine-tuning. Across the full set of generated pairs, they should collectively reflect the patient's overall cardiac state, major abnormalities, and all clinically relevant supported findings in the text.

QUESTIONS (GP):
- Questions must sound like a GP asking about the ECG of this specific patient.
- Questions must not state or assume specific ECG diagnoses, measurements, or values.
- Do not ask about information that is not present in the text.
- Do not include the findings in the question. For example, do not ask: "If sinus bradycardia is present...", "Given LV dilatation...", or "If EF is reduced...".
- Instead, ask the AI to interpret, describe, and highlight abnormalities and to identify all relevant findings.
- Questions may reflect uncertainty or clinical curiosity, for example: "Does the ECG suggest ...?", "Could this ECG indicate ...?", "Do you see any signs of ...?", or "Is there evidence of ...?".

ANSWERS (AI):
- The AI has access to all information in the text, but must formulate answers as if the insights come purely from ECG interpretation.
- Answers must contain several clear sentences, with a direct and unambiguous answer to the question.
- Answers must include as much relevant information as possible about the patient's cardiac condition and abnormalities, combining several fields where appropriate.
- All answers must be directly supported by the provided fields and must never contradict any given measurements.
- Never mention anything not present in the provided text.

OUTPUT FORMAT (STRICT):
Return only valid JSON in this exact format:

[
  {{
    "question": "Question 1?",
    "answer": "Answer 1."
  }},
  {{
    "question": "Question 2?",
    "answer": "Answer 2."
  }}
]

Text:
{text}
\end{Verbatim}
\end{tcolorbox}

\begin{tcolorbox}[
    title=Prompt for reasoning-style QA generation with UK Biobank data,
    colback=gray!5,
    colframe=gray!50,
    boxrule=0.5pt,
    arc=2pt,
    left=2pt,right=2pt,top=2pt,bottom=2pt,
    fonttitle=\bfseries,
    breakable
]
\footnotesize
\begin{Verbatim}[breaklines, breakanywhere]
Create N complex reasoning-style question-answer examples from this text for LLM training in English.

BEHAVIOR:
- Simulate an interaction via QA pairs between a General Practitioner (GP) and a perfect AI system for ECG interpretation.
- The GP asks questions about this ECG, and the AI answers.
- The GP sees only the ECG and asks clinically meaningful, open-ended questions about this patient to obtain a comprehensive understanding of the patient's cardiac condition and abnormalities.
- The AI may internally use all provided information from the text.
- The AI must always answer as if the insights come purely from ECG interpretation.
- Answers should use as much provided information as possible, combining several fields to provide a comprehensive picture of the patient's cardiac status and abnormalities.
- Do not invent diagnoses, numbers, thresholds, or advice not supported by the provided fields.
- These examples are intended for LLM fine-tuning. Across the full set of generated examples, they should collectively reflect the patient's overall cardiac state, major abnormalities, and all clinically relevant supported findings in the text.

QUESTIONS (GP):
- Questions must sound like a GP asking about the ECG of this specific patient.
- Questions must be clinically meaningful and require multi-step interpretation.
- Questions must not state or assume specific ECG diagnoses, measurements, or values.
- Do not ask about information that is not present in the text.
- Do not include the findings in the question. For example, do not ask: "If sinus bradycardia is present...", "Given LV dilatation...", or "If EF is reduced...".
- Instead, ask the AI to interpret, describe, and highlight abnormalities and to identify all relevant findings.
- Questions may reflect uncertainty or clinical curiosity, for example: "Does the ECG suggest ...?", "Could this ECG indicate ...?", "Do you see any signs of ...?", or "Is there evidence of ...?".

ANSWERS (AI):
- The AI has access to all information in the text, but must formulate answers as if the insights come purely from ECG interpretation.
- Each example must contain:
  1. a challenging ECG-focused clinical question,
  2. a concise reasoning field that links the available findings step by step,
  3. a final answer that directly addresses the question.
- The reasoning must be clinically grounded and directly supported by the provided fields.
- The final answer must be concise, direct, and unambiguous.
- All reasoning and answers must be directly supported by the provided fields and must never contradict any given measurements.
- Never mention anything not present in the provided text.

OUTPUT FORMAT (STRICT):
Return only valid JSON in this exact format:

[
  {{
    "question": "Complex question about this ECG?",
    "reasoning": "Step 1: ... Step 2: ... Step 3: ...",
    "answer": "Final answer based on the reasoning."
  }},
  {{
    "question": "Another complex question about this ECG?",
    "reasoning": "Step 1: ... Step 2: ... Step 3: ...",
    "answer": "Final answer based on the reasoning."
  }}
]

Text:
{text}
\end{Verbatim}
\end{tcolorbox}

\begin{tcolorbox}[
    title=Prompt for QA generation with MIMIC-IV-ECG data,
    colback=gray!5,
    colframe=gray!50,
    boxrule=0.5pt,
    arc=2pt,
    left=2pt,right=2pt,top=2pt,bottom=2pt,
    fonttitle=\bfseries,
    breakable
]
\footnotesize
\begin{Verbatim}[breaklines, breakanywhere]
Create N question-answer pairs from this text for LLM training in English.

BEHAVIOR:
- Simulate an interaction between a General Practitioner (GP) and a PERFECT AI for ECG interpretation.
- The GP asks questions about this ECG, and the PERFECT AI answers.
- The GP sees only the ECG and asks clinically meaningful, open-ended questions about THIS patient to obtain a comprehensive understanding of the patient's heart condition and abnormalities.
- PERFECT AI for ECG interpretation may internally use and reason with ALL provided information from the text (ECG phenotypes: + Patient demographics),
  but MUST answer as if insights come purely from perfect ECG interpretation.
- Questions and answers must be phrased as if inferred from ECG alone. Never mention records, reports, metadata, or models.
- Do not invent diagnoses, numbers, thresholds, or advice not supported by the provided fields.
- These QA pairs are intended for LLM fine-tuning. Across the full set of generated pairs, they should collectively reflect the patient's overall cardiac state, major abnormalities, and all clinically relevant supported findings in the text.

QUESTIONS (GP):
- Must sound like a GP interpreting and asking questions about the ECG of this specific patient.
- MUST NOT state or assume specific ECG diagnoses, measurements, or values in the question.
- Do not ask about information that is not present in the text.
- Do not include the findings in the question. Instead, ask the AI to interpret, describe, and highlight abnormalities and to explain all relevant findings.
- May reflect uncertainty or clinical curiosity: "Does the ECG suggest ...?", "Could this ECG
  indicate...?", "Do you see any signs of...", "Is there evidence of...", etc.

ANSWERS (AI):
- This is a PERFECT AI for ECG interpretation. It has access to all information in the text, but must formulate answers as if insights come purely from ECG interpretation.
- Answers must be several clear sentences, with a direct and unambiguous answer to the question.
- Answers must include as much relevant information as possible about the patient's heart condition and abnormalities, combining several fields.
- All answers must be directly supported by the provided fields and must never contradict any given measurements.
- Never mention anything not present in the provided text.

OUTPUT FORMAT (STRICT):
Return ONLY valid JSON in this exact format:

[
  {{
    "question": "Question 1?",
    "answer": "Answer 1."
  }},
  {{
    "question": "Question 2?",
    "answer": "Answer 2."
  }}
]

Text:
{text}
\end{Verbatim}
\end{tcolorbox}

\begin{tcolorbox}[
    title=Prompt for QA generation with PTB-XL data,
    colback=gray!5,
    colframe=gray!50,
    boxrule=0.5pt,
    arc=2pt,
    left=2pt,right=2pt,top=2pt,bottom=2pt,
    fonttitle=\bfseries,
    breakable
]
\footnotesize
\begin{Verbatim}[breaklines, breakanywhere]
Create N question-answer pairs from this text for LLM training in English.

BEHAVIOR:
- Simulate an interaction between a General Practitioner (GP) and a PERFECT AI for ECG interpretation.
- The GP asks questions about this ECG, and the PERFECT AI answers.
- The GP sees only the ECG and asks clinically meaningful, open-ended questions about THIS patient to obtain a comprehensive understanding of the patient's heart condition and abnormalities.
- PERFECT AI for ECG interpretation may internally use and reason with ALL provided information from the text.
- PERFECT AI for ECG interpretation ALWAYS answer as if insights come purely from ECG analysis.
- The output QA pairs MUST be ONLY in English.
- Do not invent diagnoses, numbers, thresholds, or advice not supported by the provided fields.
- These QA pairs are intended for LLM fine-tuning. Across the full set of generated pairs, they should collectively reflect the patient's overall cardiac state, major abnormalities, and all clinically relevant supported findings in the text.
- QA pairs must cover all findings from the text.

QUESTIONS (GP):
- Must sound like a GP interpreting and asking questions about the ECG of this specific patient.
- NEVER include the findings from the text in the question. For example, do NOT say "If sinus bradycardia is present...", "Given LV dilatation...", or "If EF is reduced...". Instead, ask the AI to interpret, describe, and highlight abnormalities and to explain all relevant findings.
- May reflect uncertainty or clinical curiosity: "Does the ECG suggest ...?", "Could this ECG
  indicate...?", "Do you see any signs of...", "Is there evidence of...", etc.
- Do not ask about information that is not present in the text.

ANSWERS (AI):
- This is a PERFECT AI for ECG interpretation. It must formulate answers as if insights come purely from ECG interpretation.
- Answers must be several clear sentences, with a direct and unambiguous answer to the question.
- Answers must include as much relevant information as possible about the patient's heart condition and abnormalities, combining several fields.
- All answers must be directly supported by the provided fields and must never contradict any given measurements.
- Never mention anything not present in the provided text.

OUTPUT FORMAT (STRICT):
Return ONLY valid JSON in this exact format:

[
  {{
    "question": "Question 1?",
    "answer": "Answer 1."
  }},
  {{
    "question": "Question 2?",
    "answer": "Answer 2."
  }}
]

Text:
{text}
\end{Verbatim}
\end{tcolorbox}

\begin{tcolorbox}[
    title=Prompt for QA generation with EchoNext data,
    colback=gray!5,
    colframe=gray!50,
    boxrule=0.5pt,
    arc=2pt,
    left=2pt,right=2pt,top=2pt,bottom=2pt,
    fonttitle=\bfseries,
    breakable
]
\footnotesize
\begin{Verbatim}[breaklines, breakanywhere]
Create N question-answer pairs from this text for LLM training in English.

BEHAVIOR:
- Simulate an interaction via QA pairs between a General Practitioner (GP) and a PERFECT AI for ECG interpretation.
- The GP asks questions about this ECG, and the PERFECT AI answers.
- The GP sees only the ECG and asks clinically meaningful, open-ended questions about THIS patient to obtain a comprehensive understanding of the patient's heart condition and abnormalities.
- PERFECT AI for ECG interpretation may internally use and reason with ALL provided information from the text.
- PERFECT AI for ECG interpretation ALWAYS answer as if insights come purely from ECG.
- Answers should cite as much provided information as possible, combing several fields to provide a comprehensive picture of the patient's cardiac status and abnormalities.
- Do not invent diagnoses, numbers, thresholds, or advice not supported by the provided fields.
- These QA pairs are intended for LLM fine-tuning. Across the full set of generated pairs, they should collectively reflect the patient's overall cardiac state, major abnormalities, and all clinically relevant supported findings in the text.

QUESTIONS (GP):
- Must sound like a GP asking questions about the ECG of this specific patient. GP asks questions.
- MUST NOT state or assume specific ECG diagnoses, measurements, or values in the question.
- Do not ask about information that is not present in the text.
- Do not include the findings in the question. For example, do NOT say "If sinus bradycardia is present...", "Given LV dilatation...", or "If EF is reduced...". 
- Instead, ask the AI to interpret, describe, and highlight abnormalities and to find all relevant abnormalities.
- Questions may reflect uncertainty or clinical curiosity: "Does the ECG suggest ...?", "Could this ECG
  indicate...?", "Do you see any signs of...", "Is there evidence of...", etc.

ANSWERS (AI):
- This is a PERFECT AI for ECG interpretation. It has access to all information in the text, and it must formulate answers as if insights come purely from ECG interpretation.
- Answers must be several clear sentences, with a direct and unambiguous answer to the question.
- Answers must include as much relevant information as possible about the patient's heart condition and abnormalities, combining several fields.
- All answers must be directly supported by the provided fields and must never contradict any given measurements.
- Never mention anything not present in the provided text.

OUTPUT FORMAT (STRICT):
Return ONLY valid JSON in this exact format:

[
  {{
    "question": "Question 1?",
    "answer": "Answer 1."
  }},
  {{
    "question": "Question 2?",
    "answer": "Answer 2."
  }}
]

Text:
{text}
\end{Verbatim}
\end{tcolorbox}

\begin{tcolorbox}[
    title=Prompt for reasoning-style QA generation with EchoNext data,
    colback=gray!5,
    colframe=gray!50,
    boxrule=0.5pt,
    arc=2pt,
    left=2pt,right=2pt,top=2pt,bottom=2pt,
    fonttitle=\bfseries,
    breakable
]
\footnotesize
\begin{Verbatim}[breaklines, breakanywhere]
Create N complex reasoning examples from this text that demonstrate chain-of-thought thinking.
        
BEHAVIOR:
- Simulate an interaction via QA pairs between a General Practitioner (GP) and a PERFECT AI for ECG interpretation.
- The GP asks questions about this ECG, and the PERFECT AI answers.
- The GP sees only the ECG and asks clinically meaningful, open-ended questions about THIS patient to obtain a comprehensive understanding of the patient's heart condition and abnormalities.
- PERFECT AI for ECG interpretation may internally use and reason with ALL provided information from the text.
- PERFECT AI for ECG interpretation ALWAYS answer as if insights come purely from ECG.
- Answers should cite as much provided information as possible, combing several fields to provide a comprehensive picture of the patient's cardiac status and abnormalities.
- Do not invent diagnoses, numbers, thresholds, or advice not supported by the provided fields.
- These QA pairs are intended for LLM fine-tuning. Across the full set of generated pairs, they should collectively reflect the patient's overall cardiac state, major abnormalities, and all clinically relevant supported findings in the text.

QUESTIONS (GP):
- Must sound like a GP asking questions about the ECG of this specific patient. GP asks questions.
- MUST NOT state or assume specific ECG diagnoses, measurements, or values in the question.
- Do not ask about information that is not present in the text.
- Do not include the findings in the question. For example, do NOT say "If sinus bradycardia is present...", "Given LV dilatation...", or "If EF is reduced...". 
- Instead, ask the AI to interpret, describe, and highlight abnormalities and to find all relevant abnormalities.
- Questions may reflect uncertainty or clinical curiosity: "Does the ECG suggest ...?", "Could this ECG
  indicate...?", "Do you see any signs of...", "Is there evidence of...", etc.

ANSWERS (AI):
- This is a PERFECT AI for ECG interpretation. It has access to all information in the text, and it must formulate answers as if insights come purely from ECG interpretation.
- Answers must be several clear sentences, with a direct and unambiguous answer to the question.
- Answers must include as much relevant information as possible about the patient's heart condition and abnormalities, combining several fields.
- All answers must be directly supported by the provided fields and must never contradict any given measurements.
- Never mention anything not present in the provided text.

Each example should have:
1. A challenging question that requires step-by-step reasoning
2. Detailed reasoning steps that break down the problem
3. A concise final answer

Return JSON format only:

[
  {{
    "question": "Complex question about the text?",
    "reasoning": "Step 1: First, I need to consider...\nStep 2: Then, I analyze...\nStep 3: Finally, I can conclude...",
    "answer": "Final answer based on the reasoning."
  }},
  {{
    "question": "Another complex question?",
    "reasoning": "First, I'll analyze... Next, I need to determine... Based on this analysis...",
    "answer": "Final answer drawn from the reasoning."
  }}
]

Text:
{text}
\end{Verbatim}
\end{tcolorbox}

\FloatBarrier

\SuppSection{Implementation Details}
\label{sec:implementation_details}
\input{tables/implementation_details}

\FloatBarrier

\SuppSection{Diagnostic report generation prompts}
\label{sec:supp_diagnostic_report_prompts}

\input{tables/supp_diagnostic_report_questions}

\FloatBarrier

\SuppSection{LLM-as-judge prompt for ECG diagnosis evaluation}
\label{sec:llm_as_judge_sec}
The following prompts were used for LLM-based evaluation of generated ECG diagnostic answers against ground-truth diagnostic statements.

\begin{tcolorbox}[
    title=System prompt for LLM-as-judge evaluation,
    colback=gray!5,
    colframe=gray!50,
    boxrule=0.5pt,
    arc=2pt,
    left=2pt,right=2pt,top=2pt,bottom=2pt,
    fonttitle=\bfseries,
    breakable
]
\label{box:supp_llm_judge_prompt}
\scriptsize
\begin{Verbatim}[breaklines, breakanywhere]
You are a strict but clinically fair ECG diagnosis evaluator.

Compare MODEL ANSWER against GROUND TRUTH. Treat clinically equivalent wording as correct. Do not require identical phrasing.
Score correctness, completeness, and specificity independently.

Clinical equivalence rules:
- Borderline ECG, abnormal ECG, nonspecific, possible, probable, normal variant, and age-undetermined findings are lower-confidence/global findings.
- Do not strongly penalize missing uncertain findings if the main ECG interpretation is correct.
- Penalize missing GT findings under completeness and unsupported extra findings under specificity. Do not assign 0 if at least one meaningful GT finding is correctly identified. But don't assign 2 if important GT findings are missing or if the model adds unsupported major findings.

For PTB-XL ground truth: #Added only for PTB-XL evaluation
- Use the Original report as the primary reference.
- Use the English report only as a translation aid.
- Do not treat the English report as an additional independent list of required findings.

Return only valid JSON.
\end{Verbatim}
\end{tcolorbox}

\begin{tcolorbox}[
    title=User prompt template for LLM-as-judge evaluation,
    colback=gray!5,
    colframe=gray!50,
    boxrule=0.5pt,
    arc=2pt,
    left=2pt,right=2pt,top=2pt,bottom=2pt,
    fonttitle=\bfseries,
    breakable
]
\scriptsize
\begin{Verbatim}[breaklines, breakanywhere]
GROUND TRUTH:
{gt}

MODEL ANSWER:
{pred}

Score independently.

Correctness:
2 = main ECG interpretation is clinically correct or nearly correct.
1 = partially correct: some main findings are correct, but there is an important wrong or missing conclusion.
0 = mostly incorrect, or the main diagnosis conclusion is wrong.

Completeness:
2 = all or nearly all important GT diagnoses are present.
1 = at least one important GT diagnosis is present, but relevant GT findings are missing.
0 = no clinically meaningful GT diagnosis is present.

Specificity:
2 = nearly all model diagnoses are supported by or compatible with GT.
1 = mixed answer: some supported diagnoses plus unsupported or wrong extras.
0 = mostly unsupported or dominated by wrong diagnoses.

Return JSON exactly:
{{
  "correctness": int,
  "completeness": int,
  "specificity": int,
  "total_score": int,
  "notes": "reason"
}}
\end{Verbatim}
\end{tcolorbox}

%% file: tables/ecg_dataset_stats.tex
\begin{center}
\centering
\small
\begin{tabularx}{\columnwidth}{l l *{5}{>{\raggedleft\arraybackslash}X}}
\toprule
\rowcolor{TblHeader}
\textbf{Split} & \textbf{Entity} & \textbf{UK Biobank} & \textbf{MIMIC-IV-ECG} & \textbf{PTB-XL} & \textbf{EchoNext} & \textbf{All} \\
\midrule
Train & ECG studies & 34{,}280 & 560{,}791 & 11{,}566 & 72{,}475 & 679{,}112 \\
Train & Patients    & 34{,}280 & 115{,}508 & 10{,}403 & 26{,}218 & 186{,}409 \\
\midrule
Validation & ECG studies & --       & 62{,}251  & 1{,}274  & 4{,}626  & 68{,}151 \\
Validation & Patients    & --       & 12{,}834  & 1{,}155  & 4{,}626  & 18{,}615 \\
\midrule
Test & ECG studies & 6{,}049  & 156{,}851 & 3{,}214  & 5{,}442  & 171{,}556 \\
Test & Patients    & 6{,}049  & 32{,}086  & 2{,}891  & 5{,}442  & 46{,}468 \\
\bottomrule
\end{tabularx}
\captionsetup{hypcap=false}\captionof{table}{
Dataset split statistics. The table reports the number of ECG studies and unique patients in the training, validation, and test splits for each ECG cohort. UK Biobank was split subject-wise into training and test sets only.
}
\label{tab:ecg_dataset_split_stats}
\end{center}

%% file: tables/ecg_instruction_corpus_stats.tex
\begin{center}
\centering
\scriptsize
\setlength{\tabcolsep}{5pt}
\renewcommand{\arraystretch}{0.95}

\begin{tabular}{lllr}
\toprule
\rowcolor{TblHeader}
\textbf{Dataset} & \textbf{QA source} & \textbf{Type} & \textbf{\# Samples} \\
\midrule

UK Biobank
& Structured-record QA
& ECG and CMR, standard
& 670{,}545 \\

& Structured-record QA
& ECG and CMR, abnormal-enriched
& 173{,}238 \\

& Structured-record QA
& ECG and CMR, chain-of-thought
& 167{,}701 \\

\midrule

EchoNext
& Structured-record QA
& ECG and echo, standard
& 546{,}156 \\

& Structured-record QA
& ECG and echo, chain-of-thought
& 129{,}625 \\

\midrule

MIMIC-IV-ECG
& Structured-record QA
& ECG interpretation, standard
& 2{,}775{,}751 \\

& ECG-QA
& Single-Verify
& 125{,}380 \\

& ECG-QA
& Single-Choose
& 107{,}860 \\

& ECG-QA
& Single-Query
& 153{,}156 \\

& ECGInstruct
& Feature (Close/Open)
& 3{,}865 \\

& ECGInstruct
& Rhythm (Close/Open)
& 38{,}124 \\

& ECGInstruct
& Morphology (Close/Open)
& 62{,}663 \\

& ECGInstruct
& Report (Open)
& 345{,}561 \\

\midrule

PTB-XL
& Structured-record QA
& ECG interpretation, standard
& 127{,}219 \\

& ECG-QA
& Single-Verify
& 62{,}554 \\

& ECG-QA
& Single-Choose
& 50{,}015 \\

& ECG-QA
& Single-Query
& 46{,}737 \\

& ECGInstruct
& Feature (Close/Open)
& 57{,}108 \\

& ECGInstruct
& Rhythm (Close/Open)
& 18{,}899 \\

& ECGInstruct
& Morphology (Close/Open)
& 108{,}379 \\

& ECGInstruct
& Report (Open)
& 7{,}890 \\

\midrule

\textbf{Total}
&
&
& \textbf{5{,}778{,}426} \\

\bottomrule
\end{tabular}

\captionsetup{hypcap=false}\captionof{table}{
Final ECG-LLM instruction-tuning corpus. Structured-record QA denotes generated examples produced from the per-study cardiac records created in this work.
}
\label{tab:ecg_instruction_corpus_composition}
\end{center}

%% file: tables/ukb_question_bank.tex
\begin{center}
\centering
\scriptsize
\setlength{\tabcolsep}{3pt}
\renewcommand{\arraystretch}{1.04}
\begin{tabularx}{0.98\textwidth}{
    >{\raggedright\arraybackslash}p{0.20\textwidth}
    >{\raggedright\arraybackslash}p{0.50\textwidth}
    X
}
\toprule
\rowcolor{TblHeader}
\textbf{Question group} & \textbf{ECG-LLM questions} & \textbf{Target phenotypes} \\
\midrule

Heart rate and rhythm
&
\begin{minipage}[t]{\linewidth}
 What does the ECG reveal about the patient's heart rate and rhythm?\\
 Review the ECG and describe the rhythm and rate.
\end{minipage}
&
ECG heart rate [bpm]; RR interval [ms]. \\

P-wave characteristics and atrioventricular intervals
&
\begin{minipage}[t]{\linewidth}
 Review this ECG and describe atrial activity and AV conduction.\\
 Review the ECG and infer the P-wave characteristics and atrioventricular intervals.
\end{minipage}
&
P duration [ms]; PP interval [ms]; PQ interval [ms]. \\

QRS duration and ventricular conduction
&
\begin{minipage}[t]{\linewidth}
 Is the QRS duration within normal limits?\\
 Review the ECG and identify any conduction delays.
\end{minipage}
&
QRS duration [ms]. \\

QT interval and ventricular repolarization
&
\begin{minipage}[t]{\linewidth}
 Are there any repolarization abnormalities on this ECG?\\
 Is the QT interval within normal limits?
\end{minipage}
&
QT interval [ms]; QTC interval [ms]. \\

Electrical axes
&
\begin{minipage}[t]{\linewidth}
 Are the electrical axes within normal limits?\\
 Does the ECG show any axis deviation?
\end{minipage}
&
P axis [degrees]; R axis [degrees]; T axis [degrees]. \\

Left ventricular systolic function
&
\begin{minipage}[t]{\linewidth}
 Review the ECG and describe LV ejection fraction.\\
 What does this ECG reveal about left ventricular systolic function?
\end{minipage}
&
LV ejection fraction. \\

Left ventricular stroke volume
&
\begin{minipage}[t]{\linewidth}
 Review the ECG and describe LV stroke volume.\\
 What does this ECG reveal about left ventricular stroke volume?
\end{minipage}
&
LV stroke volume. \\

Left ventricular volumes
&
\begin{minipage}[t]{\linewidth}
 Review the ECG and describe any abnormalities found for LV.\\
 What does this ECG reveal about left ventricular volumes?
\end{minipage}
&
LV end diastolic volume; LV end systolic volume. \\

Left ventricular myocardial mass and wall thickness
&
\begin{minipage}[t]{\linewidth}
 Is there any evidence of ventricular hypertrophy or increased myocardial mass?\\
 What does this ECG reveal about left ventricular wall thickness?
\end{minipage}
&
LV myocardial mass; LV mean myocardial wall thickness global. \\

Left ventricular strain
&
\begin{minipage}[t]{\linewidth}
 Do the strain measurements suggest any myocardial dysfunction?\\
 Review this ECG and describe any abnormalities in left ventricular strain.
\end{minipage}
&
LV circumferential strain global; LV longitudinal strain global; LV radial strain global. \\

Cardiac output and cardiac index
&
\begin{minipage}[t]{\linewidth}
 What does this ECG indicate about cardiac output and cardiac index?\\
 Examine this ECG and tell if the cardiac output or index within normal limits?
\end{minipage}
&
LV cardiac output; cardiac index [litres/min/m2]. \\

Right ventricular stroke volume
&
\begin{minipage}[t]{\linewidth}
 Review this ECG and describe RV stroke volume.\\
 What does this ECG reveal about right ventricular stroke volume?
\end{minipage}
&
RV stroke volume. \\

Right ventricular ejection fraction
&
\begin{minipage}[t]{\linewidth}
 Review this ECG and describe RV ejection fraction.\\
 What does this ECG reveal about right ventricular systolic function?
\end{minipage}
&
RV ejection fraction. \\

Right ventricular volumes and function
&
\begin{minipage}[t]{\linewidth}
 Review this ECG and describe all abnormalities for RV.\\
 Review the ECG and indicate about abnormalities in right ventricular volumes.
\end{minipage}
&
RV end diastolic volume; RV end systolic volume. \\

Left atrial volumes and function
&
\begin{minipage}[t]{\linewidth}
 What does this ECG suggest about left atrial volumes and function?\\
 Examine this ECG and tell if there is any evidence of left atrial enlargement or dysfunction?
\end{minipage}
&
LA maximum volume; LA minimum volume; LA stroke volume; LA ejection fraction. \\

Right atrial volumes and function
&
\begin{minipage}[t]{\linewidth}
 What does this ECG reveal about right atrial volumes and function?\\
 Examine this ECG and tell if there is any evidence of right atrial enlargement or dysfunction?
\end{minipage}
&
RA maximum volume; RA minimum volume; RA stroke volume; RA ejection fraction. \\

Aortic distensibility
&
\begin{minipage}[t]{\linewidth}
 Review this ECG and state whether there is any indirect evidence that could suggest abnormal aortic or vascular stiffness.
\end{minipage}
&
Ascending aorta distensibility; Descending aorta distensibility. \\

\bottomrule
\end{tabularx}
\captionsetup{hypcap=false}\captionof{table}{
UK Biobank question bank used for ECG-derived and CMR-derived phenotype elicitation from ECG alone.}
\label{tab:supp_ukb_question_bank}
\end{center}

%% file: tables/ukb_full_results.tex
\begin{center}
\centering
\scriptsize
\setlength{\tabcolsep}{2.6pt}
\renewcommand{\arraystretch}{1.04}

\begin{tabularx}{0.98\textwidth}{
    >{\raggedright\arraybackslash}X
    r r r r r r r r r
}
\toprule
\rowcolor{TblHeader}
\textbf{Field} &
\textbf{$n$} &
\textbf{$K$} &
\textbf{Cov.} &
\textbf{Miss.} &
\textbf{Acc.} &
\textbf{BAcc.} &
\textbf{wF1} &
\textbf{Maj.} &
\textbf{mF1} \\
\midrule
\multicolumn{10}{l}{\textit{ECG measurements}} \\
\midrule
ECG heart rate [bpm] & 1000 & 5 & 1.000 & 0.000 & 0.958 & 0.974 & 0.958 & 0.126 & 0.977 \\
R axis [degrees] & 1000 & 3 & 1.000 & 0.000 & 0.987 & 0.919 & 0.987 & 0.315 & 0.940 \\
RR interval [ms] & 1000 & 5 & 1.000 & 0.000 & 0.944 & 0.969 & 0.944 & 0.127 & 0.926 \\
P duration [ms] & 979 & 2 & 1.000 & 0.000 & 0.974 & 0.824 & 0.972 & 0.481 & 0.886 \\
QRS duration [ms] & 1000 & 3 & 0.987 & 0.013 & 0.976 & 0.886 & 0.976 & 0.320 & 0.874 \\
T axis [degrees] & 1000 & 4 & 1.000 & 0.000 & 0.888 & 0.681 & 0.880 & 0.212 & 0.726 \\
QTC interval [ms] & 1000 & 3 & 1.000 & 0.000 & 0.962 & 0.652 & 0.958 & 0.324 & 0.719 \\
PP interval [ms] & 999 & 5 & 1.000 & 0.000 & 0.930 & 0.605 & 0.927 & 0.126 & 0.637 \\
P axis [degrees] & 938 & 3 & 1.000 & 0.000 & 0.908 & 0.540 & 0.888 & 0.310 & 0.614 \\
PQ interval [ms] & 980 & 3 & 1.000 & 0.000 & 0.906 & 0.458 & 0.879 & 0.313 & 0.522 \\

\midrule
\multicolumn{10}{l}{\textit{CMR-derived phenotypes}} \\
\midrule
LV mean myocardial wall thickness global & 990 & 2 & 1.000 & 0.000 & 0.965 & 0.974 & 0.966 & 0.445 & 0.947 \\
RV end diastolic volume & 1000 & 3 & 0.999 & 0.001 & 0.993 & 0.708 & 0.991 & 0.327 & 0.708 \\
LV radial strain global & 833 & 3 & 0.975 & 0.025 & 0.998 & 0.667 & 0.996 & 0.311 & 0.666 \\
RV end systolic volume & 1000 & 3 & 0.999 & 0.001 & 0.995 & 0.667 & 0.993 & 0.327 & 0.641 \\
RA minimum volume & 988 & 3 & 1.000 & 0.000 & 0.910 & 0.594 & 0.885 & 0.314 & 0.592 \\
RA maximum volume & 989 & 3 & 0.999 & 0.001 & 0.906 & 0.569 & 0.894 & 0.312 & 0.560 \\
LA maximum volume & 989 & 3 & 0.999 & 0.001 & 0.950 & 0.473 & 0.937 & 0.322 & 0.547 \\
LV end systolic volume & 998 & 3 & 1.000 & 0.000 & 0.943 & 0.460 & 0.930 & 0.318 & 0.507 \\
LV myocardial mass & 1000 & 2 & 0.977 & 0.023 & 0.950 & 0.495 & 0.934 & 0.488 & 0.487 \\
Ascending aorta distensibility & 872 & 2 & 0.814 & 0.186 & 0.866 & 0.484 & 0.891 & 0.489 & 0.486 \\
RV ejection fraction & 999 & 5 & 1.000 & 0.000 & 0.981 & 0.443 & 0.974 & 0.197 & 0.469 \\
LA minimum volume & 988 & 3 & 1.000 & 0.000 & 0.881 & 0.429 & 0.850 & 0.305 & 0.459 \\
RV stroke volume & 999 & 3 & 1.000 & 0.000 & 0.961 & 0.472 & 0.955 & 0.327 & 0.458 \\
RA stroke volume & 989 & 3 & 0.999 & 0.001 & 0.884 & 0.403 & 0.866 & 0.314 & 0.416 \\
LV end diastolic volume & 1000 & 3 & 0.998 & 0.002 & 0.945 & 0.383 & 0.925 & 0.322 & 0.410 \\
Cardiac index [litres/min/m$^2$] & 843 & 5 & 1.000 & 0.000 & 0.696 & 0.396 & 0.672 & 0.114 & 0.397 \\
LA ejection fraction & 989 & 3 & 0.999 & 0.001 & 0.908 & 0.369 & 0.868 & 0.316 & 0.381 \\
LV ejection fraction & 1000 & 5 & 1.000 & 0.000 & 0.946 & 0.321 & 0.934 & 0.191 & 0.371 \\
LV stroke volume & 1000 & 3 & 1.000 & 0.000 & 0.941 & 0.360 & 0.917 & 0.323 & 0.371 \\
RA ejection fraction & 989 & 3 & 0.999 & 0.001 & 0.876 & 0.361 & 0.831 & 0.310 & 0.363 \\
LV longitudinal strain global & 815 & 3 & 0.996 & 0.004 & 0.966 & 0.346 & 0.950 & 0.327 & 0.352 \\
Descending aorta distensibility & 872 & 3 & 0.810 & 0.190 & 0.857 & 0.358 & 0.887 & 0.326 & 0.339 \\
LA stroke volume & 989 & 3 & 0.999 & 0.001 & 0.956 & 0.333 & 0.935 & 0.326 & 0.326 \\
LV circumferential strain global & 833 & 3 & 0.975 & 0.025 & 0.950 & 0.333 & 0.925 & 0.324 & 0.325 \\
LV cardiac output & 1000 & 5 & 0.843 & 0.157 & 0.766 & 0.271 & 0.695 & 0.170 & 0.277 \\
\bottomrule
\end{tabularx}

\captionsetup{hypcap=false}\captionof{table}{
Full UK Biobank phenotype extraction results for ECG-derived and CMR-derived fields.
The table reports field-level sample size ($n$), number of possible classification classes ($K$), coverage, missing rate, accuracy, balanced accuracy, weighted-F1, majority-class macro-F1, and macro-F1. Fields are ordered by macro-F1 within each group.
}
\label{tab:supp_ukb_full_results}
\end{center}

%% file: tables/echonext_questions.tex
\begin{center}
\centering
\scriptsize
\setlength{\tabcolsep}{5pt}
\renewcommand{\arraystretch}{1.08}
\begin{tabularx}{0.96\textwidth}{>{\raggedright\arraybackslash}p{0.25\textwidth} X}
\toprule
\rowcolor{TblHeader}
\textbf{Phenotype} & \textbf{ECG-LLM question} \\
\midrule
LVEF $\leq 45\%$
& Does this ECG suggest moderately or severely reduced left ventricular systolic function? \\

LVWT $\geq 1.3$ cm
& Does this ECG show moderate left ventricular hypertrophy with the maximum of the interventricular septum (IVS) or posterior wall (LVPW) thickness greater than or equal to 1.3 cm? \\

AS $\geq$ moderate
& Does this ECG show moderate or severe aortic stenosis? \\

AR $\geq$ moderate
& Does this ECG suggest moderate or severe aortic regurgitation? \\

MR $\geq$ moderate
& Does this ECG suggest moderate or severe mitral regurgitation? \\

TR $\geq$ moderate
& Does this ECG suggest moderate or severe tricuspid regurgitation? \\

PR $\geq$ moderate
& Does this ECG suggest moderate or severe pulmonary regurgitation? \\

RV systolic dysf. $\geq$ moderate
& Does this ECG suggest moderate or severe right ventricular systolic dysfunction? \\

Pericardial eff. $\geq$ moderate/large
& Does this ECG suggest a moderate or large pericardial effusion? \\

PASP $\geq 45$ mmHg
& Does this ECG suggest pulmonary hypertension, as would be expected with a pulmonary artery systolic pressure of at least 45 mmHg? \\

TR Vmax $\geq 3.2$ m/s
& Does this ECG suggest a tricuspid regurgitation jet velocity of at least 3.2 m/s? \\
\bottomrule
\end{tabularx}
\captionsetup{hypcap=false}\captionof{table}{
EchoNext phenotype-specific questions used for ECG-only yes/no question answering.
For each ECG and phenotype, ECG-LLM first answered the initial question in free text.
In the same conversation, the model was then asked: ``Give the final answer to the original question as yes or no.''
The final yes/no response was parsed and compared with the corresponding echocardiographic binary label.
}
\label{tab:supp_echonext_questions}
\end{center}

%% file: tables/supp_ptbxl_ecgqa_qtype_attribute_type.tex
\begin{longtable}{l l r r r r r}

\toprule
\rowcolor{TblHeader}
\textbf{Question format} & \textbf{Attribute type} & \textbf{$n$} & \textbf{EM} & \textbf{$\mu$P} & \textbf{$\mu$R} & \textbf{$\mu$F1} \\
\midrule
\endfirsthead

\toprule
\rowcolor{TblHeader}
\textbf{Question format} & \textbf{Attribute type} & \textbf{$n$} & \textbf{EM} & \textbf{$\mu$P} & \textbf{$\mu$R} & \textbf{$\mu$F1} \\
\midrule
\endhead

\endfoot
\endlastfoot

Single-Verify & SCP code & 8{,}243 & 79.52 & 79.52 & 79.52 & 79.52 \\
Single-Verify & Heart axis & 237 & 89.87 & 89.87 & 89.87 & 89.87 \\
Single-Verify & Stage of infarction & 168 & 82.74 & 82.74 & 82.74 & 82.74 \\
Single-Verify & Noise & 3{,}183 & 67.73 & 67.73 & 67.73 & 67.73 \\
Single-Verify & Extra systole & 230 & 77.39 & 77.39 & 77.39 & 77.39 \\
Single-Verify & Numeric feature & 1{,}020 & 77.84 & 77.84 & 77.84 & 77.84 \\

\midrule
Single-Choose & SCP code & 9{,}358 & 69.30 & 77.50 & 72.95 & 75.16 \\
Single-Choose & Heart axis & 36 & 75.00 & 75.00 & 75.00 & 75.00 \\
Single-Choose & Stage of infarction & 18 & 55.56 & 55.56 & 55.56 & 55.56 \\
Single-Choose & Noise & 317 & 43.22 & 55.21 & 46.27 & 50.35 \\
Single-Choose & Extra systole & 22 & 50.00 & 54.55 & 46.15 & 50.00 \\
Single-Choose & Numeric feature & 104 & 60.58 & 60.58 & 60.58 & 60.58 \\

\midrule
Single-Query & SCP code & 9{,}208 & 43.07 & 59.47 & 57.99 & 58.72 \\
Single-Query & Heart axis & 167 & 74.85 & 74.85 & 74.85 & 74.85 \\
Single-Query & Stage of infarction & 178 & 46.07 & 46.07 & 46.07 & 46.07 \\
Single-Query & Noise & 3{,}960 & 38.86 & 44.48 & 46.04 & 45.24 \\
Single-Query & Extra systole & 241 & 54.77 & 56.85 & 55.02 & 55.92 \\
Single-Query & Numeric feature & 4{,}403 & 25.03 & 65.75 & 65.96 & 65.85 \\

\bottomrule
\caption{
PTB-XL ECG-QA performance stratified by question format and ECG-QA attribute type.
Question format is the single-ECG ECG-QA format evaluated in this work.
Attribute type is the broad ECG-QA metadata category used to generate the question.
$n$ is the number of evaluated question instances in the corresponding question-format and attribute-type subset.
EM is exact-match accuracy over the complete predicted answer set.
$\mu$P, $\mu$R, and $\mu$F1 are micro-averaged option-level precision, recall, and F1.
}
\label{tab:supp_ptbxl_ecgqa_qtype_attribute_type}

\end{longtable}

%% file: tables/supp_mimic_ecgqa_qtype_attribute_type.tex
\begin{longtable}{l l r r r r r}

\toprule
\rowcolor{TblHeader}
\textbf{Question format} & \textbf{Attribute type} & \textbf{$n$} & \textbf{EM} & \textbf{$\mu$P} & \textbf{$\mu$R} & \textbf{$\mu$F1} \\
\midrule
\endfirsthead

\toprule
\rowcolor{TblHeader}
\textbf{Question format} & \textbf{Attribute type} & \textbf{$n$} & \textbf{EM} & \textbf{$\mu$P} & \textbf{$\mu$R} & \textbf{$\mu$F1} \\
\midrule
\endhead

\endfoot
\endlastfoot

Single-Verify & SCP code & 23{,}950 & 81.16 & 81.16 & 81.16 & 81.16 \\
Single-Verify & Heart axis & 240 & 81.67 & 81.67 & 81.67 & 81.67 \\
Single-Verify & Stage of infarction & 180 & 78.33 & 78.33 & 78.33 & 78.33 \\
Single-Verify & Noise & 148 & 79.05 & 79.05 & 79.05 & 79.05 \\
Single-Verify & Numeric feature & 1{,}020 & 79.22 & 79.22 & 79.22 & 79.22 \\

\midrule
Single-Choose & SCP code & 21{,}971 & 68.43 & 75.41 & 71.82 & 73.57 \\
Single-Choose & Heart axis & 38 & 50.00 & 55.26 & 52.50 & 53.85 \\
Single-Choose & Stage of infarction & 18 & 77.78 & 77.78 & 77.78 & 77.78 \\
Single-Choose & Numeric feature & 104 & 60.58 & 60.58 & 60.58 & 60.58 \\

\midrule
Single-Query & SCP code & 54{,}167 & 19.25 & 55.34 & 48.35 & 51.61 \\
Single-Query & Heart axis & 200 & 64.50 & 64.50 & 64.50 & 64.50 \\
Single-Query & Stage of infarction & 250 & 44.80 & 44.80 & 44.80 & 44.80 \\
Single-Query & Noise & 27 & 7.41 & 12.50 & 12.90 & 12.70 \\
Single-Query & Numeric feature & 7{,}617 & 20.90 & 69.54 & 66.92 & 68.21 \\

\bottomrule
\caption{
MIMIC-IV-ECG ECG-QA performance stratified by question format and ECG-QA attribute type.
The question format is the single-ECG ECG-QA format evaluated in this work.
Attribute type is the broad ECG-QA metadata category used to generate the question.
$n$ is the number of evaluated question instances in the corresponding question-format and attribute-type subset.
EM is exact-match accuracy over the complete predicted answer set.
$\mu$P, $\mu$R, and $\mu$F1 are micro-averaged option-level precision, recall, and F1.
}
\label{tab:supp_mimic_ecgqa_qtype_attribute_type}\\

\end{longtable}

%% file: tables/supp_ecgqa_single_verify_attribute.tex
\begingroup
\scriptsize
\setlength{\LTpre}{0pt}
\setlength{\LTpost}{0pt}
\setlength{\LTleft}{0pt}
\setlength{\LTright}{\fill}
\setlength{\tabcolsep}{3pt}
\renewcommand{\arraystretch}{0.86}
\captionsetup{justification=raggedright,singlelinecheck=false}

\newcommand{\Attr}[1]{%
  \hangindent=0.8em
  \hangafter=1
  \hspace*{0.8em}#1%
}

\newcommand{\GrayGroup}[1]{%
\addlinespace[3pt]
\rowcolor{TblGroup}
\multicolumn{5}{@{}l}{\textbf{#1}}\\[-1pt]
\addlinespace[1pt]
}

\begin{longtable}{
@{}
>{\raggedright\arraybackslash}p{0.62\linewidth}
>{\raggedleft\arraybackslash}p{0.055\linewidth}
>{\raggedleft\arraybackslash}p{0.060\linewidth}
>{\raggedleft\arraybackslash}p{0.055\linewidth}
>{\raggedleft\arraybackslash}p{0.060\linewidth}
@{}
}

\caption{
ECG-QA Single-Verify performance by attribute for PTB-XL and MIMIC-IV-ECG.
EM is reported as percent.
}
\label{tab:supp_ecgqa_single_verify_attribute}\\

\toprule
\raisebox{-0.65\baselineskip}{\textbf{Attribute}} &
\multicolumn{2}{c}{\makebox[0pt][c]{\textbf{MIMIC}}} &
\multicolumn{2}{r}{\makebox[0pt][r]{\textbf{PTB-XL}}} \\
\cmidrule(lr){2-3}\cmidrule(lr){4-5}
&
\textbf{$n$} & \textbf{EM} &
\multicolumn{2}{r}{\textbf{$n$}\hspace{0.7em}\textbf{EM}} \\
\midrule
\endfirsthead

\toprule
\raisebox{-0.65\baselineskip}{\textbf{Attribute}} &
\multicolumn{2}{c}{\makebox[0pt][c]{\textbf{MIMIC}}} &
\multicolumn{2}{r}{\makebox[0pt][r]{\textbf{PTB-XL}}} \\
\cmidrule(lr){2-3}\cmidrule(lr){4-5}
&
\textbf{$n$} & \textbf{EM} &
\multicolumn{2}{r}{\textbf{$n$}\hspace{0.7em}\textbf{EM}} \\
\midrule
\endhead

\bottomrule
\endlastfoot

\GrayGroup{SCP-code diagnostic statements}
\Attr{2:1 AV block} & 60 & 93.3 & -- & -- \\
\Attr{2:1 sinoatrial block} & 42 & 92.9 & -- & -- \\
\Attr{3:1 AV block} & 60 & 70.0 & -- & -- \\
\Attr{4:1 AV block} & 60 & 73.3 & -- & -- \\
\Attr{aberrant supraventricular complexes} & 60 & 75.0 & -- & -- \\
\Attr{aberrant ventricular complex} & 60 & 90.0 & -- & -- \\
\Attr{abnormal qrs} & -- & -- & 60 & 71.7 \\
\Attr{abnormal R wave progression} & 60 & 83.3 & -- & -- \\
\Attr{abnormal ventricular conduction pathways} & 60 & 73.3 & -- & -- \\
\Attr{accelerated idioventricular rhythm} & 60 & 88.3 & -- & -- \\
\Attr{accelerated junctional rhythm} & 80 & 65.0 & -- & -- \\
\Attr{any diagnostic symptoms} & 120 & 75.8 & 120 & 80.8 \\
\Attr{any form-related symptoms} & 60 & 80.0 & 60 & 66.7 \\
\Attr{any kind of abnormal symptoms} & 60 & 81.7 & 60 & 71.7 \\
\Attr{any rhythm-related symptoms} & 60 & 75.0 & 60 & 65.0 \\
\Attr{atrial arrhythmia} & 12 & 100.0 & -- & -- \\
\Attr{atrial bigeminy} & 60 & 66.7 & -- & -- \\
\Attr{atrial couplet} & 60 & 65.0 & -- & -- \\
\Attr{atrial fibrillation} & 60 & 90.0 & 60 & 86.7 \\
\Attr{atrial flutter} & 80 & 63.7 & 36 & 94.4 \\
\Attr{atrial premature complex} & -- & -- & 60 & 71.7 \\
\Attr{atrial tachycardia} & 60 & 91.7 & -- & -- \\
\Attr{atypical left bundle branch block} & 60 & 91.7 & -- & -- \\
\Attr{atypical right bundle branch block} & 60 & 85.0 & -- & -- \\
\Attr{AV dissociation} & 60 & 76.7 & -- & -- \\
\Attr{AV sequential pacemaker} & 60 & 90.0 & -- & -- \\
\Attr{bi-atrial overload/enlargement} & 80 & 68.8 & -- & -- \\
\Attr{bigeminal pattern (unknown origin, supraventricular, or ventricular)} & -- & -- & 50 & 80.0 \\
\Attr{Biventricular hypertrophy} & 60 & 91.7 & -- & -- \\
\Attr{Broad R wave in lateral leads} & 60 & 86.7 & -- & -- \\
\Attr{complete left bundle branch block} & 60 & 93.3 & 60 & 100.0 \\
\Attr{complete right bundle branch block} & 60 & 98.3 & 61 & 96.7 \\
\Attr{conduction disturbance} & 60 & 71.7 & 60 & 83.3 \\
\Attr{Deep S wave} & 581 & 91.7 & -- & -- \\
\Attr{dextrocardia} & 56 & 78.6 & -- & -- \\
\Attr{digitalis effect} & -- & -- & 67 & 76.1 \\
\Attr{dual chamber electronic pacing} & 60 & 98.3 & -- & -- \\
\Attr{early R wave transition} & 60 & 95.0 & -- & -- \\
\Attr{early repolarization} & 7 & 85.7 & -- & -- \\
\Attr{ectopic atrial bradycardia} & 60 & 88.3 & -- & -- \\
\Attr{ectopic atrial rhythm} & 60 & 73.3 & -- & -- \\
\Attr{ectopic atrial tachycardia} & 60 & 91.7 & -- & -- \\
\Attr{electronic atrial pacing} & 60 & 83.3 & -- & -- \\
\Attr{extreme tachycardia} & 59 & 93.2 & -- & -- \\
\Attr{first degree AV block} & 60 & 83.3 & 63 & 66.7 \\
\Attr{fusion complexes} & 60 & 86.7 & -- & -- \\
\Attr{high amplitude T-waves} & 60 & 91.7 & -- & -- \\
\Attr{high grade AV block} & 58 & 77.6 & -- & -- \\
\Attr{high P-voltages} & 426 & 82.4 & -- & -- \\
\Attr{high QRS voltage} & 600 & 86.8 & 145 & 86.2 \\
\Attr{hypertrophy} & 60 & 83.3 & 60 & 88.3 \\
\Attr{idioventricular rhythm} & 60 & 86.7 & -- & -- \\
\Attr{incomplete left bundle branch block} & 60 & 91.7 & 51 & 92.2 \\
\Attr{incomplete right bundle branch block} & 60 & 88.3 & 69 & 71.0 \\
\Attr{intermittent second degree AV block} & 60 & 80.0 & -- & -- \\
\Attr{intraventricular conduction disturbance} & 60 & 88.3 & -- & -- \\
\Attr{inverted T-waves} & 60 & 70.0 & 621 & 82.0 \\
\Attr{ischemic in anterior leads} & -- & -- & 32 & 84.4 \\
\Attr{ischemic in anterolateral leads} & -- & -- & 80 & 61.3 \\
\Attr{ischemic in anteroseptal leads} & -- & -- & 63 & 84.1 \\
\Attr{ischemic in inferior leads} & -- & -- & 66 & 74.2 \\
\Attr{ischemic in inferolateral leads} & -- & -- & 67 & 62.7 \\
\Attr{ischemic in lateral leads} & -- & -- & 55 & 65.5 \\
\Attr{ischemic ST-T changes} & 60 & 85.0 & -- & -- \\
\Attr{ischemic ST-T changes in anterior leads} & 80 & 66.2 & -- & -- \\
\Attr{ischemic ST-T changes in anteroseptal leads} & 80 & 68.8 & -- & -- \\
\Attr{ischemic ST-T changes in diffuse leads} & 60 & 65.0 & -- & -- \\
\Attr{ischemic ST-T changes in inferior leads} & 80 & 62.5 & -- & -- \\
\Attr{ischemic ST-T changes in inferoseptal leads} & 80 & 77.5 & -- & -- \\
\Attr{ischemic ST-T changes in lateral leads} & 14 & 71.4 & -- & -- \\
\Attr{ischemic ST-T changes in septal leads} & 80 & 76.2 & -- & -- \\
\Attr{junctional bradycardia} & 50 & 80.0 & -- & -- \\
\Attr{junctional rhythm} & 80 & 67.5 & -- & -- \\
\Attr{junctional tachycardia} & 60 & 93.3 & -- & -- \\
\Attr{late R wave transition} & 30 & 86.7 & -- & -- \\
\Attr{left anterior fascicular block} & 80 & 77.5 & 68 & 82.4 \\
\Attr{left atrial overload/enlargement} & 80 & 43.8 & 80 & 55.0 \\
\Attr{left posterior fascicular block} & 60 & 85.0 & 60 & 98.3 \\
\Attr{Left ventricular hypertrophy} & 80 & 65.0 & 80 & 78.8 \\
\Attr{long QT interval} & 60 & 88.3 & 52 & 78.8 \\
\Attr{low amplitude t-wave} & -- & -- & 636 & 78.0 \\
\Attr{low QRS voltage} & 1260 & 93.7 & -- & -- \\
\Attr{low qrs voltages in the frontal and horizontal leads} & -- & -- & 126 & 88.1 \\
\Attr{low R} & 438 & 75.3 & -- & -- \\
\Attr{Mobitz type 1 second-degree AV block} & 60 & 73.3 & -- & -- \\
\Attr{Mobitz type 2 second-degree AV block} & 49 & 83.7 & -- & -- \\
\Attr{multifocal premature ventricular complexes} & 60 & 88.3 & -- & -- \\
\Attr{Myocardial infarction} & 60 & 85.0 & 60 & 78.3 \\
\Attr{Myocardial infarction in anterior leads} & 80 & 71.2 & 80 & 67.5 \\
\Attr{Myocardial infarction in anterolateral leads} & 80 & 63.7 & 76 & 78.9 \\
\Attr{Myocardial infarction in anteroseptal leads} & 80 & 57.5 & 80 & 66.2 \\
\Attr{Myocardial infarction in inferior leads} & 80 & 67.5 & 80 & 75.0 \\
\Attr{Myocardial infarction in inferolateral leads} & 80 & 73.8 & 79 & 68.4 \\
\Attr{myocardial infarction in inferoposterior leads} & -- & -- & 7 & 71.4 \\
\Attr{myocardial infarction in inferoposterolateral leads} & -- & -- & 7 & 85.7 \\
\Attr{Myocardial infarction in lateral leads} & 80 & 66.2 & 49 & 81.6 \\
\Attr{Myocardial infarction in posterior leads} & 80 & 75.0 & 7 & 85.7 \\
\Attr{Myocardial infarction in septal leads} & 80 & 72.5 & -- & -- \\
\Attr{non-diagnostic T abnormalities} & 1260 & 87.2 & 743 & 79.3 \\
\Attr{non-specific intraventricular conduction disturbance (block)} & -- & -- & 63 & 76.2 \\
\Attr{non-specific ischemic} & -- & -- & 80 & 70.0 \\
\Attr{non-specific ST changes} & 1080 & 77.6 & 456 & 77.4 \\
\Attr{non-specific ST depression} & 198 & 85.9 & 656 & 80.5 \\
\Attr{non-specific st elevation} & -- & -- & 24 & 83.3 \\
\Attr{non-specific t abnormality} & 1260 & 78.4 & -- & -- \\
\Attr{non-specific T-wave changes} & 1080 & 77.9 & 684 & 75.4 \\
\Attr{non-sustained ventricular tachycardia} & 60 & 93.3 & -- & -- \\
\Attr{Normal ECG} & 60 & 85.0 & 60 & 86.7 \\
\Attr{normal functioning artificial pacemaker} & -- & -- & 48 & 100.0 \\
\Attr{notched P wave} & 49 & 83.7 & -- & -- \\
\Attr{P wave abnormality} & 60 & 73.3 & -- & -- \\
\Attr{pacemaker activity} & 60 & 85.0 & -- & -- \\
\Attr{pacemaker rhythm} & 60 & 96.7 & -- & -- \\
\Attr{paired ventricular premature complexes} & 60 & 80.0 & -- & -- \\
\Attr{paroxysmal idioventricular rhythm} & 60 & 73.3 & -- & -- \\
\Attr{poor R wave progression} & 420 & 76.0 & -- & -- \\
\Attr{premature atrial complexes} & 60 & 73.3 & -- & -- \\
\Attr{premature ventricular complexes} & 60 & 85.0 & -- & -- \\
\Attr{premature ventricular interpolated complexes} & 60 & 83.3 & -- & -- \\
\Attr{Prolonged PR interval} & 60 & 68.3 & 60 & 81.7 \\
\Attr{prolonged QRS duration} & 60 & 85.0 & -- & -- \\
\Attr{Q waves present} & 1039 & 80.2 & 647 & 83.2 \\
\Attr{QRS changes in anteroseptal leads} & 60 & 76.7 & -- & -- \\
\Attr{QTc prolongation} & 60 & 73.3 & -- & -- \\
\Attr{rapid ventricular response} & 60 & 88.3 & -- & -- \\
\Attr{regular rhythm} & 60 & 76.7 & -- & -- \\
\Attr{repolarization abnormality} & 1260 & 81.0 & -- & -- \\
\Attr{Reversed R wave progression} & 52 & 82.7 & -- & -- \\
\Attr{right atrial overload/enlargement} & 60 & 85.0 & 58 & 91.4 \\
\Attr{right ventricular conduction delay} & 59 & 71.2 & -- & -- \\
\Attr{Right ventricular hypertrophy} & 80 & 70.0 & -- & -- \\
\Attr{rSr' type in V1 or V2} & 60 & 95.0 & -- & -- \\
\Attr{S1 S2 S3 type QRS pattern} & 36 & 91.7 & -- & -- \\
\Attr{second degree AV block} & 60 & 78.3 & -- & -- \\
\Attr{second-degree SA block type II} & 60 & 68.3 & -- & -- \\
\Attr{Short PR interval} & 60 & 70.0 & -- & -- \\
\Attr{short QT interval} & 60 & 78.3 & -- & -- \\
\Attr{short QTc} & 30 & 86.7 & -- & -- \\
\Attr{Significant repolarization change} & 60 & 73.3 & -- & -- \\
\Attr{sinus arrhythmia} & 60 & 60.0 & 60 & 71.7 \\
\Attr{sinus bradycardia} & 60 & 95.0 & 60 & 83.3 \\
\Attr{sinus pause} & 60 & 71.7 & -- & -- \\
\Attr{sinus rhythm} & 60 & 80.0 & 60 & 76.7 \\
\Attr{sinus tachycardia} & 80 & 92.5 & 60 & 98.3 \\
\Attr{slow ventricular response} & 60 & 91.7 & -- & -- \\
\Attr{ST Depression} & 1260 & 84.0 & -- & -- \\
\Attr{ST Elevation} & 1020 & 77.3 & -- & -- \\
\Attr{ST(-T) change} & 1260 & 79.1 & -- & -- \\
\Attr{st/t change} & -- & -- & 60 & 78.3 \\
\Attr{subendocardial injury in anterolateral leads} & -- & -- & 61 & 88.5 \\
\Attr{subendocardial injury in anteroseptal leads} & -- & -- & 63 & 88.9 \\
\Attr{subendocardial injury in inferolateral leads} & -- & -- & 6 & 83.3 \\
\Attr{subendocardial injury in lateral leads} & -- & -- & 13 & 84.6 \\
\Attr{supraventricular bigeminy BIGU bigeminal pattern} & 60 & 60.0 & -- & -- \\
\Attr{supraventricular extrasystoles} & 60 & 61.7 & -- & -- \\
\Attr{supraventricular premature complex} & 60 & 88.3 & -- & -- \\
\Attr{supraventricular rhythm} & 60 & 71.7 & -- & -- \\
\Attr{supraventricular tachycardia} & 80 & 76.2 & 18 & 77.8 \\
\Attr{T-wave abnormality} & 1260 & 79.9 & 150 & 80.7 \\
\Attr{tall R wave in V1 or V2} & 60 & 93.3 & -- & -- \\
\Attr{tall R wave in V5 or V6} & 57 & 86.0 & -- & -- \\
\Attr{third degree AV block} & 60 & 81.7 & -- & -- \\
\Attr{TU fusion} & 52 & 78.8 & -- & -- \\
\Attr{unclassified aberrantly conducted complexes} & 60 & 78.3 & -- & -- \\
\Attr{uncontrolled ventricular response} & 60 & 98.3 & -- & -- \\
\Attr{ventricular bigeminy} & 60 & 86.7 & -- & -- \\
\Attr{ventricular couplet} & 60 & 80.0 & -- & -- \\
\Attr{ventricular escape rhythm} & 24 & 79.2 & -- & -- \\
\Attr{ventricular premature complex} & -- & -- & 60 & 88.3 \\
\Attr{ventricular trigeminy} & 60 & 95.0 & -- & -- \\
\Attr{ventricular-paced complexes or rhythm} & 60 & 95.0 & -- & -- \\
\Attr{Voltage criteria (QRS) for left ventricular hypertrophy} & 60 & 88.3 & 180 & 88.3 \\
\Attr{wandering pacemaker} & 36 & 77.8 & -- & -- \\
\Attr{wide QRS tachycardia} & 60 & 95.0 & -- & -- \\
\Attr{Wolff-Parkinson type A} & 6 & 83.3 & -- & -- \\
\Attr{Wolff-Parkinson type B} & 50 & 82.0 & -- & -- \\
\Attr{Wolff-Parkinson-White syndrome} & 60 & 85.0 & -- & -- \\

\GrayGroup{Heart axis}
\Attr{extreme axis deviation} & 60 & 90.0 & 57 & 93.0 \\
\Attr{left axis deviation} & 60 & 66.7 & 60 & 80.0 \\
\Attr{normal heart axis} & 60 & 76.7 & 60 & 93.3 \\
\Attr{right axis deviation} & 60 & 93.3 & 60 & 93.3 \\

\GrayGroup{Stage of infarction}
\Attr{early stage of myocardial infarction} & 60 & 83.3 & 59 & 89.8 \\
\Attr{middle stage of myocardial infarction} & 60 & 80.0 & 60 & 80.0 \\
\Attr{old stage of myocardial infarction} & 60 & 71.7 & 49 & 77.6 \\

\GrayGroup{Noise}
\Attr{any kind of noises} & -- & -- & 780 & 65.9 \\
\Attr{baseline drift} & -- & -- & 780 & 66.7 \\
\Attr{baseline wander} & 148 & 79.1 & -- & -- \\
\Attr{burst noise} & -- & -- & 777 & 68.1 \\
\Attr{electrodes problems} & -- & -- & 66 & 81.8 \\
\Attr{static noise} & -- & -- & 780 & 69.1 \\

\GrayGroup{Numeric features}
\Attr{above the normal range of p duration} & 60 & 68.3 & 60 & 68.3 \\
\Attr{above the normal range of pr interval} & 60 & 81.7 & 60 & 83.3 \\
\Attr{above the normal range of qrs duration} & 60 & 78.3 & 60 & 78.3 \\
\Attr{above the normal range of qt corrected} & 60 & 70.0 & 60 & 75.0 \\
\Attr{above the normal range of qt interval} & 60 & 81.7 & 60 & 70.0 \\
\Attr{above the normal range of rr interval} & 60 & 91.7 & 60 & 98.3 \\
\Attr{below the normal range of pr interval} & 60 & 66.7 & 60 & 61.7 \\
\Attr{below the normal range of qrs duration} & 60 & 85.0 & 60 & 85.0 \\
\Attr{below the normal range of qt corrected} & 60 & 78.3 & 60 & 81.7 \\
\Attr{below the normal range of qt interval} & 60 & 81.7 & 60 & 83.3 \\
\Attr{below the normal range of rr interval} & 60 & 100.0 & 60 & 93.3 \\
\Attr{within the normal range of p duration} & 60 & 73.3 & 60 & 65.0 \\
\Attr{within the normal range of pr interval} & 60 & 68.3 & 60 & 71.7 \\
\Attr{within the normal range of qrs duration} & 60 & 78.3 & 60 & 63.3 \\
\Attr{within the normal range of qt corrected} & 60 & 70.0 & 60 & 73.3 \\
\Attr{within the normal range of qt interval} & 60 & 73.3 & 60 & 76.7 \\
\Attr{within the normal range of rr interval} & 60 & 100.0 & 60 & 95.0 \\

\GrayGroup{Extra systole}
\Attr{any kind of extra systoles} & -- & -- & 50 & 82.0 \\
\Attr{extrasystoles} & -- & -- & 60 & 81.7 \\
\Attr{supraventricular extrasystoles} & -- & -- & 60 & 66.7 \\
\Attr{ventricular extrasystoles} & -- & -- & 60 & 80.0 \\

\end{longtable}
\endgroup

%% file: tables/supp_training_fields.tex
\begingroup
\scriptsize
\setlength{\tabcolsep}{3pt}
\renewcommand{\arraystretch}{0.92}
\setlength{\LTpre}{4pt}
\setlength{\LTpost}{4pt}

\begin{longtable}{@{}>{\raggedright\arraybackslash}p{0.30\linewidth}
                  >{\raggedright\arraybackslash}p{0.23\linewidth}
                  >{\raggedright\arraybackslash}p{0.41\linewidth}@{}}
\caption{Dataset-specific fields used to construct structured per-study records for synthetic QA generation.}
\label{tab:supp_training_fields}\\

\toprule
\rowcolor{TblHeader}
\textbf{Field} & \textbf{Dataset} & \textbf{Comment} \\
\midrule
\endfirsthead

\toprule
\rowcolor{TblHeader}
\textbf{Field} & \textbf{Dataset} & \textbf{Comment} \\
\midrule
\endhead

\bottomrule
\endfoot

\multicolumn{3}{@{}l}{\textbf{Patient context}} \\
\midrule

Patient age
& UK Biobank, MIMIC-IV-ECG, EchoNext, PTB-XL
& MIMIC-IV-ECG uses approximate anchor age. \\

Patient sex
& UK Biobank, MIMIC-IV-ECG, EchoNext, PTB-XL
& -- \\

BMI
& UK Biobank
& -- \\

Height
& UK Biobank
& -- \\

Body surface area
& UK Biobank
& Used for indexed CMR reference ranges. \\

Chest pain or discomfort
& UK Biobank
& -- \\

Shortness of breath on level ground
& UK Biobank
& -- \\

Race or ethnicity
& EchoNext
& -- \\

\midrule
\multicolumn{3}{@{}l}{\textbf{ECG measurements}} \\
\midrule

ECG heart rate
& MIMIC-IV-ECG, UK Biobank
& In MIMIC-IV-ECG, derived from RR interval if heart rate is unavailable. \\

Ventricular rate
& EchoNext
& -- \\

Atrial rate
& EchoNext
& -- \\

RR interval
& MIMIC-IV-ECG, UK Biobank
& -- \\

PP interval
& MIMIC-IV-ECG, UK Biobank
& -- \\

PR interval
& MIMIC-IV-ECG, EchoNext
& -- \\

PQ interval
& MIMIC-IV-ECG, UK Biobank
& Derived from P onset and Q onset when needed. \\

P duration
& MIMIC-IV-ECG, UK Biobank
& Derived from P onset and P offset when needed. \\

QRS duration
& MIMIC-IV-ECG, UK Biobank, EchoNext
& In MIMIC-IV-ECG, derived from QRS onset and QRS end when needed. \\

QT interval
& MIMIC-IV-ECG, UK Biobank
& Used to derive QTc when needed. \\

QTc interval
& MIMIC-IV-ECG, UK Biobank, EchoNext
& In MIMIC-IV-ECG and UK Biobank, derived from QT and RR when needed. \\

P axis
& MIMIC-IV-ECG, UK Biobank
& -- \\

R axis
& MIMIC-IV-ECG, UK Biobank, PTB-XL
& In PTB-XL, heart-axis labels are mapped to unified R-axis categories. MID: normal; LAD and ALAD: leftward; RAD and ARAD: rightward. \\

T axis
& MIMIC-IV-ECG, UK Biobank
& -- \\

\midrule
\multicolumn{3}{@{}l}{\textbf{ECG report-derived fields}} \\
\midrule

ECG diagnosis and machine-report diagnosis text
& MIMIC-IV-ECG, UK Biobank
& ECG diagnosis is retained as text. For MIMIC-IV-ECG, diagnosis text is built from machine-measurement report columns. \\

German ECG report
& PTB-XL
& Original PTB-XL report text. \\

English ECG report
& PTB-XL
& English translation of the PTB-XL report generated using GPT-OSS-20B. \\

SCP-ECG diagnostic codes
& PTB-XL
& SCP codes are decoded into textual diagnostic descriptions. \\

Infarction stage
& PTB-XL
& -- \\

\midrule
\multicolumn{3}{@{}l}{\textbf{CMR-derived fields}} \\
\midrule

LV end diastolic volume
& UK Biobank
& -- \\

LV end systolic volume
& UK Biobank
& -- \\

LV stroke volume
& UK Biobank
& -- \\

LV myocardial mass
& UK Biobank
& -- \\

LV ejection fraction
& UK Biobank
& -- \\

LV cardiac output
& UK Biobank
& -- \\

Cardiac index
& UK Biobank
& -- \\

LV mean myocardial wall thickness global
& UK Biobank
& -- \\

LV circumferential strain global
& UK Biobank
& -- \\

LV longitudinal strain global
& UK Biobank
& -- \\

LV radial strain global
& UK Biobank
& -- \\

RV end diastolic volume
& UK Biobank
& -- \\

RV end systolic volume
& UK Biobank
& -- \\

RV stroke volume
& UK Biobank
& -- \\

RV ejection fraction
& UK Biobank
& -- \\

LA maximum volume
& UK Biobank
& -- \\

LA minimum volume
& UK Biobank
& -- \\

LA stroke volume
& UK Biobank
& -- \\

LA ejection fraction
& UK Biobank
& -- \\

RA maximum volume
& UK Biobank
& -- \\

RA minimum volume
& UK Biobank
& -- \\

RA stroke volume
& UK Biobank
& -- \\

RA ejection fraction
& UK Biobank
& -- \\

Ascending aorta distensibility
& UK Biobank
& -- \\

Descending aorta distensibility
& UK Biobank
& -- \\

\midrule
\multicolumn{3}{@{}l}{\textbf{EchoNext echocardiographic categorical fields}} \\
\midrule

Aortic stenosis
& EchoNext
& Severity category. \\

Aortic regurgitation
& EchoNext
& Severity category. \\

Mitral regurgitation
& EchoNext
& Severity category. \\

Tricuspid regurgitation
& EchoNext
& Severity category. \\

Pulmonary regurgitation
& EchoNext
& Severity category. \\

Right ventricular systolic function
& EchoNext
& Qualitative systolic function category. \\

Pericardial effusion
& EchoNext
& Presence and size category. \\

\midrule
\multicolumn{3}{@{}l}{\textbf{EchoNext echocardiographic numeric fields}} \\
\midrule

Interventricular septal thickness
& EchoNext
& -- \\

Left ventricular posterior wall thickness
& EchoNext
& -- \\

Left ventricular systolic function
& EchoNext
& Derived from LVEF value in percent. \\

\midrule
\multicolumn{3}{@{}l}{\textbf{EchoNext binary structural heart disease fields}} \\
\midrule

Left ventricular ejection fraction is 45\% or lower
& EchoNext
& Binary descriptor derived from LVEF. \\

Left ventricular wall thickness is at least 1.3 cm
& EchoNext
& Binary descriptor derived from LV wall thickness. \\

Aortic stenosis is moderate or greater
& EchoNext
& Binary descriptor derived from aortic stenosis severity. \\

Aortic regurgitation is moderate or greater
& EchoNext
& Binary descriptor derived from aortic regurgitation severity. \\

Mitral regurgitation is moderate or greater
& EchoNext
& Binary descriptor derived from mitral regurgitation severity. \\

Tricuspid regurgitation is moderate or greater
& EchoNext
& Binary descriptor derived from tricuspid regurgitation severity. \\

Pulmonary regurgitation is moderate or greater
& EchoNext
& Binary descriptor derived from pulmonary regurgitation severity. \\

Right ventricular systolic dysfunction is moderate or greater
& EchoNext
& Binary descriptor derived from right ventricular systolic function. \\

Pericardial effusion is moderate or large
& EchoNext
& Binary descriptor derived from pericardial effusion size. \\

Pulmonary artery systolic pressure is at least 45 mmHg
& EchoNext
& Binary descriptor derived from PASP. \\

Tricuspid regurgitation max velocity is at least 3.2 m/s
& EchoNext
& Binary descriptor derived from maximum TR jet velocity. \\

\end{longtable}

\endgroup

%% file: tables/supp_categorization_rules.tex
\begingroup
\scriptsize
\setlength{\tabcolsep}{3pt}
\renewcommand{\arraystretch}{0.92}
\setlength{\LTpre}{4pt}
\setlength{\LTpost}{4pt}

\begin{longtable}{@{}>{\raggedright\arraybackslash}p{0.30\linewidth}
                  >{\raggedright\arraybackslash}p{0.66\linewidth}@{}}
\caption{
Clinical categorisation rules used to convert numeric measurements into textual descriptors before synthetic QA generation.
}
\label{tab:supp_categorization_rules}\\

\toprule
\rowcolor{TblHeader}
\textbf{Measurement} & \textbf{Categorisation rule} \\
\midrule
\endfirsthead

\toprule
\rowcolor{TblHeader}
\textbf{Measurement} & \textbf{Categorisation rule} \\
\midrule
\endhead

\bottomrule
\endfoot

\multicolumn{2}{@{}l}{\textbf{ECG measurements}} \\
\midrule

ECG heart rate [bpm], ventricular rate, atrial rate
& $\leq50$: marked bradycardia; $>50$ to $\leq60$: bradycardia; $>60$ to $\leq100$: normal; $>100$ to $\leq120$: mild tachycardia; $>120$: marked tachycardia. \\

RR interval [ms], PP interval [ms]
& Convert interval to rate as $60000/\mathrm{interval}$. Rate $>120$: markedly short; $>100$ to $\leq120$: short; $\geq60$ to $\leq100$: normal; $\geq50$ to $<60$: prolonged; $<50$: markedly prolonged. \\

P duration [ms]
& If missing, derive as P offset $-$ P onset. Then categorise as: $<120$: normal; $\geq120$: prolonged. \\

PQ interval [ms], PR interval [ms]
& If PQ is missing, derive as Q onset $-$ P onset. Then categorise as: $<120$: short; $120$--$200$: normal; $>200$: prolonged. \\

QRS duration [ms]
& $<110$: normal; $\geq110$ to $<120$: mildly prolonged; $\geq120$: prolonged. \\

QTC interval [ms]
& If QTC is unavailable, derive it from QT and RR using Bazett correction:
$\mathrm{QTc}=\mathrm{QT}/\sqrt{\mathrm{RR}/1000}$.
Female: $\leq470$: normal; $>470$ to $\leq490$: borderline; $>490$: prolonged.
Male or unavailable sex: $\leq450$: normal; $>450$ to $\leq480$: borderline; $>480$: prolonged. \\

P axis [degrees]
& $<0$: leftward; $0$--$75$: normal; $>75$: rightward. \\

R axis [degrees]
& $<-30$: leftward; $-30$--$90$: normal; $>90$: rightward. \\

T axis [degrees]
& $<-15$: leftward; $-15$ to $<15$: borderline; $15$--$75$: normal; $>75$ to $\leq105$: borderline; $>105$: rightward. \\

\midrule
\multicolumn{2}{@{}l}{\textbf{Ventricular function and cardiac output}} \\
\midrule

LV ejection fraction, left ventricular systolic function
& Female: $<30$: severely reduced; $\geq30$ to $<40$: moderately reduced; $\geq40$ to $<52$: mildly reduced; $52$--$79$: normal; $>79$: hyperdynamic.
Male or unavailable sex: $<30$: severely reduced; $\geq30$ to $<40$: moderately reduced; $\geq40$ to $<49$: mildly reduced; $49$--$79$: normal; $>79$: hyperdynamic. \\

RV ejection fraction
& Female: $<30$: severely reduced; $\geq30$ to $<40$: moderately reduced; $\geq40$ to $<46$: mildly reduced; $46$--$74$: normal; $>74$: hyperdynamic.
Male or unavailable sex: $<30$: severely reduced; $\geq30$ to $<40$: moderately reduced; $\geq40$ to $<42$: mildly reduced; $42$--$72$: normal; $>72$: hyperdynamic. \\

cardiac index [litres/min/m2]
& $<2.2$: low output; $2.2$ to $\leq2.5$: borderline low; $>2.5$ to $\leq4.0$: normal; $>4.0$ to $\leq4.5$: borderline high; $>4.5$: high output. \\

LV cardiac output
& If body surface area is available, first compute cardiac index and apply the cardiac-index rule. Otherwise, use absolute ranges. Male $3.4$--$7.8$: normal; female $2.7$--$6.3$: normal; unknown sex $2.7$--$7.8$: normal. Below range: low output; above range: high output. \\

LV mean myocardial wall thickness global
& $<5.0$: thinned; $5.0$--$11.0$: normal; $>11.0$: hypertrophied. \\

Echocardiographic LV wall thickness [cm], interventricular septal thickness, left ventricular posterior wall thickness
& $<0.6$: thinned. Male: $\leq1.0$: normal; $>1.0$ to $\leq1.3$: mildly increased; $>1.3$ to $\leq1.6$: moderately increased; $>1.6$: severely increased. Female or unknown sex: $\leq0.9$: normal; $>0.9$ to $\leq1.2$: mildly increased; $>1.2$ to $\leq1.5$: moderately increased; $>1.5$: severely increased. \\

\midrule
\multicolumn{2}{@{}l}{\textbf{CMR ventricular volumes, stroke volume, and mass}} \\
\midrule

LV end diastolic volume
& If body surface area is available, use indexed range: male $50$--$108$: normal; female $50$--$96$: normal. Otherwise use absolute range: male $95$--$215$: normal; female $78$--$167$: normal. Below range: small; above range: dilated. \\

LV end systolic volume
& If body surface area is available, use indexed range: male $11$--$47$: normal; female $10$--$40$: normal. Otherwise use absolute range: male $23$--$100$: normal; female $18$--$73$: normal. Below range: small; above range: dilated. \\

LV stroke volume
& If body surface area is available, use indexed range: male $33$--$72$: normal; female $33$--$64$: normal. Otherwise use absolute range: male $60$--$135$: normal; female $47$--$99$: normal. Below range: reduced; above range: increased. \\

LV myocardial mass
& If body surface area is available, use indexed range: male $39$--$85$: normal; female $30$--$68$: normal. Otherwise use absolute range: male $73$--$171$: normal; female $61$--$121$: normal. Below range: decreased; above range: increased. \\

RV end diastolic volume
& If body surface area is available, use indexed range: male $53$--$123$: normal; female $48$--$104$: normal. Otherwise use absolute range: male $105$--$258$: normal; female $84$--$193$: normal. Below range: small; above range: dilated. \\

RV end systolic volume
& If body surface area is available, use indexed range: male $17$--$59$: normal; female $13$--$48$: normal. Otherwise use absolute range: male $34$--$123$: normal; female $22$--$86$: normal. Below range: small; above range: dilated. \\

RV stroke volume
& If body surface area is available, use indexed range: male $28$--$75$: normal; female $29$--$66$: normal. Otherwise use absolute range: male $57$--$141$: normal; female $44$--$116$: normal. Below range: reduced; above range: increased. \\

\midrule
\multicolumn{2}{@{}l}{\textbf{CMR atrial volumes and function}} \\
\midrule

LA maximum volume
& If body surface area is available, use indexed range: male $17$--$59$: normal; female $17$--$61$: normal. Otherwise use absolute range: male $31$--$112$: normal; female $28$--$100$: normal. Below range: small; above range: dilated. \\

LA minimum volume
& If body surface area is available, use indexed range: male $3$--$24$: normal; female $4$--$23$: normal. Otherwise use absolute range: male $6$--$44$: normal; female $7$--$38$: normal. Below range: small; above range: dilated. \\

LA stroke volume
& If body surface area is available, use indexed range: male $10$--$34$: normal; female $10$--$34$: normal. Otherwise use absolute range: male $21$--$67$: normal; female $21$--$62$: normal. Below range: reduced; above range: increased. \\

LA ejection fraction
& Male $43$--$75$: normal; female $47$--$75$: normal. Below range: reduced; above range: hyperdynamic. \\

RA maximum volume
& If body surface area is available, use indexed range: male $32$--$79$: normal; female $31$--$69$: normal. Otherwise use absolute range: male $59$--$158$: normal; female $49$--$122$: normal. Below range: small; above range: dilated. \\

RA minimum volume
& If body surface area is available, use indexed range: male $10$--$42$: normal; female $8$--$32$: normal. Otherwise use absolute range: male $16$--$84$: normal; female $11$--$55$: normal. Below range: small; above range: dilated. \\

RA stroke volume
& If body surface area is available, use indexed range: male $14$--$46$: normal; female $14$--$42$: normal. Otherwise use absolute range: male $26$--$90$: normal; female $23$--$71$: normal. Below range: reduced; above range: increased. \\

RA ejection fraction
& Male $34$--$74$: normal; female $41$--$77$: normal. Below range: reduced; above range: hyperdynamic. \\

\midrule
\multicolumn{2}{@{}l}{\textbf{Global strain}} \\
\midrule

LV circumferential strain global
& Male $-27.2$ to $-14.6$: normal; female $-29.2$ to $-16.2$: normal. Value above the upper limit: reduced magnitude; value below the lower limit: increased magnitude. \\

LV longitudinal strain global
& Male $-26.1$ to $-12.7$: normal; female $-28.7$ to $-14.2$: normal. Value above the upper limit: reduced magnitude; value below the lower limit: increased magnitude. \\

LV radial strain global
& Male and female $20.0$--$55.0$: normal. Below range: reduced magnitude; above range: increased magnitude. \\

\midrule
\multicolumn{2}{@{}l}{\textbf{Aortic distensibility}} \\
\midrule

Ascending aorta distensibility
& Male age $\leq54$: $0.7$--$5.1$: normal; male age $55$--$64$: $0.0$--$4.2$: normal; male age $>64$: $0.0$--$2.4$: normal. Female age $\leq54$: $0.5$--$5.7$: normal; female age $55$--$64$: $0.0$--$3.9$: normal; female age $>64$: $0.0$--$2.7$: normal. Below range: decreased; above range: increased. \\

Descending aorta distensibility
& Male age $\leq54$: $1.6$--$6.0$: normal; male age $55$--$64$: $0.7$--$5.1$: normal; male age $>64$: $0.4$--$4.0$: normal. Female age $\leq54$: $1.6$--$6.0$: normal; female age $55$--$64$: $0.6$--$4.6$: normal; female age $>64$: $0.4$--$3.6$: normal. Below range: decreased; above range: increased. \\

\midrule
\multicolumn{2}{@{}l}{\textbf{EchoNext binary echocardiography rules}} \\
\midrule

LVEF $\leq45\%$
& Binary descriptor: yes if left ventricular ejection fraction is 45\% or lower; otherwise no. \\

LV wall thickness $\geq1.3$ cm
& Binary descriptor: yes if left ventricular wall thickness is at least 1.3 cm; otherwise no. \\

Aortic stenosis $\geq$ moderate
& Binary descriptor: yes if aortic stenosis is moderate or greater; otherwise no. \\

Aortic regurgitation $\geq$ moderate
& Binary descriptor: yes if aortic regurgitation is moderate or greater; otherwise no. \\

Mitral regurgitation $\geq$ moderate
& Binary descriptor: yes if mitral regurgitation is moderate or greater; otherwise no. \\

Tricuspid regurgitation $\geq$ moderate
& Binary descriptor: yes if tricuspid regurgitation is moderate or greater; otherwise no. \\

Pulmonary regurgitation $\geq$ moderate
& Binary descriptor: yes if pulmonary regurgitation is moderate or greater; otherwise no. \\

RV systolic dysfunction $\geq$ moderate
& Binary descriptor: yes if right ventricular systolic dysfunction is moderate or greater; otherwise no. \\

Pericardial effusion moderate or large
& Binary descriptor: yes if pericardial effusion is moderate or large; otherwise no. \\

PASP $\geq45$ mmHg
& Binary descriptor: yes if pulmonary artery systolic pressure is at least 45 mmHg; otherwise no. \\

TR maximum velocity $\geq3.2$ m/s
& Binary descriptor: yes if tricuspid regurgitation maximum velocity is at least 3.2 m/s; otherwise no. \\

\end{longtable}

\endgroup

%% file: tables/implementation_details.tex
\begin{table*}[!htbp]
\centering
\scriptsize
\setlength{\tabcolsep}{2.5pt}
\renewcommand{\arraystretch}{0.92}

\begin{minipage}[t]{0.485\textwidth}
\centering
\begin{tabularx}{\linewidth}{@{}>{\raggedright\arraybackslash}p{0.39\linewidth} X@{}}
\toprule
\rowcolor{TblHeader}
\textbf{Parameter} & \textbf{Value} \\
\midrule

Input & 12 leads $\times$ 5,000 samples \\
Patch size & $1 \times 100$ \\
ECG tokens & 300 before masking \\
Masking ratio & 0.75 \\
Loss & MSE over masked patches \\
Encoder & ViT, 12 layers, hidden size 768, 12 heads \\
Decoder & Transformer, 2 layers, hidden size 512, 8 heads \\
Activation & GELU \\
Positional embedding & Learnable \\
Drop-path rate & 0.1 \\
Optimiser & AdamW \\
Learning rate & $2 \times 10^{-4}$ \\
Weight decay & 0.05 \\
Batch size & 768 \\
Precision & bfloat16 mixed precision \\
Training duration & 120 epochs \\
Warm-up & 1{,}000 steps \\
Minimum LR factor & 0.01 \\
Augmentations & Random crop, Fourier surrogate perturbation, jitter, rescaling \\

\bottomrule
\end{tabularx}
\caption{ECG ViT-MAE pretraining configuration.}
\label{tab:supp_ecg_mae_pretraining_config}
\end{minipage}
\hfill
\begin{minipage}[t]{0.485\textwidth}
\centering
\begin{tabularx}{\linewidth}{@{}>{\raggedright\arraybackslash}p{0.39\linewidth} X@{}}
\toprule
\rowcolor{TblHeader}
\textbf{Parameter} & \textbf{Value} \\
\midrule

ECG encoder & Pretrained ViT-MAE, frozen \\
Language model & Llama-3.1-8B-Instruct, frozen backbone \\
ECG crop & Random crop to 2{,}500 time-steps \\
ECG augmentations & Disabled \\
Projector & Linear$(d_{\mathrm{ecg}}, d_{\mathrm{llm}})$, SiLU, Linear$(d_{\mathrm{llm}}, d_{\mathrm{llm}})$ \\
Aggregation & Identity; all ECG tokens preserved \\
Projector bias & Disabled \\
LoRA modules & \texttt{q\_proj}, \texttt{k\_proj}, \texttt{v\_proj}, \texttt{o\_proj} \\
LoRA rank / alpha & 128 / 128 \\
LoRA dropout & 0.1 \\
Curriculum & First 1{,}500 steps update only the projector, then projector + LoRA training \\
Optimiser & Paged AdamW 8-bit \\
Learning rates & Projector $10^{-4}$; LoRA $10^{-5}$ \\
Weight decay & $10^{-3}$ \\
Schedule & Linear warm-up from $0.1\times$ to peak LR over 3\% of optimiser steps, then cosine decay to $0.3\times$ peak LR \\
Precision & bfloat16 mixed precision \\
Training duration & 1 epoch \\
Batching & 4 GPUs, batch size 8/GPU, gradient accumulation 4 \\
Effective batch size & 128 \\
Max sequence length & 1200 tokens \\

\bottomrule
\end{tabularx}
\caption{ECG-LLM instruction-tuning configuration.}
\label{tab:supp_ecg_llm_instruction_config}
\end{minipage}

\end{table*}

%% file: tables/supp_diagnostic_report_questions.tex
\begin{table*}[!htbp]
\centering
\scriptsize
\setlength{\tabcolsep}{5pt}
\renewcommand{\arraystretch}{1.08}
\begin{tabularx}{0.96\textwidth}{>{\raggedright\arraybackslash}p{0.16\textwidth} X}
\toprule
\rowcolor{TblHeader}
\textbf{Model} & \textbf{Diagnostic report generation prompt} \\
\midrule
\multirow{8}{*}{ECG-LLM} & Review the ECG signal and produce a detailed report on your diagnostic observations, ending with the final diagnosis. \\
& Describe this ECG in detail and tell if there are any abnormalities present. \\
& Review this ECG and provide an ECG diagnosis. \\
& What is wrong with this ECG? \\
& Provide an ECG report for this tracing. \\
& What are the key diagnostic findings in this ECG? \\
& Analyze this ECG carefully and provide a detailed ECG report. \\
& What is the summary for this ECG? \\
\midrule
ECG-Chat & Could you please help me explain my ECG? \\
PULSE-7B & Please write a clinical report based on this ECG image. \\
MedGemma-4B & Please interpret this 12-lead ECG image. Provide a concise diagnostic report, including rhythm, rate if assessable, conduction or axis abnormalities if present, ST-T or infarction-related findings if present, and an overall ECG impression. If a finding is uncertain or not clearly visible, say so rather than inventing it. \\
\bottomrule
\end{tabularx}
\caption{
Prompts used for ECG diagnostic summary generation.
For ECG-LLM, one question was randomly selected from the ECG-LLM prompt pool for each ECG.
For baseline models, the model-specific prompt shown in the table was used.
}
\label{tab:supp_diagnostic_report_questions}
\end{table*}